\documentclass[12pt]{article}
\usepackage[hyperfootnotes=false]{hyperref}
\usepackage{epsfig}
\usepackage{amsmath}
\usepackage{amssymb}
\usepackage{caption}

\usepackage{graphicx}
\usepackage{bm}
\usepackage{amsmath}
\usepackage{amssymb}
\usepackage{amsfonts}
\usepackage{mathtools}
\usepackage{braket}
\usepackage{anyfontsize}
\usepackage[normalem]{ulem}
\usepackage{wrapfig}
\usepackage[left=1in,right=1in,top=1in,bottom=1in]{geometry}

\usepackage[ascii]{inputenc}
\usepackage{bbm,braket,microtype,mathtools}
\usepackage[T1]{fontenc}
\usepackage{lmodern}
\usepackage[USenglish]{babel}

\DeclareMathOperator{\tr}{\text{tr}}

\hypersetup{colorlinks=true,urlcolor=[rgb]{0,0,0.5},citecolor=[rgb]{0.5,0,0},linkcolor=[rgb]{0,0,0.4}}

\renewcommand{\title}[1]{\vbox{\center\bf{\Large{#1}}}\vspace{5mm}}
\renewcommand{\author}[1]{\vbox{\center#1}\vspace{5mm}}
\newcommand{\address}[1]{\vbox{\center\em#1}}

\def\tr{{\rm tr}}

\def\C{\mathbb{C}}

\def\1den{\hbox{$1\hskip -1.2pt\vrule depth 0pt height 1.53ex width 0.7pt
                  \vrule depth 0pt height 0.3pt width 0.12em$}}

\def\and{\quad {\rm and} \quad}

\def\bea{\begin{eqnarray}}
\def\eea{\end{eqnarray}}

\begin{document}

\begin{titlepage}

\begin{center}
\vspace*{2cm}
\title{{\fontsize{0.75cm}{0.75cm}\selectfont Quantum Causal Influence}}
\author{Jordan Cotler, Xizhi Han, Xiao-Liang Qi, and Zhao Yang}
\address{{\fontsize{0.4cm}{.4cm}\selectfont
Stanford Institute for Theoretical Physics,\\ Stanford University, Stanford, California 94305, USA
}}


\end{center}

\begin{abstract}
We introduce a framework to study the emergence of time and causal structure in quantum many-body systems.  In doing so, we consider quantum states which encode spacetime dynamics, and develop information theoretic tools to extract the causal relationships between putative spacetime subsystems.  Our analysis reveals a quantum generalization of the thermodynamic arrow of time and begins to explore the roles of entanglement, scrambling and quantum error correction in the emergence of spacetime.  For instance, exotic causal relationships can arise due to dynamically induced quantum error correction in spacetime: there can exist a spatial region in the past which does not causally influence any small spatial regions in the future, but yet it causally influences the union of several small spatial regions in the future.  We provide examples of quantum causal influence in Hamiltonian evolution, quantum error correction codes, quantum teleportation, holographic tensor networks, the final state projection model of black holes, and many other systems.  We find that the quantum causal influence provides a unifying perspective on spacetime correlations in these seemingly distinct settings.  In addition, we prove a variety of general structural results and discuss the relation of quantum causal influence to spacetime quantum entropies.
\end{abstract}

\end{titlepage}

\tableofcontents
\newpage

\section{Introduction}

Causal structure is an essential property of spacetime geometry. In relativistic classical mechanics, the causal structure is determined by the behavior of null geodesics. The future light cone of a point $x$ comprises all of the points that may be influenced by an arbitrary perturbation at $x$.  In relativistic quantum field theory, we usually treat the causal structure as classical, with well-defined light cones. In more general quantum many-body systems which may be non-relativistic or do not posses quasiparticles resembling massless excitations, there is still a generalization of the causal structure so long as there is an upper-bound on the speed of information propagation. For example, for lattice models with a local Hamiltonian, the Lieb-Robinson bound \cite{lieb1972finite} gives a velocity $v_{LR}$ which defines an analog of the speed of light.  In particular, a local perturbation can only influence the region inside its future Lieb-Robinson cone. 

\begin{figure}[t]
\center
\includegraphics[width=6in]{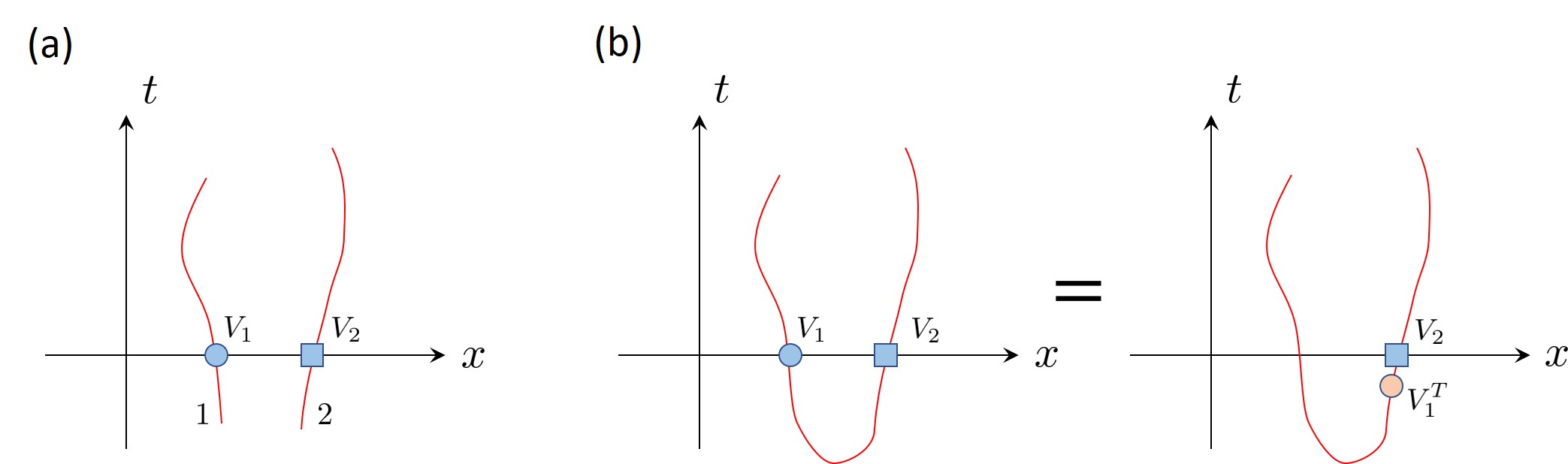}
\caption{(a) The world lines of two spin-$\frac12$ particles $1,2$ in spacetime (red curves). Two operators $V_1$ and $V_2$ probe the spins of the two particles at time $t=0$. (b) When the initial state of the spins of the two particles forms an EPR pair, the effect of $V_1\otimes V_2$ on particles $1$ and $2$ is equivalent to applying $V_2V_1^T$ to particle $2$ alone. \label{fig:simplest}}
\end{figure}

However, beyond these familiar cases, the causal structure in quantum mechanics can be much richer. As a simple example, consider two spin-$\frac12$ particles $1,2$ in Figure \ref{fig:simplest} above. At time $t=0$, particles $1$ and $2$ are at location $x_1$ and $x_2$. On a fixed time slice $t=0$, suppose we probe the spin degrees of freedom of particles $1$ and $2$ with separate Hermitian operators $V_1$ and $V_2$, respectively.  These two probe events are clearly spacelike separated.  Now if we prepare the spin degrees of freedom of particles $1$ and $2$ in an EPR pair state $\frac1{\sqrt{2}}\left(\ket{\uparrow}_1\ket{\uparrow}_2+\ket{\downarrow}_1\ket{\downarrow}_2\right)$ at an earlier time $t_i<0$, applying $V_1$ and $V_2$ to particles $1$ and $2$ at time $t=0$ is equivalent to applying $V_2V_1^T$ only to particle $2$ at $t=0$. ($V_1^T$ is an operator defined by the matrix transpose of $V_1$ in the $S_z$ basis.) Therefore, for our particular initial state of the spin degrees of freedom, it becomes ambiguous whether the two probe events are spacelike or time-like separated.

Following the spirit of Einstein's theory of relativity, one would like an {\it observable} way to define the causal relation between events in a quantum many-body system, which is uniquely determined by physical correlation functions and has an unambiguous interpretation. This is the goal of the current paper. We propose a measure of quantum causal influence that determines whether a spacetime region $A$ has nontrivial influence on another spacetime region $B$.  The measure reproduces the ordinary causal structure for the familiar case of relativistic classical systems, but also unveils various unconventional causal structures that are unique to quantum mechanics.

Our emphasis on correlation functions and many-body states differs from previous work on causality in quantum mechanics which emphasize few-body systems and causal inference on data from decoherent measurements \cite{oreshkov2012quantum, brukner2014quantum, aharonov2014each, fitzsimons2015quantum, ried2015quantum, pienaar2015graph, brukner2015bounding, costa2016quantum, oreshkov2016causal, ringbauer2016experimental, allen2017quantum, maclean2017quantum, castro2018dynamics}.  We are primarily interested in the emergence of causal structure in quantum systems with many degrees of freedom, and the flow of time experienced by observers inside the systems.  For related work in this direction using the quantum process tensor formalism and related formalisms, see \cite{oreshkov2015operational, oreshkov2016operational, hardy2012operator, hardy2016operational, jia2017generalizing, jia2018tensor, jia2018quantum}.

\begin{figure}[t]
\center
\includegraphics[width=3in]{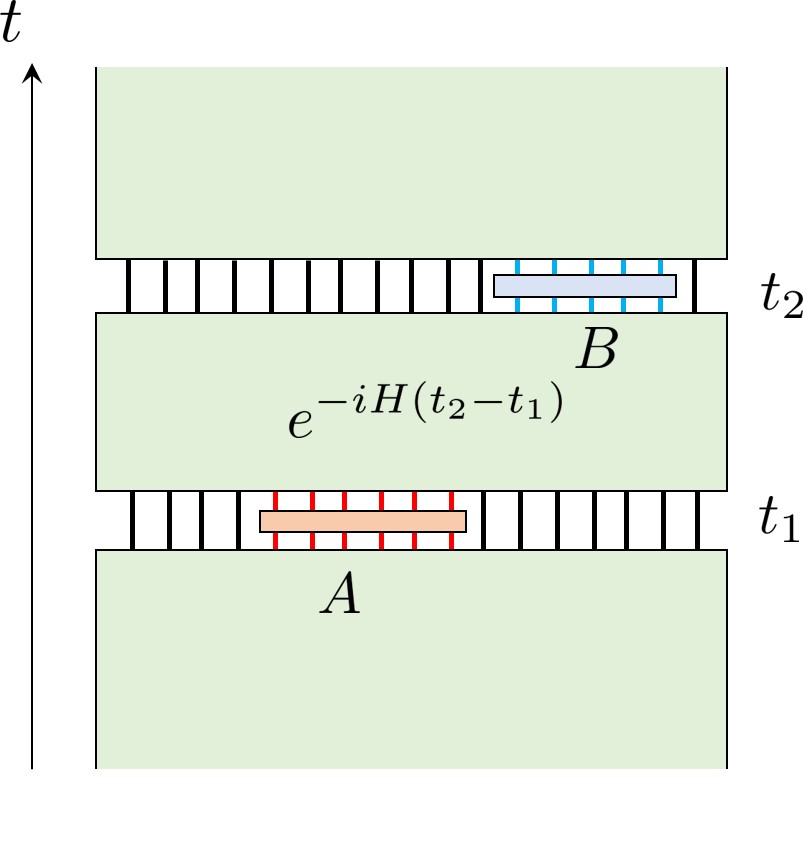}
\caption{\label{fig:1d} Depiction of regions $A$ at time $t_1$ and $B$ at time $t_2$ for a spin chain. The causal influence is measured by inserting a unitary operator $U_A$ in region $A$ (orange box) and studying its effect on the measurement of an arbitrary operator $O_B$ in $B$ (blue box). }
\end{figure}

To illustrate the idea of our proposal, let us first consider time evolution with a local Hamiltonian. For concreteness, we can consider a $(1+1)$-dimensional model of $N$ spins labeled by $x=1,2,...,N$ with a Hamiltonian that couples neighboring spins.  We will refer to this system as the ``main system.''  Starting from an initial state $\ket{\psi_i}$ at time $t=0$, the main system evolves as $\ket{\psi(t)}=e^{-iHt}\ket{\psi_i}$ in the Schr\"{o}dinger picture. Consider two spatial regions, $A$ at time $t_1$ and $B$ at time $t_2$, as shown in Figure \ref{fig:1d}. Now suppose there is an experimentalist who can only access the two spacetime regions $A$ and $B$, but can otherwise perform arbitrary operations.
In particular, the experimentalist is a superobserver who can couple her external apparatus to region $A$ by performing a joint unitary on $A$ and her apparatus at time $t_1$, and similarly for $B$ at time $t_2$. We also assume that the experimentalist has the ability to reset the whole system to the initial state $\ket{\psi_i}$ and run the experiment an unlimited number of times. Now the question is, how can the experimentalist determine whether physical operations in $A$ causally influence the region $B$? Naturally, the experimentalist can run different experiments with different perturbations on region $A$ (by coupling her external apparatus to $A$ in different ways) and measure some physical quantity at region $B$. If the result of the measurement at $B$ depends nontrivially on the pertubation at $A$, we conclude that $A$ causally influences region $B$.

However, it is important to distinguish causal influence and correlation. Even if $A$ and $B$ are spacelike separated, operators in $A$ and $B$ can certainly have a nontrivial connected correlation function. Measures of connected correlation (such as the quantum mutual information between the two regions, if they are spacelike separated) are symmetric between the two regions, and thus do not probe the causal structure.  For instance, it may be the case that $A$ causally influences $B$ but $B$ does not causally influence $A$, and so causal influence is necessarily an asymmetric relation. It turns out that a simple modification of the setup can distinguish causal influence from other kinds of correlation. The experimentalist can apply a unitary gate $U_A$ to region $A$, which changes the state of the system but does not introduce entanglement with her apparatus. Then, the experimentalist can couple her apparatus to region $B$ in the ordinary way, which generically entangles the main system with the apparatus. If $B$ has no overlap with the future light cone (or for a lattice model, the Lieb-Robinson cone) of $A$, the unitary operator $U_A$ does not change the reduced density matrix of $B$ and therefore does not change any physical property there. 

The procedure described above may sound a bit trivial since it is exactly how we do response theory in many-body systems. If we consider an infinitesimal unitary $U_A=\exp\left(-i\,\epsilon J_A\right)$, and measure an operator $J_B$ at $B$, the linear response function is determined by the commutator $-i\left[J_A(t_1),J_B(t_2)\right]\theta(t_2-t_1)$, which vanishes outside the light cone. However, the commutator expression depends on the Heisenberg picture, which relies on picking a choice of time slicing (i.e., Cauchy surfaces). Since we want a measure of the causal structure that is not predicated on pre-defined time slices, it is more natural to work with tensor networks, which are not endowed with a pre-defined causal structure.  Indeed, our proposal allows us to study causal structure in systems with no obvious time slicings.  For example, in a hyperbolic ``perfect tensor network'' \cite{pastawski2015holographic}, there are isometry relations between operators acting on different subsets of links, but there is no light cone or preferred time-like direction. Our proposal allows us to start from scratch and probe causal influence between different degrees of freedom in the system, without any {\it a priori} knowledge of a time direction. In particular, there is no need to distinguish whether some qubits (or more generally, degrees of freedom) in $A$ and $B$ are ``the same qubits evolved in time'' or ``independent qubits that are entangled.'' 

The remainder of the paper is organized as follows: We start by presenting the general setup. For concreteness, we use the  language of tensor networks to describe a general quantum system, \textit{without} needing to designate how degrees of freedom sit in a putative spacetime.  This is a very useful framework for ``spacetime agnostic'' descriptions of quantum systems.
Even if we have a continuum of degrees of freedom, as long as we assume that accessible regions $A$ and $B$ comprise of discrete spacetime points, the system can be described by a tensor network. We show how a general quantum system can be considered as a tensor network with insertions of operators in links, and with a given boundary condition. For example, in the more familiar setting of a quantum system with unitary time evolution, the boundary conditions of the tensor network correspond to an initial density operator (i.e., an initial state) and optionally a final density operator (i.e., a final state).  Ordinary quantum mechanics without a final state density operator is equivalent to having a \textit{maximally mixed} final state density operator.  We will discuss this in detail later. 

Next, we provide the definition of quantum causal influence in the general setup.  With this probe of quantum causal structure at hand, we investigate various examples and identify some key features of causal influence that are unique to quantum systems. One feature is that the causal structure generically depends on the initial state, or more generally the boundary conditions of the tensor network. In the familiar case of a quantum system with unitary time evolution, the direction of the ``future'' is determined by the fact that the final state is maximally mixed but the initial state is not.  If the initial state contains a region with a maximally mixed reduced density operator, the future light cone of points in the domain of dependence of that region will be ``erased.''  Another example of causal structure which is sensitive to the initial state is quantum teleportation. We show how quantum teleportation corresponds to ``erasing'' part of the future light cone of the teleportee due to a special initial state containing EPR pairs.

The other unique feature of quantum causal influence is that it is generically nonlocal.  In classical mechanics, causal structure is determined by the causal relationships of pairs of points.  Classically, a spacetime region $B$ is influenced by a spacetime region $A$ if and only if some points in $B$ are in the future light cone of some points in $A$.  This is not the case for quantum systems. To fully understand the quantum causal structure of a system, it is essential to consider the influence between regions $A,B$ of generic size.  In fact, the quantum causal influence between subsystems of $A$ and $B$ do not generically determine the quantum causal influence between $A$ and $B$ themselves.  For instance, it is possible to have smaller regions $B_1$ and $B_2$ which are not \textit{individually} influenced by $A$, but for which the union $B_1\cup B_2$ is influenced by $A$. Such nonlocal influence is a key feature of quantum erasure codes. The encoding map of a quantum erasure code takes quantum information in a region $A$ and maps it to $B=B_1\cup B_2$ nonlocally. If the influence of $A$ to each subregion $B_1$, $B_2$ is trivial, the code is immune to local errors that occur in only one of $B_1$ or $B_2$.

The nonlocality of quantum causal influence provides a new perspective on the exotic causal structures underlying holographic duality. In holographic tensor networks such as perfect tensor networks or large bond dimension random tensor networks \cite{hayden2016holographic}, all pairs of small regions appear ``spacelike separated'' since no small region influences any other small region.  However, a small region (or more precisely, code subspace operators in a small bulk region) can influence \textit{large} regions and ultimately influence the boundary in a nonlocal way, as is required by the reconstruction of bulk operators on the boundary.  Using quantum causal influence, we find that holographic tensor networks can admit exotic quantum analogs of Cauchy slices comprising of concentric spheres.  Another example we study is the final state projection model of black holes \cite{horowitz2004black}, which utilizes post-selected quantum mechanics. We discuss how causal influence between small regions does not know about a post-selected random final state, while regions that are large enough have abnormal causal relations and do detect the violation of unitarity by the final state.

After discussing various features and examples of quantum causal influence, we turn to some more quantitative properties.  We define a ``superdensity operator'' \cite{cotler2017superdensity} of regions $A,B$ which determines all correlation functions involving these two regions. With this tool, we investigate the averaged quantum causal influence by averaging over unitaries in $A$ and generic operators in $B$. The averaged causal influence is a quantum information theoretic property of the superdensity operator. As two examples, we numerically computed the averaged causal influence in quantum Ising spin chains and stabilizer code models.

We find that quantum causal influence provides a new probe of many-body chaos since the influence between two small regions decays in a chaotic system even if the regions are time-like separated. This is a consequence of operator scrambling and thermalization -- a local perturbation becomes non-local and at a later time has little effect on local regions except by contributing to conserved quantities such as energy.  We also discuss an upper bound of the causal influence by spacetime quantum mutual information (which is again defined for the superdensity operator) \cite{cotler2017superdensity}. Finally, we discuss some open questions and future directions.

Below is a brief summary, section by section:
\begin{itemize}
\item
In Section \ref{sec:setup}, we provide definitions of general tensor networks, graphical tensor networks,
and quantum causal influence.
\item 
In Section \ref{sec:BCdependence}, we explore how quantum causal influence depends on boundary conditions.
We provide many examples, and prove general, structural results.
\item 
In Section \ref{sec:nonlocalityQCI}, we discuss the nonlocality of quantum causal influence in the context of
quantum error correction codes, scrambling, and quantum teleportation.
\item 
In Section \ref{sec:QGexamples}, we give examples in the context of quantum gravity, specifically for holographic tensor networks and models
of a black hole final state.
\item
In Section \ref{sec:measures}, we establish the relationship between the averaged quantum causal
influence and spacetime quantum entropies and mutual information.  We use our results to analyze quantum causal influence in quantum spin chains and stabilizer tensor networks.
\item 
In Section \ref{sec:conclusion}, we make concluding remarks and discuss future directions.
\item
In the Appendices, we provide classical and quantum generalizations of causal influence,
review the superdensity operator formalism, and also review stabilizer tensor networks.
\end{itemize}

\section{General Setup}\label{sec:setup}

\subsection{General tensor networks}

\begin{figure}[t]
\center
\includegraphics[width=6.5in]{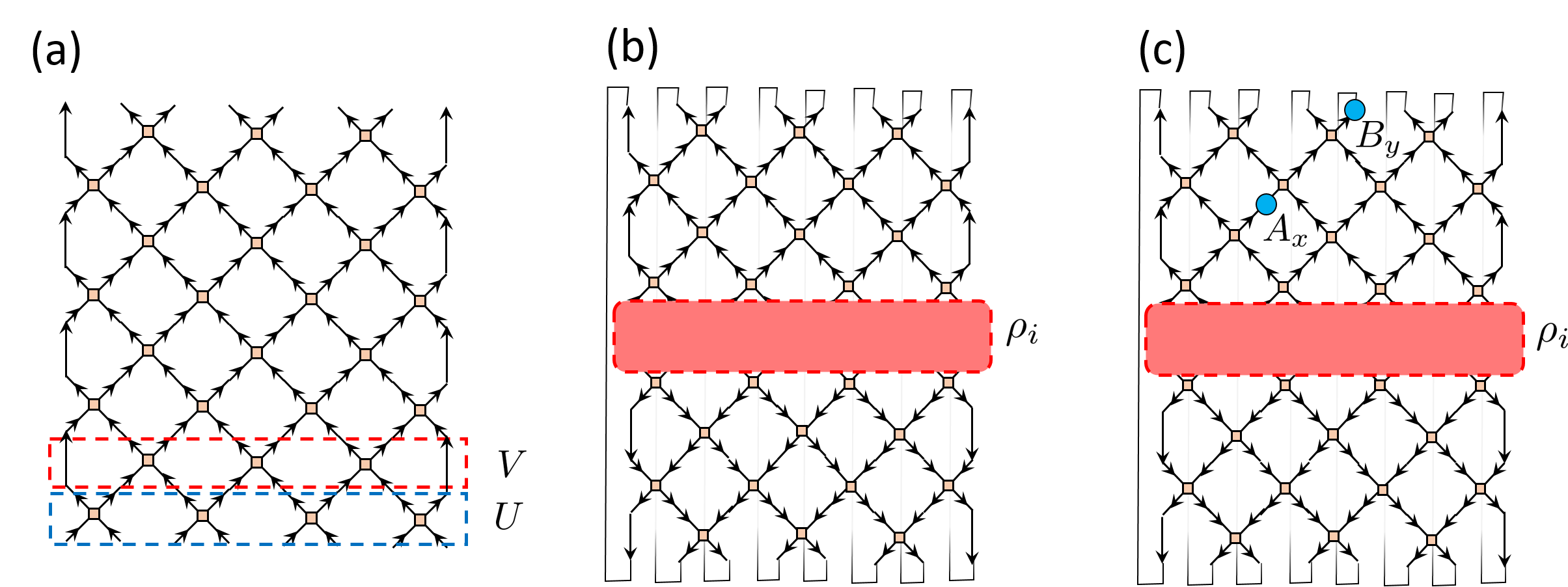}
\caption{(a) An example of a tensor network describing a unitary operator $W = (VU)^M$. Each vertex is a two-qubit unitary gate with the inputs and outputs indicated by arrows pointing toward or away from the vertex, respectively. (b) This is the tensor network obtained by contracting $W$ and $W^\dagger$ with an initial state $\rho_i$ (the red box), and then taking a trace.  In other words, the tensor network computes $\text{tr}(W \rho_i W^\dagger) = \text{tr}(\rho_i) = 1$. (c) The tensor network representation of a two-point function defined in Eqn.~\eqref{twoptfunc}.
\label{fig:Trotter0}}
\end{figure}

In order to define characteristics of quantum causal structure, we need to start from a description of a quantum many-body system that does not pick out a time direction.  A suitable framework is general tensor networks \cite{vidal2003efficient,verstraete2004renormalization,levin2007tensor,cotler2017superdensity, qi2018space}.  Even though popular examples of tensor networks often have a constrained form, the framework of general tensor networks is far broader and encompasses the entire scope of familiar (and unfamiliar) quantum many-body systems.

We start from a simple example of a tensor network, before providing the most general definition. Consider $N$ qubits, where $N$ is even, arranged in a line. First, we apply in parallel two-qubit gates to adjacent qubits via the unitary $U=U_{12}\otimes U_{34}\otimes \cdots \otimes U_{N-1,N}$.  Next, we apply another unitary on a different pairing of adjacent qubits, namely $V=V_{23}\otimes V_{45}\otimes \cdots \otimes V_{N-2,N-1}$.  Afterwards, we again apply $U$ followed by $V$, and so on a total of $M$ times, as illustrated in Figure \ref{fig:Trotter0}(a). This procedure yields the unitary operator $W=(VU)^M$.  The discrete time evolution implemented by sequential applications of $U$ and $V$ can be considered as a discretization of a continuous time evolution operator $e^{-iHt}$ where $H$ is a local Hamiltonian.  Indeed, we can find $U$ and $V$ via a Suzuki-Trotter decomposition of $e^{-iHt}$. 

Mathematically, the matrix $W$ is obtained by contracting indices of small matrices $U_{2k-1,2k}$ and $V_{2k,2k+1}$ along all internal links of the network in Figure \ref{fig:Trotter0}(a). We can then contract $W$ and a $W^\dagger$ with some initial state $\rho_i$, and then take a trace.  This yields the tensor network in Figure \ref{fig:Trotter0}(b), which computes $\text{tr}(W \rho_i W^\dagger) = \text{tr}(\rho_i) = 1$.  The tensor network is a discrete analog of a partition function, which can be used to compute physical correlation functions. For example, the time-ordered two-point function
\bea
\langle \mathcal{T} \,B_y(t_2)A_x(t_1)\rangle={\rm tr}\left[B_y(VU)^{t_2-t_1}A_x(VU)^{t_1}\rho_i (U^\dagger V^\dagger)^{t_2}\right]\,,\label{twoptfunc}
\eea
where for concreteness we suppose $t_2 > t_1$, can be computed from the tensor network in Figure \ref{fig:Trotter0}(b) by inserting the operators $A_x, B_y$ into links corresponding to $x$ and $y$ which yields the tensor network Figure \ref{fig:Trotter0}(c).  Indeed, the tensor network in Figure \ref{fig:Trotter0}(c) evaluates to the two-point function in Eqn.~\eqref{twoptfunc} above.

\begin{figure}[t]
\center
\includegraphics[width=2.5in]{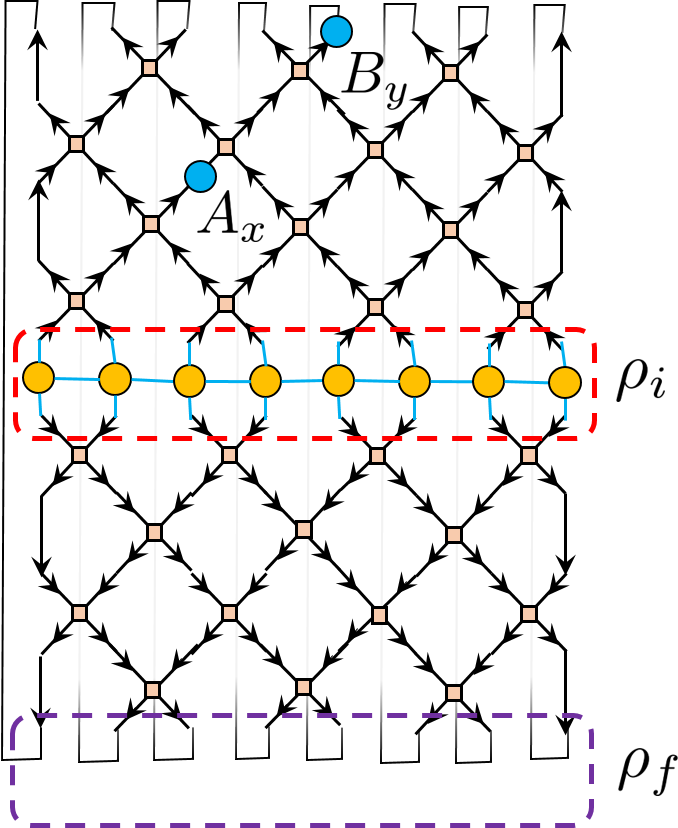}
\caption{This tensor network is a special case of the one in Figure \ref{fig:Trotter0}(c).  The network here specifies a particular choice of $\rho_i$, namely a matrix product operator (MPO), which is depicted within the dashed red lines.  We have also put in purple dashed lines to illustrate the fact that taking a trace is equivalent to taking an inner product with a maximally mixed density matrix $\rho_f= \textbf{1}/d$, up to a normalization $d$ (i.e., the Hilbert space dimension of a spatial slice). \label{fig:TrotterParticular}}
\end{figure}

For concreteness, in Figure \ref{fig:TrotterParticular} we have chosen an initial density matrix $\rho_i$ which is a matrix product operator (MPO).  We will not use MPO's later in the paper, but it suffices to say that the state $\rho_i$ is represented by the partially contracted tensors in the red dashed box in Figure \ref{fig:TrotterParticular}. The tensor network representation of Figure \ref{fig:Trotter0}(b) also highlights the fact that taking the trace in Eqn.~\eqref{twoptfunc} is, up to a normalization, equivalent to taking an inner product with another density matrix $\rho_f = \textbf{1}/d$, which is the maximal entropy state on a spatial slice (here, we suppose that the Hilbert space dimension of a spatial slice is $d$).  This is just to say that correlation functions, such as the two-point function in Eqn.~\eqref{twoptfunc}, can be written as
\bea
{\rm tr}\left[\rho_f B_y(VU)^{t_2-t_1}A_x(VU)^{t_1}\rho_i (U^\dagger V^\dagger)^{t_2}\right] \propto {\rm tr}\left[B_y(VU)^{t_2-t_1}A_x(VU)^{t_1}\rho_i (U^\dagger V^\dagger)^{t_2}\right]\label{twoptfunc2}
\eea
since $\rho_f$ is proportional to the identity.  Although this may seem like a trivial rewriting, we will see later that it is significant.

By making both $\rho_i$ and $\rho_f$ explicit, we see that $\rho_i$ and $\rho_f$ play symmetric roles.  More general tensor networks with insertions on links provide a powerful framework for describing physical processes of quantum many-body systems.  Much like a partition function, a tensor network is an object into which operators can be inserted to compute correlation functions.  However, partition functions require a Hamiltonian or action that implicitly or explicitly specifies spatial and temporal degrees of freedom.  For instance, Hamiltonians and actions specify dynamical degrees of freedom such as spins, particles or fields, and designate both spatial and temporal coordinates.  By contrast, a tensor network is a completely general contraction of quantum operators which is a priori agnostic to distinctions of space and time. 

Going back to our example, we have so far viewed the network in Figure \ref{fig:Trotter0}(b) as an initial state with unitary time evolution vertically and two operator insertions at $A_x, B_y$.  However, the tensor network is agnostic to the words we use to describe it: we could instead equivalently say that the tensor network implements non-unitary evolution \textit{horizontally}, and that what we formerly called \textit{spatial} open boundary conditions correspond here to \textit{temporal} boundary conditions (such as initial and final states).  From this perspective, $\rho_i$ and $\rho_f$ now play the role of \textit{spatial} boundary conditions.  Also from this point of view, the operator insertions $A_x, B_y$ compute a two-point correlation function in a different physical system.

This example may seem somewhat contrived, since we intuitively know that viewing the tensor network as implementing evolution vertically yields the familiar form of unitary time evolution, whereas viewing the tensor network as implementing evolution horizontally leads to peculiar non-unitary evolution.  Thinking carefully about this distinction, we might ask: what precisely makes the ``vertical'' point of view more natural than the ``horizontal'' point of view, for this example?  More generally, we may have a tensor network that does not have an obvious causal structure.  So then we may ask, how do we diagnose the causal structure of a general tensor network?  Which tensor networks yield familiar causal structures, either exactly or approximately?  Are there new kinds of causal structures which are natural but specific to quantum systems?  These are the questions which we begin to study in this paper.

Now, let us give the most general definition of a tensor network:  \\ \\
\noindent \textbf{Definition (general tensor network):} \textit{A tensor network is specified by a triple }$\{\{\mathcal{H}_i\}, |L\rangle, \rho_P\}$ \textit{comprised of:}
\begin{enumerate}
\item \textit{A set of Hilbert spaces} $\{\mathcal{H}_i\}$ \textit{which each correspond to a spacetime subsystem} $i$\,,
\item \textit{A link state} $|L\rangle \in \mathcal{H} = \bigotimes_i \mathcal{H}_i$\,,
\item \textit{A density operator} $\rho_P$ \textit{acting on the same Hilbert space} $\mathcal{H}$.
\end{enumerate}
\textit{The most general correlation function of the tensor network is computed by} $\langle L | Q_1 \, \rho_P \, Q_2 |L\rangle$ \textit{where} $Q_1, Q_2$ \textit{are operators acting on} $\mathcal{H}$.
\\ \\
In other words, a general tensor network is like a quantum many-body state given by $|L\rangle$, except that the inner product is defined by positive semi-definite quadratic form $\rho_P$ instead of the ordinary inner product in the Hilbert space.  Furthermore, a general tensor network can encode correlations in time, since we regard each tensor factor $\mathcal{H}_i$ as a subsystem in spacetime.  For instance, if our tensor network described standard unitary time evolution, the contracted tensor network would have unitary time evolution operators connecting subsystems corresponding to adjacent times.

\subsection{Tensor networks based on graphs}
\label{sec:GraphTensorNetworks}
Here we explain a useful type of tensor network, called a \textit{graphical tensor network} (GTN).  We will utilize GTN's throughout the paper.  A GTN is defined for an undirected graph $G = (V,E)$ where $V$ is the set of vertices and $E$ is the set of edges.  For a given vertex $v$, let $\text{deg}(v)$ (i.e., the degree of $v$) denote the number of edges which attach to it.  The GTN corresponding to $G$ has a Hilbert space
\begin{equation}
\mathcal{H} = \bigotimes_{v \in V} \mathcal{H}_v
\end{equation}
where $\mathcal{H}_v \simeq (\mathbb{C}^d)^{\otimes \text{deg}(v)}$.  In words, each Hilbert space $\mathcal{H}_v$ corresponding to a vertex $v$ comprises of $\text{deg}(v)$ tensored copies of $\mathbb{C}^d$, also known as $\text{deg}(v)$ qudits.\footnote{A qudit is a $d$-level system (hence qu\textit{d}it), whereas a qubit is a $2$-level system.}  It will be convenient to write the full Hilbert space as
\begin{equation}
\label{GTNfullHilbert1}
\mathcal{H} = \bigotimes_{v \in V} \bigotimes_{j=1}^{\text{deg}(v)} \mathcal{H}_{v_j}
\end{equation}
where $\mathcal{H}_{v_j} \simeq \mathbb{C}^d$, and $v_j$ denotes the $j$th qudit of $\mathcal{H}_v$\,.

\begin{figure}[t]
\center
\includegraphics[width=6in]{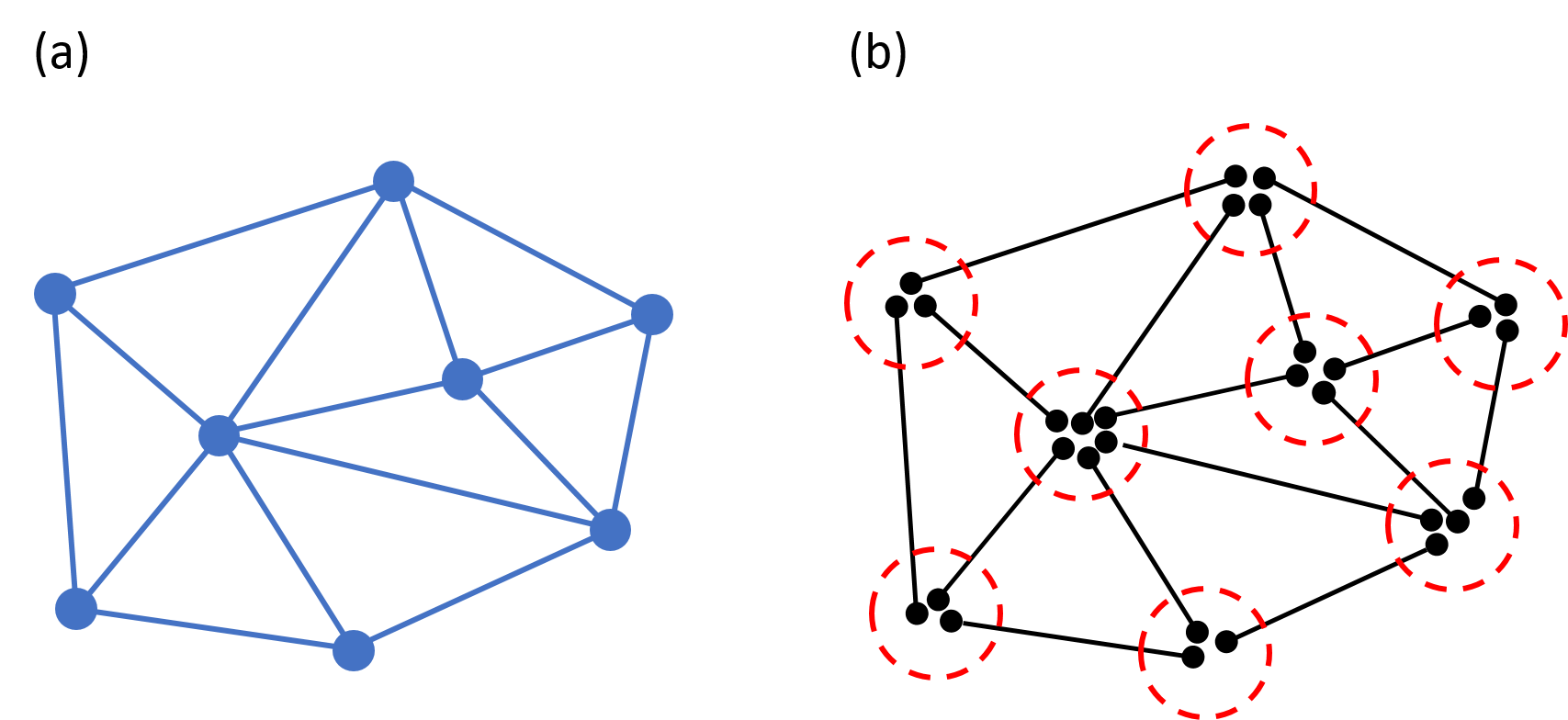}
\caption{(a) A graph $G = (V,E)$ is shown in blue. (b) A representation of the link state $|L\rangle$.  Each line with a dot at each end represents an EPR pair, with the dots corresponding to qudits. The dotted red circles designate the collections of qudits corresponding to vertices $v$ of the graph $G$.  The number of qudits at a vertex of the graph is the same as the degree of that vertex.  \label{fig:graph1}}
\end{figure}

Then $|L\rangle$ is a ``link state'' comprised of a tensor product of EPR pair states as follows.  (The explanation of the construction of $|L\rangle$ is slightly involved, but has a simple pictographic interpretation given in Figure \ref{fig:graph1} above).  Let us denote by $(v,w)$ an edge $e$ of the graph which connects the vertices $v$ and $w$.  Since our graph $G$ is undirected, $(v,w)$ is an unordered pair.  Now we define a function $f$ which assigns a pair of qudits to each edge $e$.  The function $f$ has two properties:
\begin{enumerate}
\item $f\left((v,w)\right) = \{v_m, w_n\}$ for some $m,n$ with $1 \leq m \leq \text{deg}(v)$ and $1 \leq n \leq \text{deg}(w)$.  In words, in this case $f$ assigns $(v,w)$ to the $m$th qudit of $\mathcal{H}_v$ and the $n$th qudit of $\mathcal{H}_w$.
\item For every pair of distinct edges $e,e'$, we have $f(e) \cap f(e') = \emptyset$.  In words, $f$ assigns to each edge $e$ a unique pair of qudits which does not intersect with the qudits assigned to any other edge.  
\end{enumerate}
Let $|\text{EPR}_{v_m w_n}\rangle$ denote some EPR state, say $\frac{1}{\sqrt{d}}\sum_{i=1}^d |i\rangle |i\rangle$, between the $m$th qudit of $\mathcal{H}_v$ and the $n$th qudit of $\mathcal{H}_w$.  Then $|L\rangle$ is given by
\begin{equation}
|L\rangle = \bigotimes_{e \in E} |\text{EPR}_{f(e)}\rangle\,.
\end{equation}

For clarity, consider the graph in Figure \ref{fig:graph1}(a) above.  Then we can visualize $|L\rangle$ by EPR pairs organized as in Figure \ref{fig:graph1}(b) above.  Indeed, we can imagine that the edges of the graph have been ``replaced'' by EPR pairs.  Finally, the state $\rho_P$ has the structure
\begin{equation}
\label{rhoP1}
\rho_P = \bigotimes_{v \in V} P_v 
\end{equation}
where $P_v$ is a projector on $\mathcal{H}_v$.  Hence, $\rho_P$ is furnished with a subscript $P$ (for ``projector'').  In some graph-based tensor networks, $\rho_P$ is not restricted to comprise of a tensor product of projectors, and can instead be any density matrix on $\bigotimes_{v \in V} \mathcal{H}_v$.

\begin{figure}[t]
\center
\includegraphics[width=6.5in]{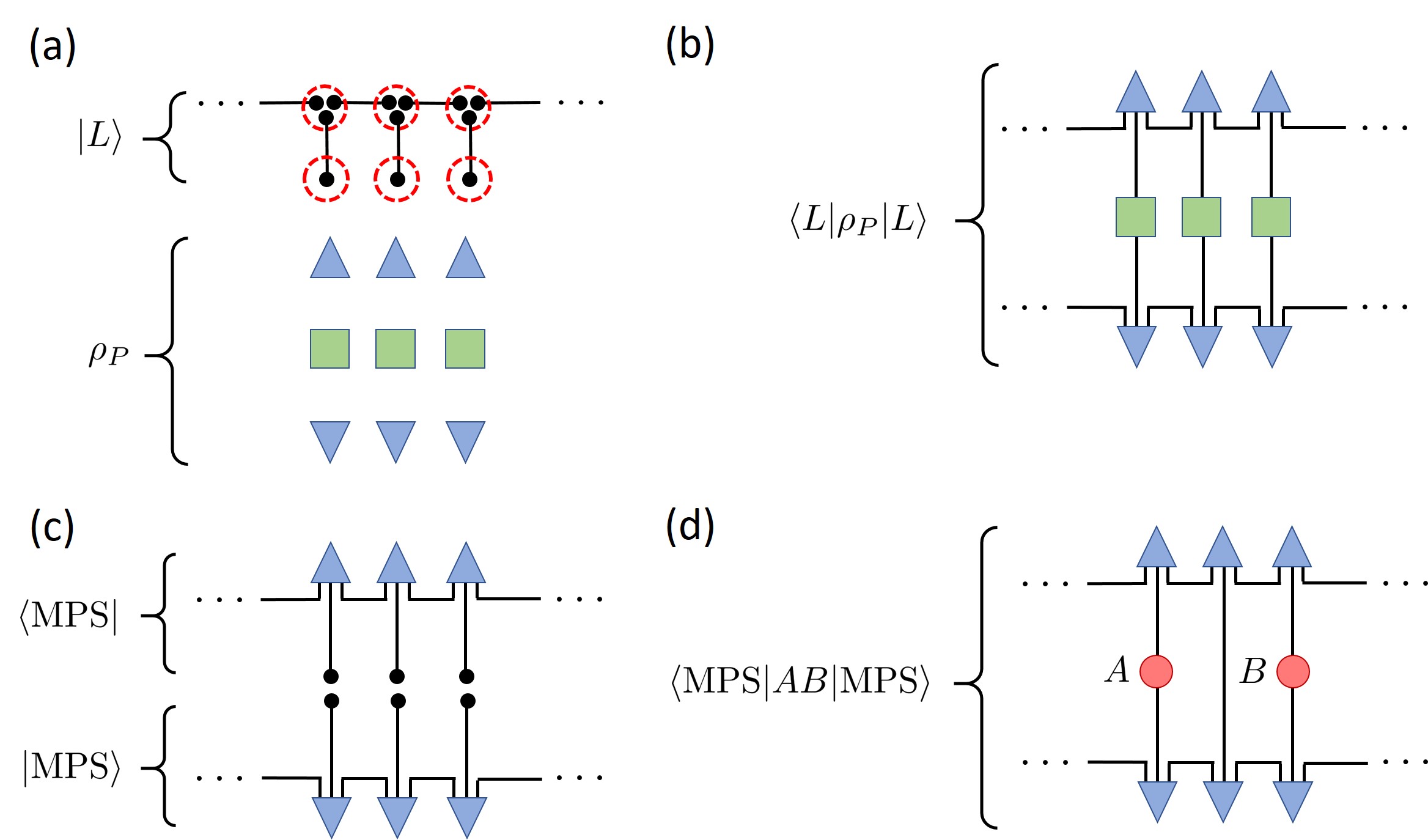}
\vskip.75cm
\caption{(a) A diagrammatic representation of $|L\rangle$ and $\rho_P$ for a nascent MPS tensor network.  The blue triangles represent the 3-qudit pure states $\langle \varphi|$ (each upper triangle) and $|\varphi\rangle$ (each lower triangle), and the green boxes are 1-qudit identity operators.  Therefore, $\rho_P = (|\varphi\rangle \langle \varphi| \otimes \textbf{1})^{\otimes N}$ for some $N$. (b) A diagrammatic representation of $\langle L| \rho_P |L\rangle$.  The green boxes can be omitted since they are identity operators. (c) If we split $\langle L| \rho_P |L\rangle$ by cutting through the vertical links, we obtain two MPS states.  (d) A diagram of the two-point function $\bra{\text{MPS}}AB\ket{\text{MPS}}$. \label{fig:MPS}}
\end{figure}

As an example of a GTN, we consider correlation functions in a matrix product state (MPS) tensor network.  To construct the MPS tensor network, we start with a link state $|L\rangle$ and density operator $\rho_P = (|\varphi\rangle \langle \varphi| \otimes \textbf{1})^{\otimes N}$ for some $N$, as depicted in Figure \ref{fig:MPS}(a).  Here, $\langle \varphi|$ and $|\varphi\rangle$ are 3-qudit states, and are each represented, respectively, by an upper and lower blue triangle in Figure \ref{fig:MPS}(a).  The identity operator $\textbf{1}$ acts on one qudit, and is depicted as a blue box in Figure \ref{fig:MPS}(a).  Contracting $\langle L|$ and $\rho_P$ and $|L\rangle$ as $\langle L| \rho_P |L\rangle$, we obtain the tensor network in Figure \ref{fig:MPS}(b).  Here, the green boxes can be omitted since they are just identity operators.  We can sever the vertical links to obtain two MPS states $|\text{MPS}\rangle$ and $\langle \text{MPS}|$, as in Figure \ref{fig:MPS}(c).  Indeed, we have $\langle \text{MPS} | \text{MPS} \rangle = \langle L | \rho_P |L\rangle$.  Finally, to compute correlation functions of the MPS state $|\text{MPS}\rangle$, we contract $\langle \text{MPS}|$ and $A$ and $B$ and $|\text{MPS}\rangle$ to obtain $\langle \text{MPS}| A B |\text{MPS}\rangle$, which is depicted by the tensor network in Figure \ref{fig:MPS}(d).

The Trotter networks in Fig.'s \ref{fig:Trotter0}(b), \ref{fig:Trotter0}(c),  and \ref{fig:TrotterParticular} are also examples of GTN's.  For these GTN's, the state $\rho_P$ is
\begin{equation}
\rho_P = \rho_i \otimes \bigotimes_{t = 1}^M \left(\bigotimes_i |U_{i,i+1}\rangle \langle U_{i,i+1}| \otimes \bigotimes_j |V_{j,j+1}\rangle \langle V_{j,j+1}|\right)
\end{equation}
where $|U_{i,i+1}\rangle$ and $|V_{j,j_1}\rangle$ are Choi-Jamiolkowski representations of the local unitary operators $U_{i,i+1}$ and $V_{j,j+1}$.  For instance, for a unitary two-qubit gate $U_{i,i+1}$ with matrix elements $[U_{i,i+1}]_{\alpha\beta}^{\gamma\delta}$ in some basis, one can define its Choi-Jamiolkowski representation which is the four-qubit state 
$$\ket{U_{i,i+1}}=\frac12 \sum_{\alpha\beta\gamma\delta}[U_{i,i+1}]_{\alpha\beta}^{\gamma\delta}\ket{\alpha}\ket{\beta}\ket{\gamma}\ket{\delta}\,.$$
The states $|V_{j,j+1}\rangle$ are represented similarly.

Then $|L\rangle$ comprises of qubit EPR pairs which link together the Choi-Jamiolkowski representations of the local unitary operators $\{U_{i,i+1}\}$ and $\{V_{j,j+1}\}$, as well as the initial state $\rho_i$, to form the tensor networks in Fig.'s \ref{fig:Trotter0}(b), \ref{fig:Trotter0}(c),  and \ref{fig:TrotterParticular}.  Here, the role of $|L\rangle$ is to ``unwrap'' the Choi-Jamiolkowski isomorphism and glue the appropriate unitaries together in space (for instance, $U_{i,i+1}$ should linked on the right with $U_{i+1,i+2}$) and in time (for instance, $U$'s are followed in the next time step by $V$'s).

Although much of the tensor network literature is centered around GTN's, our discussion of quantum causal influence below applies to general tensor networks.

\subsection{Defining quantum causal influence}

In the framework of general tensor networks, we now define our measures of quantum causal influence. Roughly speaking, the key idea is to distinguish causal influence from other forms of correlation by using unitary operators.  The causal influence of a region $R_1$ on a region $R_2$ is characterized by how correlations within $R_2$ can be changed by arbitrarily varying a unitary operator acting on $R_1$. As a prerequisite for this discussion, a unitary acting on $R_1$ has to preserve the norm of the tensor network, namely
\begin{equation}
\langle L | U_{R_1} \, \rho_P \, U_{R_1}^\dagger |L\rangle = \langle L | \rho_P |L\rangle\,,
\end{equation}
which is generically not true due to the ``metric'' $\rho_P$.   Therefore we define the concept of {\it unitary regions}.

Consider a tensor network with a Hilbert space decomposition into subsystems as $\mathcal{H} = \bigotimes_{i \in \Omega} \mathcal{H}_i$, where $\Omega$ indexes the subsystems.  Let us call the subsystems indexed by $\Omega$ the fundamental subsystems, since they are prescribed by the definition of the tensor network.  A \textit{unitary region} is a subsystem $R$, with $R \subseteq \Omega$, and an associated Hilbert space $\mathcal{H}_R = \bigotimes_{i \in R} \mathcal{H}_i$ such that
\begin{eqnarray}
\label{unitaryregion1}
\bra{L} U_R \, \rho_P \, U_R^\dagger \ket{L}=\bra{L}\rho_P\ket{L}
\end{eqnarray}
for arbitrary unitaries $U_R$ supported on $R$.  In other words, a unitary region is a subsystem for which acting with local unitaries preserves the norm of the tensor network.  We also say that two regions $R_1$, $R_2$ are \textit{mutually unitary regions} if
\begin{eqnarray}
\label{unitaryregion2}
\bra{L} U_{R_1} \, U_{R_2} \, \rho_P \, U_{R_2}^\dagger \, U_{R_1}^\dagger \ket{L}=\bra{L}\rho_P\ket{L}
\end{eqnarray}
for arbitrary unitaries $U_{R_1}$ supported on $R_1$ and arbitrary unitaries $U_{R_2}$ supported on $R_2$.  Notice that if $R_1$, $R_2$ are mutually unitary regions, then they are each unitary regions individually.  The converse is not generally true.

For concreteness, in the Trotter networks in Fig.'s \ref{fig:Trotter0}(a), \ref{fig:Trotter0}(b), \ref{fig:Trotter0}(c) and \ref{fig:TrotterParticular}, we can define $45^o$ lines as ``light cones.''  Using these light cones, it is easy to see that all regions that only contain only ``spacelike'' separated points are unitary regions.  All pairs of such regions are in fact mutually unitary regions.   In contrast, a region with two time-like separated points $x,y$ is not a unitary region. As another example, for a general MPS tensor network as depicted in Figure \ref{fig:MPS}(d), only the sites obtained by breaking apart vertical links are unitary regions.

Given a unitary region $R_1$, its causal influence on another region $R_2$ is reflected in the following quantity:
\begin{eqnarray}
\label{Meq1}
M(U_{R_1}: O_{R_2}): =\bra{L} (U_{R_1} \otimes O_{R_2})\,\rho_P \, (U_{R_1}^\dagger \otimes O_{R_2}^\dagger)\ket{L}
\end{eqnarray}
If $M(U_{R_1}: O_{R_2})$ has nontrivial dependence on $U_{R_1}$, this means that physical operations on region $R_1$ have a nontrivial causal influence on physical observables in region $R_2$. 

Using $M(U_{R_1}:O_{R_2})$, one can define different measures of quantum causal influence that are independent from the choice of operators $U_{R_1}, O_{R_2}$. For example, one can define the \textit{maximal quantum causal influence} (henceforth, mQCI)
\begin{eqnarray}
\text{CI}(R_1:R_2)=\sup_{U_{R_1}, O_{R_2}}\frac1{||O_{R_2}||_2^2}\left|M(U_{R_1}:O_{R_2})-\int dU_{R_1} \, M(U_{R_1}:O_{R_2})\right|,\label{maximalinfluence}
\end{eqnarray}
and the \textit{averaged quantum causal influence} (henceforth, aQCI)
\begin{eqnarray}
\overline{\text{CI}}(R_1:R_2)=\int dU_{R_1}\int_{||O_{R_2}||_2^2=1} dO_{R_2} \,\left|M(U_{R_1}:O_{R_2})-\int dU_{R_1} \, M(U_{R_1}:O_{R_2})\right|^2\label{averagedinfluence}
\end{eqnarray}
where in Eqn.'s~\eqref{maximalinfluence} and~\eqref{averagedinfluence}, $U_{R_1}$ is integrated via the Haar measure, and in Eqn.~\eqref{averagedinfluence} $O_{R_2}$ is averaged with the uniform measure on the unit sphere defined by $||O_{R_2}||_2^2=1$ in the linear space of operators $O_{R_2}$.  In the rest of the paper, when we discuss whether the quantum causal influence is zero or non-zero, we do not need to distinguish between the mQCI and aQCI, and so will refer to the QCI more broadly. In Section \ref{sec:measures}, we will discuss more quantitative properties of the aQCI. 
Variations of quantum causal influence for non-unitary regions can be found in Appendix A.  A discussion of causal influence for classical systems is in Appendix C.

With our definitions at hand, we would like to gain more intuition about quantum causal influence by studying some of its key features through various examples.

\section{Boundary condition dependence of quantum causal influence}
\label{sec:BCdependence}

Before discussing more abstract properties of quantum causal influence for general tensor networks, we first present examples which exhibit interesting causal features.  Our examples in Fig.'s \ref{fig:Trotter0}(b), \ref{fig:Trotter0}(c) and \ref{fig:TrotterParticular} in the previous section have a natural form which can be abstracted as follows.  They comprise of some initial state $\rho_i$ conjugated by some (not necessarily unitary) operator $W$ which implements evolution, followed by a trace.

\begin{figure}[t]
\center
\includegraphics[width=2in]{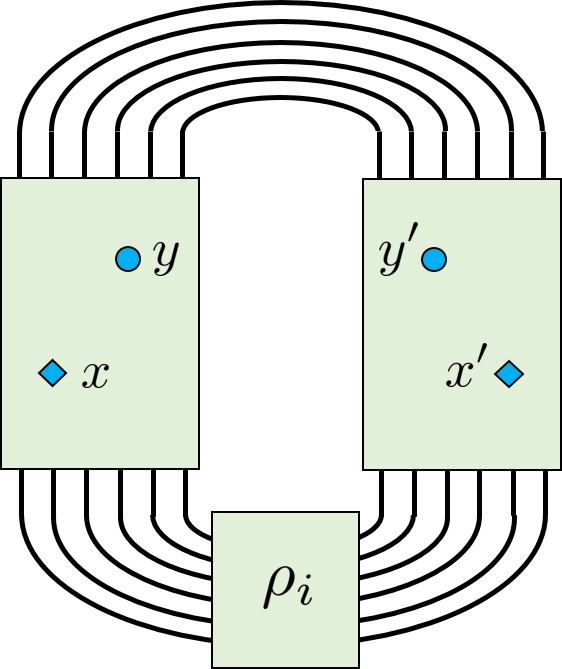}
\caption{A spacetime state with initial state $\rho_i$.  Two spacetime points $x$ and $y$ are designated, along with their mirror copies $x'$ and $y'$. \label{fig:initialstate1}}
\end{figure}

A more abstract representation is drawn in Figure \ref{fig:initialstate1}.  We call such a representation a ``spacetime state'' to distinguish it from other kinds of tensor networks.  The green boxes on either side of $\rho_i$ represent $W$ (on the left) and $W^\dagger$ (on the right).  The tensor contractions at the top of the diagram represent a trace.  Analogously to Fig.'s \ref{fig:Trotter0}(b), \ref{fig:Trotter0}(c) and \ref{fig:TrotterParticular} which comprise of a mesh of links (i.e., EPR pairs), we treat the $W$ and $W^\dagger$ boxes in Figure \ref{fig:initialstate1} as comprised of a mesh of links which we can break open to insert operators.  For instance, in Figure \ref{fig:initialstate1} we label the positions of two (hidden) links $x$ and $y$, which can be broken to insert operators.  We imagine that $x$ and $y$ are spacetime points.  Likewise, $x'$ and $y'$ are mirroring spacetime points.  By inserting $A$ into $x$, $B$ into $y$, $A^\dagger$ into $x'$ and $B^\dagger$ into $y'$, the tensor network computes
\begin{equation}
\label{Wcorr1}
\langle \mathcal{P} \, B_y\, A_x\, \rho_i \, A_x^\dagger \, B_y^\dagger \rangle
\end{equation} 
where the path ordering $\mathcal{P}$ is defined by the contracted tensor network.  Indeed, if $W$ corresponds to Hamiltonian time evolution or some discrete-time analog thereof, then Eqn.~\eqref{Wcorr1} is merely a standard correlation function with an initial state $\rho_i$.  In this case, we imagine that slicing the $W$ or $W^\dagger$ boxes along a horizontal line and contracting operators with the exposed links corresponds to operator insertions at a fixed intermediate time.  This is directly analogous to Fig.'s \ref{fig:Trotter0}(b), \ref{fig:Trotter0}(c) and \ref{fig:TrotterParticular}.

The causal structure of a spacetime state can depend on its boundary conditions -- namely the initial state $\rho_i$, and the trace taken over $W \rho_i W^\dagger$.  In this section, we illustrate the boundary condition dependence of causal influence in spacetime states in several examples. Our results suggest an explanation of ``time's arrow'' in a quantum many-body system.  

\subsection{Initial state dependence}

Suppose we have a spacetime state comprised of an initial state $\rho_i = |\psi\rangle \langle \psi|$ which is then unitarily evolved in time.  In other words, $W$ implements unitary time evolution.  As mentioned above, slicing the $W$ or $W^\dagger$ boxes along a horizontal line and contracting operators with the exposed links corresponds to operator insertions at a fixed intermediate time.  In Figure \ref{fig:initialstate1}, we allow insertions of operators into the spacetime points $x$ and $y$, and then contract the spacetime state (i.e., take its trace) at some later time.  Unpacking Eqn.~\eqref{unitaryregion1} for our case, we find that $x$ is a unitary region if
\begin{equation}
\langle \mathcal{P} \, U_x \, \rho_i \, U_x^\dagger \rangle = \langle \mathcal{P} \rho_i \rangle\,,
\end{equation}
and similarly for $y$,
\begin{equation}
\langle \mathcal{P} \, U_y \, \rho_i \, U_y^\dagger \rangle = \langle \mathcal{P} \rho_i \rangle\,.
\end{equation}
Each of the above equations is satisfied, and so any such points $x$ and $y$ are unitary regions.  In fact, we have also
\begin{equation}
\langle \mathcal{P} \, U_x \,U_y \rho_i \, U_y^\dagger \, U_x^\dagger \rangle = \langle \mathcal{P} \rho_i \rangle\,,
\end{equation}
for all such pairs $x$, $y$, and so all pairs of points $x$, $y$ form mutually unitary regions.

\begin{figure}[t]
\center
\includegraphics[width=3in]{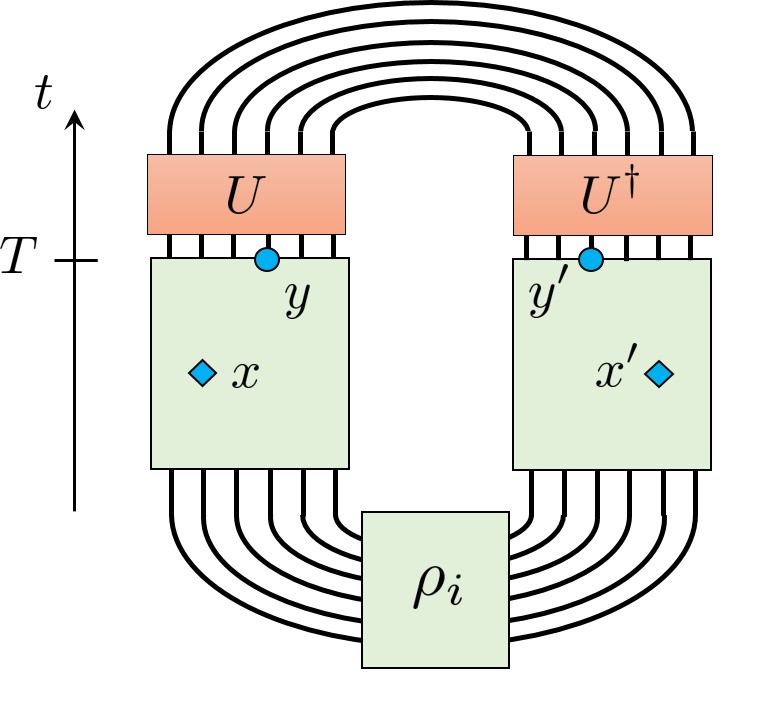}
\caption{A spacetime state, such that operators are not inserted later than a time $T$.  Then the unitary evolution $U$ after time $T$ cancels out with the corresponding unitary evolution $U^\dagger$. \label{fig:maxtime1}}
\end{figure}

Say that we insert a unitary $U_y$ at $y$ and $U_y^\dagger$ at $y'$.  This $U_y$ and $U_y^\dagger$ will cancel one another along the upper contraction of the spacetime state in Figure \ref{fig:initialstate1}.  The reason is that the unitary evolution that occurs after $y$ and $y'$ cancels across the trace -- see, for instance, the red boxes in Figure \ref{fig:maxtime1}.  These red boxes clearly cancel across the trace (i.e., the upper contracted legs), and so allow $U_y$ at $y$ and $U_y^\dagger$ at $y'$ to similarly cancel.  If we insert some Hermitian operator $O_x$ at $x$ and $O_x^\dagger$ at $x'$, then these operators will be unaffected by the cancellation of $U_y$ and $U_y^\dagger$.  Therefore,
\begin{equation}
M(U_y : O_x) = \langle \mathcal{P} \, U_y \, O_x \, \rho_i \, O_x^\dagger \, U_y^\dagger \rangle
\end{equation}
is independent of $U_y$, and thus
$$\text{CI}(y \, : \, x) = 0\,$$
meaning that $y$ does not influence $x$.  Similarly, $\overline{\text{CI}}(y \, : \, x) = 0$, although we will focus on the mQCI in this section.

But now suppose that we insert $U_x$ at $x$ and $O_y$ at $y$, and $U_x^\dagger$ at $x'$ and $O_y^\dagger$ at $y'$.  We cannot cancel out $U_x$ with $U_x^\dagger$ along the lower contraction of the spacetime state, since we are obstructed by the boundary condition $\rho_i$ (i.e., the initial state).  We might be able to cancel $U_x$ with $U_x^\dagger$ along the upper contraction of the spacetime state, but the operator insertions $O_y$ and $O_y^\dagger$ may obstruct us.  If $O_y$ and $O_y^\dagger$ \textit{obstruct} the cancellation of $U_x$ and $U_x^\dagger$ along the upper contraction, then $M(U_x : O_y)$ \textit{would} depend on $U_x$, and thus $\text{CI}(x \, : \, y) \not = 0$.  In summary, we would have
$$\text{CI}(y \, : \, x) = 0 \quad \text{and} \quad \text{CI}(x \, : \, y) \not = 0 \quad \text{implies }y\text{ is in the }\textit{future}\text{ of }x\,.$$
If instead $O_y$ and $O_y^\dagger$ do \textit{not} obstruct cancellation of $U_x$ and $U_x^\dagger$ along the upper contraction, then $M(U_x : O_y)$ would \textit{not} depend on $U_x$, and so $\text{CI}(x \, : \, y) = 0$.  Then in this case, we would have
$$\text{CI}(y \, : \, x) = 0 \quad \text{and} \quad \text{CI}(x \, : \, y) = 0 \quad \text{implies }x\text{ and }y\text{ are }\textit{spacelike}\text{ separated.}$$

The interesting feature here is that the state $\rho_i$ induces a causal structure in which time flows \textit{away from} $\rho_i$ via the unitary evolution comprising the spacetime state.  In other words, the initial state has picked out a preferred arrow of time.  Crucially, there is not a ``final state'' at the top contraction of the spacetime state.  This is perfectly physical, since we often start in an initial state and evolve it up to some time, perhaps making operator insertions intermediately.  If we only consider operator insertions up to a finite time $T$, then we only have to consider the spacetime state evolved up until that $T$.  If we evolve the state further thereafter, when computing expectation values this additional time evolution would cancel out, as depicted in Figure \ref{fig:maxtime1}.  In the Figure, the time evolution $U$ in the left red box cancels out the time evolution $U^\dagger$ in the right red box.

There is another complementary perspective which is useful.  Instead of thinking of the upper end of the spacetime state (where the trace is) as a ``cutoff time'' after which we do not care about making operator insertions, we can instead imagine that we are inserting a \textit{maximally mixed state} $\textbf{1}/d$ as a final-time state.  Here, $d$ is the Hilbert space dimension of a spatial slice.  As far as any of our analysis is concerned, these two perspectives are mathematically equivalent, up to an overall multiplicative rescaling of the spacetime state by $d$.  The benefit of this change of perspective is that we can think about $\rho_i$ and $\textbf{1}/d$ on more equal footing.  In particular, we can say:
\begin{itemize}
\item The initial state $\rho_i$ can obstruct unitary cancellation across the initial-time boundary.
\item The final state $\textbf{1}/d$ can allow unitary cancellation across the final-time boundary.
\end{itemize}
In this manner, the initial state $\rho_i$ acts as a barrier and a \textit{source} of causal flow, and the final state $\textbf{1}/d$ acts as a passageway or \textit{sink} of causal flow.  It is no coincidence that the flow of time coincides with the disparity between the entropy of the initial and final states: namely, we have the von Neumann entropies $S[\rho_i] = 0$ and $S[\textbf{1}/d] = \log(d)$ and so time is flowing from a lower entropy state to a higher entropy state.  One might na\"{i}vely guess that more generally, given an initial state $\rho_i$ and final state $\rho_f$, there would be a forward arrow of time if $S[\rho_i] < S[\rho_f]$, but this is not generally true.  There needs to be additional relations between $\rho_i$ and $\rho_f$ to get a forward arrow of time, but we will leave this for future work.

Now suppose that we choose both the initial state $\rho_i$ and the final state $\rho_f$ to be the maximally mixed state, namely $\rho_i = \textbf{1}/d$, and that we multiplicatively rescale the resulting spacetime state by $d$.  Then we have
$$\text{CI}(x \, : \, y) = 0$$
meaning that $x$ does not causally influence $y$.  Similarly, we also have
$$\text{CI}(y \, : \, x) = 0$$
meaning that $y$ does not causally influence $x$.  Then $x$ and $y$ are spacelike separated.  Indeed, when the past and future are maximally mixed states, the unitary evolution in between does not impose a particular directionality of time.

\subsection{Conceptual remarks}

In standard discussions of the arrow of time, a key ingredient is that the initial conditions of the universe provide a low-entropy initial state.\footnote{In our universe, it seems that cosmic inflation provides us with such a low-entropy initial state.}  Tied to the arrow of time is the production of coarse-grained entropy, and ultimately the universe becomes a high-entropy equilibrium state.  Once the universe has reached equilibrium, there ceases to be an arrow of time in any conventional sense, since there is no longer entropy growth.  In blunt terms, there are no local clocks in thermal equilibrium.

In the context of this paper, we find a new twist on these ideas.  Above, we found that when both boundaries of a spacetime state are maximally mixed, which we can think of as infinite temperature (or maximum entropy) states, all pairs of spacetime points in between are spacelike separated.  If we attach the word ``past'' to one of the boundaries and attach the word ``future'' to the other boundary, we can say: \textit{If the putative past and future have maximal entropy, then all spacetime points in between are spacelike separated and there is no flow of time.}

We also saw that by fixing one of the boundaries to be a low-entropy state, such as a pure state, we can induce an arrow of time.  We will later show that by imposing more interesting boundary conditions on \textit{both} boundaries, we can have even richer causal structures and local arrows of time.  Intuitively, we will see that for fine-tuned boundary conditions, regions of boundary states which have higher and lower entropies act as sinks and sources for causal flow, respectively, which is consistent with more conventional intuitions from thermodynamics.  Presumably some version of our analysis applies to more general initial and final states, but such a generalization is beyond the scope of this work.

\subsection{Trotterized tensor network}

\begin{figure}[t]
\center
\includegraphics[width=2.7in]{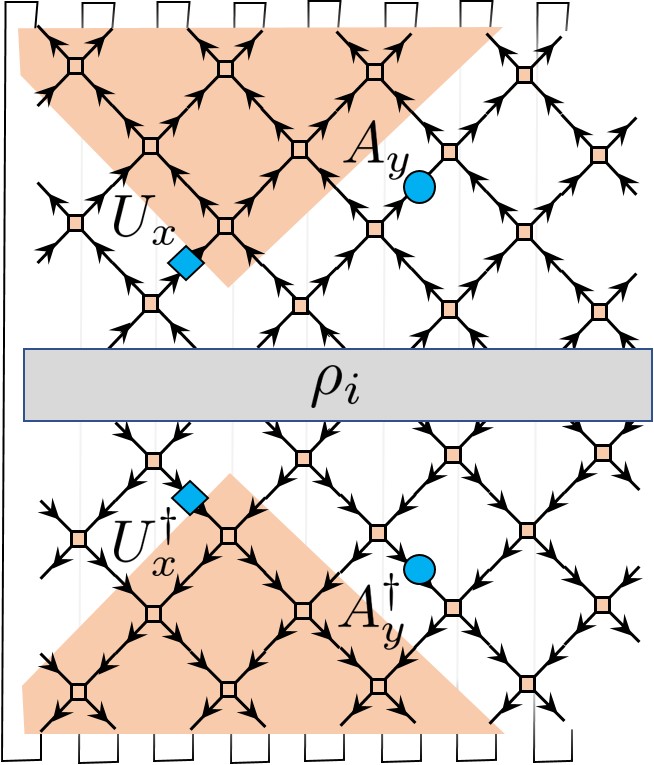}
\caption{In the Trotter network, $\text{CI}(x \, : \, y)=0$ unless $y$ is in the future light cone of $x$.
 \label{fig:Trotter1}}
\end{figure}

A nice example of a spacetime state which implements the above constructions is a Trotterized tensor network, such as in Fig.'s \ref{fig:Trotter0}(b), \ref{fig:Trotter0}(c) and \ref{fig:TrotterParticular} above.  For example, consider Figure \ref{fig:Trotter1} below which is a spacetime state with Trotterized time evolution and initial state $\rho_i$.  We see that in the contracted network, $\text{CI}(x\, : \, y) = 0$ unless $y$ is in a future cone of $x$, which is in fact the future light cone of $x$.  Notice that Figure \ref{fig:Trotter1} is folded relative to the spacetime states in Fig.'s \ref{fig:initialstate1} and \ref{fig:maxtime1} -- in particular, $\rho_i$ is in the middle, $W$ is on top, $W^\dagger$ is on the bottom, and the trace is looped behind.

\begin{figure}[h]
\center
\includegraphics[width=6in]{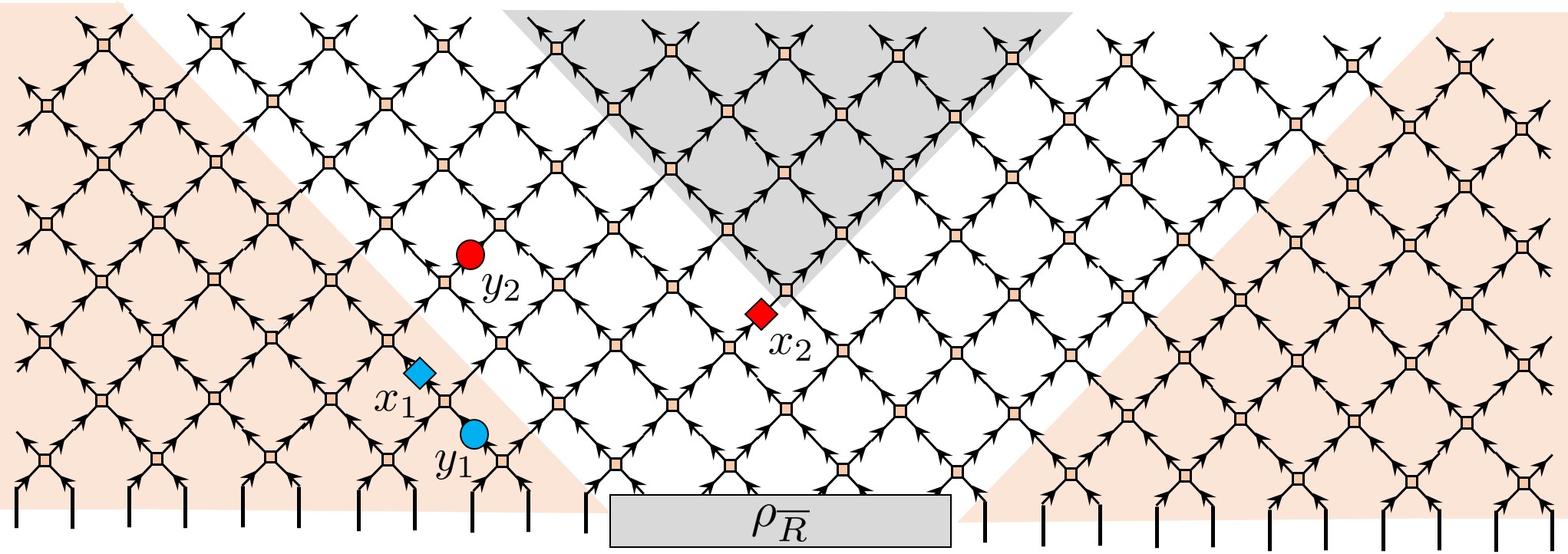}
\caption{In a Trotter network with a special initial state, some causal influence can be lost. 
 \label{fig:Trotter2}}
 
 \includegraphics[width=6in]{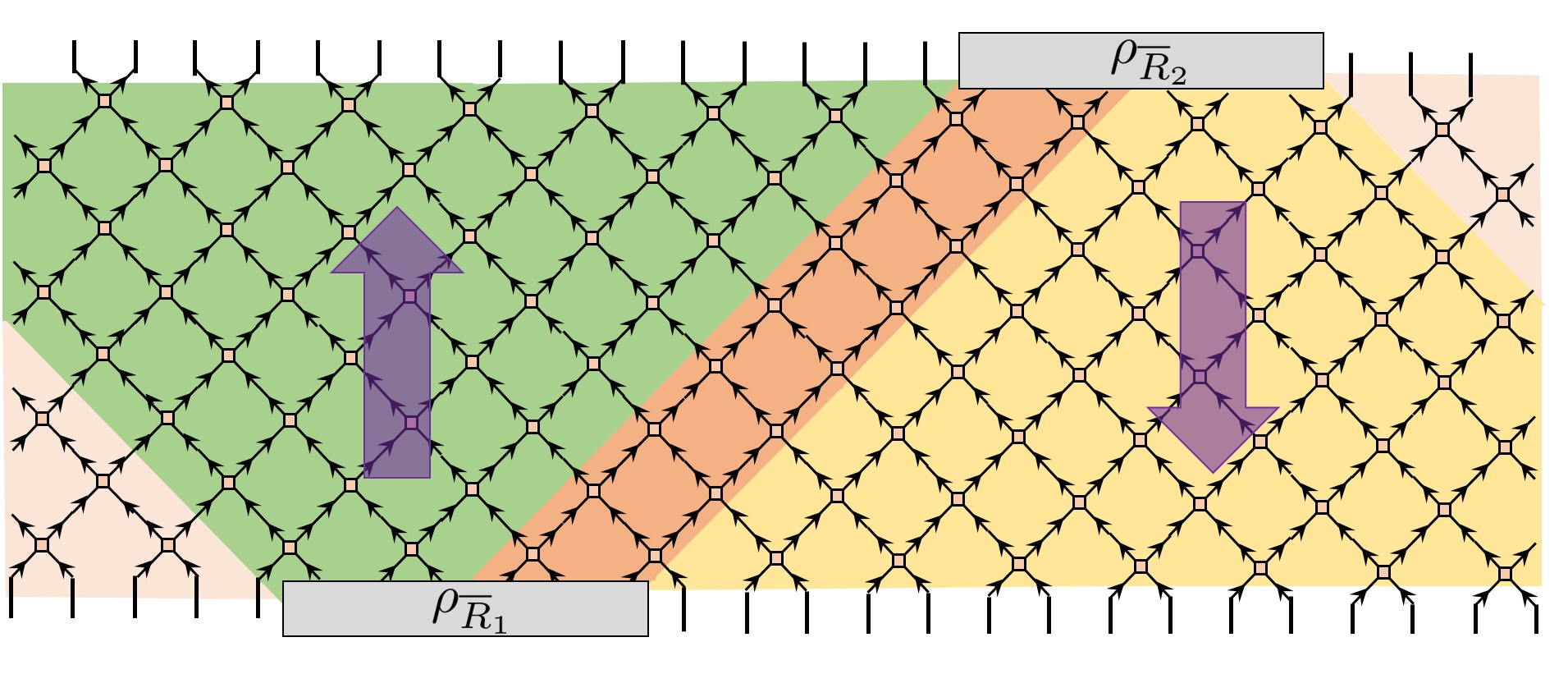}
\caption{In a Trotter network with a special initial \textit{and} final state, there can be multiple arrows of time, as depicted by the arrows. 
 \label{fig:Trotter3}}
\end{figure}

As we discussed earlier, the quantum causal structure generically depends on the initial state. For example, consider the spacetime state in Figure \ref{fig:Trotter2}, which has an initial state $\textbf{1}_R/d_R \otimes \rho_{\overline{R}}$.  The figure only displays part of the tensor network, namely $W \, (\textbf{1}_R/d_R \otimes \rho_{\overline{R}})$, and we have not depicted $W^\dagger$ or the trace.\footnote{The full diagram would give us $\text{tr}(W \, (\textbf{1}_R/d_R \otimes \rho_{\overline{R}}) \, W^\dagger)$.}  Since the initial state is maximally mixed on a subregion $R$, the spacetime has an interesting causal structure. For instance, applying a unitary $U$ to $x_1$ can cancel with a $U^\dagger$ applied to $x_1'$ across the $R$ region at the initial time, rather than canceling across the trace at the final time.  Consequently, the quantum causal influence of $x_1$ on any point in its usual future light cone\footnote{The usual future light cone of a point is defined by extending $45^o$ lines from that point, as per Figure \ref{fig:Trotter1}.  This ``usual'' future light cone is in fact the region which a point can causally influence if the initial state is pure.} vanishes.  Similarly, $x_1$ does not causally influence any point in its usual past light cone because unitaries acting at $x_1$ can still be canceled at the future boundary.  Therefore, $x_1$ does not causally influence any {\it single site} regions. However, $x_1$ can have a quantum causal influence on larger regions.  When we consider a spacetime region that overlaps with both the usual future light cone and usual past light cone of $x_1$, such as $y_1\cup y_2$, the quantum causal influence ${\rm CI}(x_1,y_1\cup y_2)$ is generically non-zero since it is not possible to push a unitary operator at $x_1$ to either the future boundary or the past boundary (since it is obstructed by the operators inserted at both $y_1$ and $y_2$) to cancel with a corresponding Hermitian conjugate unitary.

More generally, any region $A$ in the domain of dependence of $R$ (the red shaded region in Figure \ref{fig:Trotter2}) does not causally influence its usual causal future $I^+(A)$.
The only regions that are causally influenced by $A$ are those that overlap with both the usual causal future $I^+(A)$ and the usual causal past $I^-(A)$.  Thus, we see that specifying a special initial state may erase some regions from the causal future of a given region.  Although some of the causal future of a given region may be erased (such as $y_2$), nonlocal regions can still remain in the causal future (such as $y_1\cup y_2$). These observations are quite general, and we will see them instantiated in many contexts throughout the paper.

\subsection{Final state dependence (post-selection)}\label{sec:postselect}

There are many possibilities for including both initial and final states (i.e., pre-selection and post-selection), but we will only examine one case here to give a general flavor for the sorts of causal structures that can occur.  Consider the spacetime state comprised of Trotterized time evolution in Figure \ref{fig:Trotter3}, with initial state $\textbf{1}_{R_1}/d_{R_1} \otimes \rho_{\overline{R}_1}$ and final state $\textbf{1}_{R_2}/d_{R_2} \otimes \rho_{\overline{R}_2}$.  Similar to the previous figure, this figure only displays part of the tensor network, namely
$$(\textbf{1}_{R_2}/d_{R_2} \otimes \rho_{\overline{R}_2}) \, W \, (\textbf{1}_{R_1}/d_{R_1} \otimes \rho_{\overline{R}_1})\,.$$
 Accordingly, we have not depicted $W^\dagger$ or the trace.\footnote{Here, the full diagram would give us $\text{tr}\left[(\textbf{1}_{R_2}/d_{R_2} \otimes \rho_{\overline{R}_2}) \, W \, (\textbf{1}_{R_1}/d_{R_1} \otimes \rho_{\overline{R}_1}) \, W^\dagger \right]$.}   Suppose that $\overline{R}_1$ and $\overline{R}_2$ are regions of the same size, and that $\rho_{\overline{R}_1} = \rho_{\overline{R}_2}$ are pure states.  Then we see than there is a flow of time from bottom to top in the region shaded in green, but there is a flow of time from top to bottom in the region shaded in yellow.  Then every pair of points in the pink regions are spacelike separated, and the region in orange is not even a unitary region (and so, in a sense, does not have any preferred direction of time at all).  (See Appendix A for diagnostics quantum causal influence within nonunitary regions.)  This example emphasizes that pure states act as sources of causal flow, and maximally mixed states act as sinks of causal flow.  The pink regions are created by two sinks of causal flow (i.e., the maximally mixed states on each boundary), whereas the orange region is due to the interplay of two sources of causal flow (i.e., the pure states $\rho_{\overline{R}_1}$ and $\rho_{\overline{R}_2}$).

\subsection{General results}

In this subsection we summarize some generic features that can be observed from examples above, and describe them more quantitatively.

\subsubsection{Sinks of causal flow}

Having worked through explicit examples of the interplay between the initial and final states of a spacetime state and its causal structure, we now move towards more general and abstract results.  First, we present a result about GTN's that has played a role in all of the above examples.  The result generalizes the observed fact that in spacetime states, maximally mixed subsystems of initial and final states act as sinks of causal flow.

\begin{figure}[h!]
\center
\includegraphics[width=5.5in]{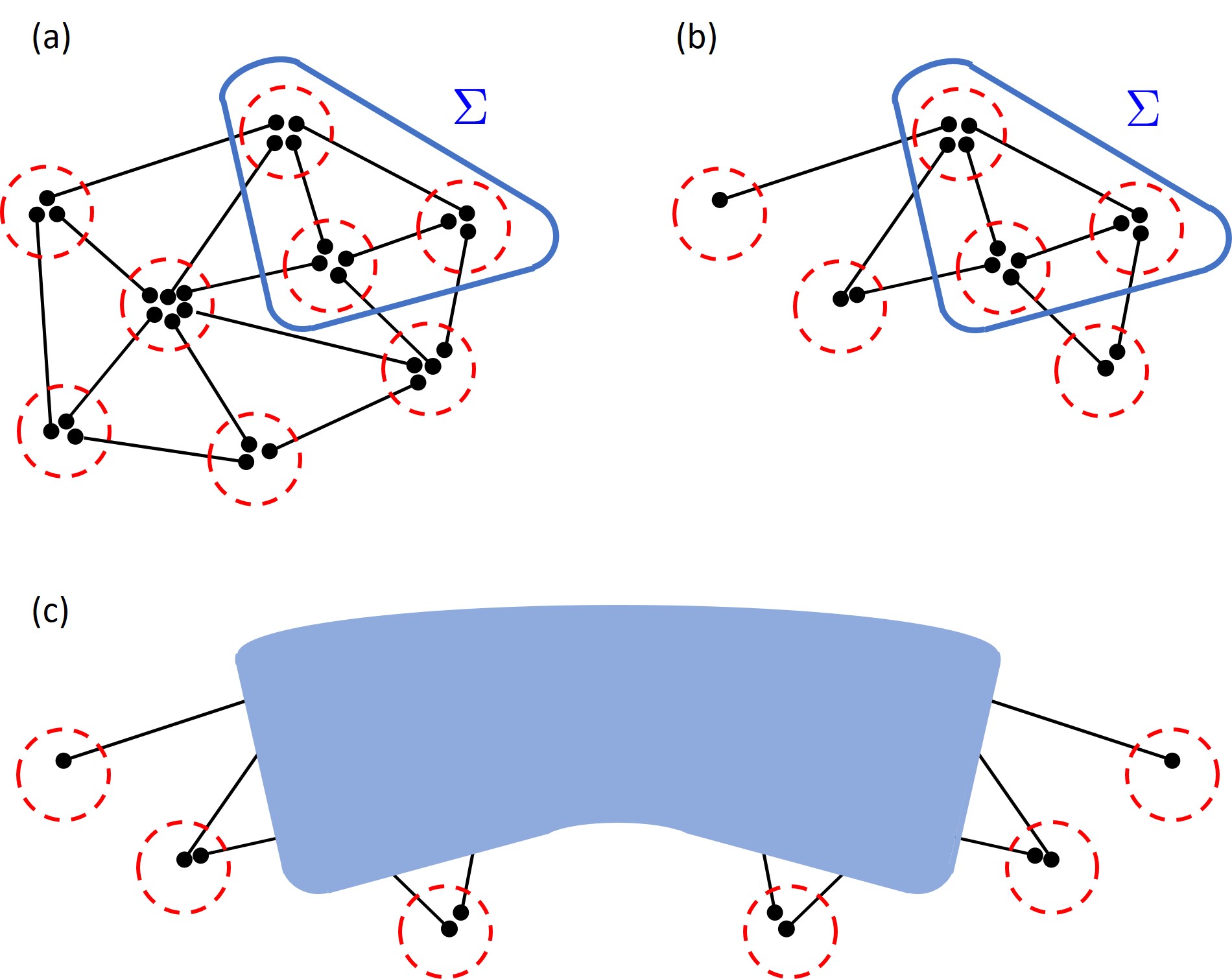}
\vskip.5cm
\caption{(a) The link state $|L\rangle$ is depicted.  The region $\Sigma$ is outlined in blue, and in this case contains 3 vertices and 11 qudits.  (b) The state $|L_{\Sigma \leftrightarrow \Sigma}\rangle \otimes |L_{\Sigma \leftrightarrow \overline{\Sigma}}\rangle$ is shown.  The EPR pairs lying within $\Sigma$ form $|L_{\Sigma \leftrightarrow \Sigma}\rangle$, and the EPR pairs crossing the boundary of $\Sigma$ form $|L_{\Sigma \leftrightarrow \overline{\Sigma}}\rangle$.  The qudits lying outside of $\Sigma$ form the Hilbert space $\mathcal{H}_{\partial \Sigma}$.  (c) By taking two copies of $|L_{\Sigma \leftrightarrow \Sigma}\rangle \otimes |L_{\Sigma \leftrightarrow \overline{\Sigma}}\rangle$ and partially contracting their $\Sigma$ regions with the state $\rho_P^\Sigma$, we obtain the density matrix $\sigma^{\partial \Sigma}$, which is depicted in the Figure.  The light blue region represents the contraction of the $\Sigma$ regions of $|L_{\Sigma \leftrightarrow \Sigma} \rangle \otimes |L_{\Sigma \leftrightarrow \overline{\Sigma}}\rangle$ and $\langle L_{\Sigma \leftrightarrow \Sigma} | \otimes \langle L_{\Sigma \leftrightarrow \overline{\Sigma}}|$ with $\rho_P^\Sigma$.  We see that the density matrix $\sigma^{\partial \Sigma}$ maps $\mathcal{H}_{\partial \Sigma}^* \otimes \mathcal{H}_{\partial \Sigma} \to \mathbb{C}$, since a state on $\mathcal{H}_{\partial \Sigma}$ can be contracted with the exposed legs on the right-hand side, and a dual state on $\mathcal{H}_{\partial \Sigma}$ can be contracted with the exposed legs on the left-hand side.
 \label{fig:graphregion1}}
\end{figure}

Suppose we have a GTN on a graph $G = (V,E)$, with the structure specified in Section \ref{sec:GraphTensorNetworks}.  As per Eqn.~\eqref{GTNfullHilbert1}, the corresponding Hilbert space is
$$\mathcal{H} = \bigotimes_{v \in V} \bigotimes_{j=1}^{\text{deg}(v)} \mathcal{H}_{v_j}\,.$$
Let $\Sigma \subseteq V$ be a subset of the vertices (which may correspond to a subregion in a putative spacetime), and partition $V$ as $V = \Sigma \cup \overline{\Sigma}$.  We can write the link state $|L\rangle$ as
\begin{equation}
|L\rangle = |L_{\Sigma \leftrightarrow \Sigma}\rangle \otimes  |L_{\Sigma \leftrightarrow \overline{\Sigma}}\rangle \otimes  |L_{\overline{\Sigma} \leftrightarrow \overline{\Sigma}}\rangle\,.
\end{equation}
In the above equation,
\begin{itemize}
\item $|L_{\Sigma \leftrightarrow \Sigma}\rangle$ are the EPR pairs associated with edges $e = (v,w)$ with $v,w \in \Sigma$;
\item $|L_{\Sigma \leftrightarrow \overline{\Sigma}}\rangle$ are the EPR pairs associated with edges $e = (v,w)$ with $v \in \Sigma$ and $w \in \overline{\Sigma}$;
\item $|L_{\overline{\Sigma} \leftrightarrow \overline{\Sigma}}\rangle$ are the EPR pairs associated with edges $e = (v,w)$ with $v,w \in \overline{\Sigma}$.
\end{itemize}
See Fig.'s \ref{fig:graphregion1}(a) and \ref{fig:graphregion1}(b) for a diagrammatic depiction.  So, for instance, each EPR pair in $|L_{\Sigma \leftrightarrow \overline{\Sigma}}\rangle$ comprises of one qudit in $\Sigma$ and one qudit in $\overline{\Sigma}$.  Let the Hilbert space of the qudits in $|L_{\Sigma \leftrightarrow \overline{\Sigma}}\rangle$ which lie in $\overline{\Sigma}$ be denoted by $\mathcal{H}_{\partial \Sigma}$.  Then the total Hilbert space $\mathcal{H}$ decomposes as
\begin{equation}
\mathcal{H} = \mathcal{H}_{\Sigma} \otimes \mathcal{H}_{\partial \Sigma} \otimes \mathcal{H}_{\overline{\Sigma \cup \partial \Sigma}}\,.
\end{equation}
Now, let $\rho_P^\Sigma := \text{tr}_{\overline{\Sigma}}(\rho_P)$, and consider the state
\begin{equation}
\sigma^{\partial \Sigma} := \text{tr}_\Sigma \left[(\rho_P^\Sigma \otimes \textbf{1}_{\partial \Sigma})\,\,  \left(|L_{\Sigma \leftrightarrow \Sigma}\rangle \langle L_{\Sigma \leftrightarrow \Sigma}| \otimes  |L_{\Sigma \leftrightarrow \overline{\Sigma}}\rangle \langle L_{\Sigma \leftrightarrow \overline{\Sigma}} |\right) \right]\,.
\end{equation}
This state $\sigma^{\partial \Sigma}$ is a density matrix on $\mathcal{H}_{\partial \Sigma}$\,.  Now we make the following proposition: \\ \\
\noindent \textbf{Proposition:} \textit{Suppose we decompose} $\mathcal{H}_{\partial \Sigma}$ \textit{into subsystems as}
\begin{equation}
\mathcal{H}_{\partial \Sigma} = \mathcal{H}_R \otimes \mathcal{H}_{\overline{R}}\,.
\end{equation}
\textit{If we have}
\begin{equation}
\sigma^{\partial \Sigma} = \frac{\textbf{1}_R}{d_R} \otimes \rho_{\overline{R}}
\end{equation}
\textit{for some} $\rho_{\overline{R}}$\textit{, then}
\begin{equation}
\text{CI}(R : S) = 0
\end{equation}
\textit{for any region} $S$ \textit{such that} $S \cap (\Sigma \cup \partial \Sigma)=\emptyset$, \textit{i.e.,} $S$ \textit{does not intersect} $\Sigma$ \textit{or} $\partial \Sigma$.
\\
\newpage\noindent \textit{Proof.} Let us compute $M(U_R : O_S)$, where $U_R$ is a unitary on $R$ and $O_S$ is some Hermitian operator on $S$.  Let $\rho_P^{\overline{\Sigma \cup \partial \Sigma}} = \text{tr}_{\Sigma \cup \partial \Sigma}(\rho_P)$.  Then
\begin{align}
\langle L| U_R \, O_S \, \rho_P \, O_S^\dagger \, U_R^\dagger |L\rangle  &= \text{tr}\left[U_R \, O_S \, \rho_P \, O_S^\dagger \, U_R^\dagger \, |L\rangle \langle L| \right] \nonumber \\ \nonumber \\
&= \text{tr}_{\overline{\Sigma \cup \partial \Sigma}}\left[U_R \, O_S \, \rho_P^{\overline{\Sigma \cup \partial \Sigma}} \, O_S^\dagger \, U_R^\dagger \,\, \left(\sigma^{\partial \Sigma} \otimes |L_{\overline{\Sigma} \leftrightarrow \overline{\Sigma}}\rangle \langle L_{\overline{\Sigma} \leftrightarrow \overline{\Sigma}} | \right) \right] \nonumber \\ \nonumber \\
&= \text{tr}_{\overline{\Sigma \cup \partial \Sigma}}\left[O_S \, \rho_P^{\overline{\Sigma \cup \partial \Sigma}} \, O_S^\dagger \,\, \left(U_R \sigma^{\partial \Sigma} U_R^\dagger \otimes |L_{\overline{\Sigma} \leftrightarrow \overline{\Sigma}}\rangle \langle L_{\overline{\Sigma} \leftrightarrow \overline{\Sigma}} | \right) \right]\,. \nonumber
\end{align}
But since $\sigma^{\partial \Sigma} = \frac{\textbf{1}_R}{d_R} \otimes \rho_{\overline{R}}$ we have $U_R \sigma^{\partial \Sigma} U_R^\dagger = \sigma^{\partial \Sigma}$ and so the $U_R$ dependence drops out of the above equation.  Then
\begin{align}
\langle L| U_R \, O_S \, \rho_P \, O_S^\dagger \, U_R^\dagger |L\rangle  &= \langle L| O_S \, \rho_P \, O_S^\dagger |L\rangle \nonumber
\end{align}
and so $M(U_R : O_S)$ does not depend on $U_R$.  Therefore, $\text{CI}(R:S) = 0$, as claimed. $\square$ \\ \\

The proposition is a technical way of saying that we can cancel out a $U_R$ with a $U_R^\dagger$ in a GTN if there is a bridge (built out of tensor contractions) between them which is a maximally mixed state.  Thus, the proposition specifies how maximally mixed states are sinks of causal flow in GTN's.  In the special case of spacetime states, we see that initial and final states with maximally mixed subsystems act as sinks of causal flow since they provide a pathway for unitary cancellation. 

\subsubsection{Structure theorem}

It is interesting to consider how causal relationships between regions of spacetime points affect the structure of correlation functions comprised of operator insertions at those points.  A particular question along these lines is: \\ \\
\textit{Suppose we have two spacetime points $x$ and $y$, where $x$ is a unitary region.  If $x$ does not causally influence $y$ so that }$\text{CI}(x : y) = 0$\textit{, then what restrictions does this impose on the structure of spacetime correlation functions of the form} $\langle L | A_x \, B_y \, \rho_P \, B_y^\dagger \, A_x^\dagger |L\rangle$ \textit{for a general tensor network, or as a special case} $\langle \mathcal{P}\, A_x \, B_y \, \rho_i \, B_y^\dagger \, A_x^\dagger \rangle$\textit{ for a spacetime state?} \\ \\
To answer such a question, we need to utilize a formalism which organizes the data of spacetime correlation functions for spacetime states.  This is called the ``superdensity operator formalism'' \cite{cotler2017superdensity}, which is reviewed in Appendix B.  In short, a superdensity operator $\varrho$ is a multilinear map taking operators to correlation functions (which evaluate to complex numbers).  In our question of interest, we will use a superdensity operator
\begin{equation}
\varrho : \mathcal{B}^*(\mathcal{H}_x) \otimes \mathcal{B}^*(\mathcal{H}_y) \otimes \mathcal{B}(\mathcal{H}_x) \otimes \mathcal{B}(\mathcal{H}_y) \longrightarrow \mathbb{C}
\end{equation}
defined by
\begin{equation}
\label{superdensity1}
\varrho[A_x^\dagger\,, B_y^\dagger \, ; \, A_x \,, B_y] := \langle L| A_x \, B_y \,\rho_P \, B_y^\dagger \, A_x^\dagger |L\rangle\,.
\end{equation}
In the special case of spacetime states, the right-hand side of the above equation becomes $\langle \mathcal{P}\, A_x \, B_y \, \rho_i \, B_y^\dagger \, A_x^\dagger \rangle$.

\begin{figure}[h]
\center
\includegraphics[width=6in]{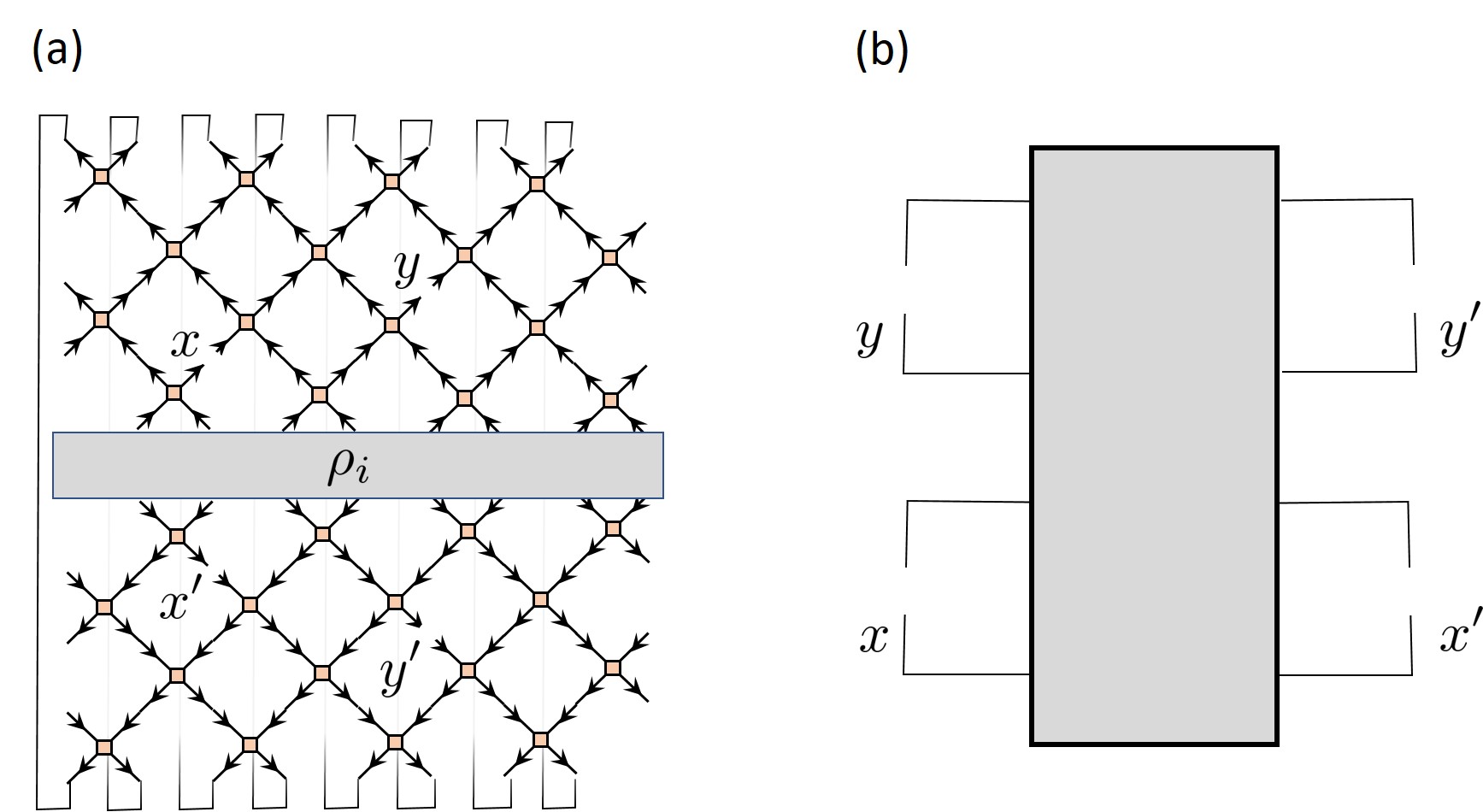}
\caption{(a) A Trotterized network comprised of a spacetime state contracted with its Hermitian conjugate with initial state $\rho_i$, and broken legs to allow the insertion of operators into $x$ and $y$ as well as $x'$ and $y'$. (b) A more abstract superdensity operator, allowing for operator insertions at $x$ and $y$ as well as $x'$ and $y'$. 
 \label{fig:TrotterSuper}}
\end{figure}

As an example, in Figure \ref{fig:TrotterSuper}(a), we depict $\varrho$ diagramatically for a spacetime state with Trotterized time evolution.  This tensor network can be more abstractly represented by the diagram in Figure \ref{fig:TrotterSuper}(b).  The diagram in Figure \ref{fig:TrotterSuper}(b) is completely general for spacetime states, and simply expresses that the superdensity operator is a multilinear object which takes as input operators on $\mathcal{B}(\mathcal{H}_x)\otimes \mathcal{B}(\mathcal{H}_y)$ as well as dual operators on the dual space $\mathcal{B}^*(\mathcal{H}_x)\otimes \mathcal{B}^*(\mathcal{H}_y)$, and outputs a complex number.

Using the superdensity setup, we prove the following structure theorem about general tensor networks: \\ \\
\textbf{Structure theorem:} \textit{If and only if} $\text{CI}(x\,:\,y) = 0$\textit{, then for fixed} $B_y$\,\textit{, the spacetime correlation function} $\langle L| A_x \, B_y \, \rho_P \, B_y^\dagger \, A_x^\dagger | L \rangle$ \textit{can be written as}
\begin{equation}
\label{structureEqn1}
\langle L| A_x \, B_y \, \rho_P \, B_y^\dagger \, A_x^\dagger | L \rangle = \alpha \, \text{tr}(O_1 \, A_x \, A_x^\dagger) + \beta \, \text{tr}(A_x^\dagger \, A_x \, O_2)
\end{equation} 
\textit{for all} $A_x$\textit{, where} $\alpha$ \textit{and} $\beta$ \textit{are complex numbers and} $O_1$ \textit{and} $O_2$ \textit{are operators which are independent of} $A_x$\,. \\ \\

\begin{figure}[t]
\center
\includegraphics[width=6in]{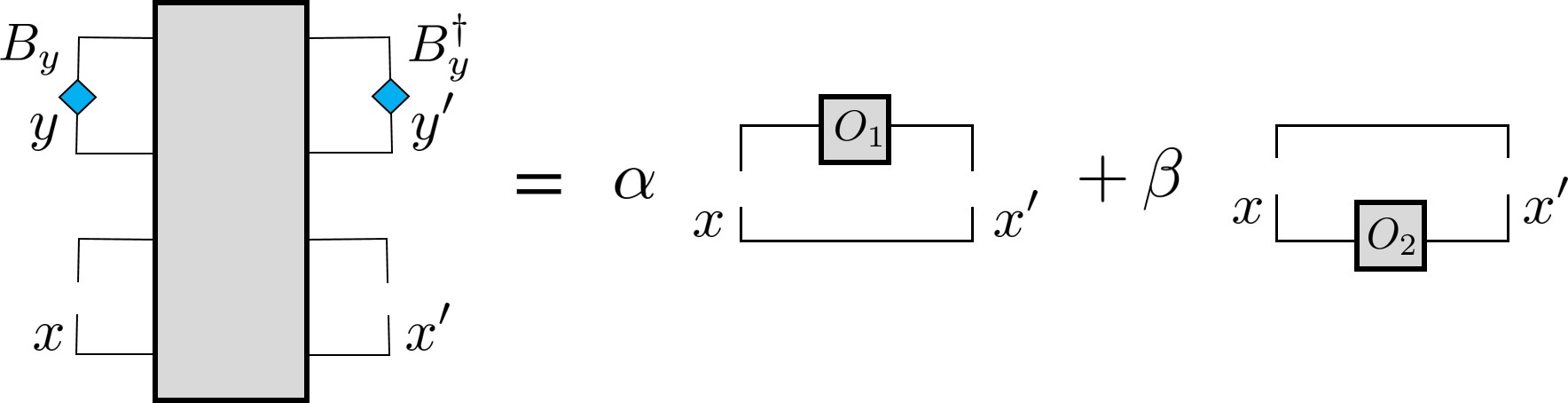}
\caption{If $\text{CI}(x \, : \, y) = 0$, then the superdensity operator with fixed $y$ insertions can be written as a linear combination of a tensor network with a maximally mixed past, and a tensor network with a maximally mixed future. 
 \label{fig:structurethm1}}
\end{figure}

Let us give a more intuitive interpretation of this theorem.  First, we note that we can rewrite Eqn.~\eqref{structureEqn1} in terms of the superdensity operator $\varrho$ given in Eqn.~\eqref{superdensity1} as
\begin{equation}
\varrho[A_x^\dagger\,,\,B_y^\dagger\,;\,A_x\,,\,B_y] =  \alpha \, \text{tr}(O_1 \, A_x \, A_x^\dagger) + \beta \, \text{tr}(A_x^\dagger \, A_x \, O_2)\,.
\end{equation}
This equivalence is depicted diagrammatically in Figure \ref{fig:structurethm1}.  We see from the figure a nice interpretation of the result: \textit{the causal influence is trivial if and only if the two-site superdensity operator is a linear superposition of a tensor network with the final state being maximally mixed and another tensor network with the initial state being maximally mixed.}  With this in mind, we prove the theorem.
 \\ \\
\textit{Proof.} For fixed $B_y$, we can generically write
\begin{equation}
\langle L| A_x \, B_y \, \rho_P \, B_y^\dagger \, A_x^\dagger | L \rangle = \sum_{i,j=0}^{d^2 - 1} K_{ij} \, \text{tr}(M^i \, A_x \, M^{j\,\dagger} \, A_x^\dagger)
\end{equation}
where $K_{ij}$ are complex numbers, $\{M^i\}$ is a complete set of orthonormal operators satisfying $\text{tr}(M^i M^{j\,\dagger}) = \delta_{ij}$, and $\mathcal{H}_x$ is a $d$-dimensional Hilbert space.  Note that the $K_{ij}$'s depend on $B_y$\,, but not on $A_x$.

If $\text{CI}(x \, : \, y) = 0$, then
$$\langle L| U_x \, B_y \, \rho_P \, B_y^\dagger \, U_x^\dagger |L\rangle = \langle L|\widetilde{U}_x \, B_y \, \rho_P \, B_y^\dagger \, \widetilde{U}_x^\dagger |L \rangle$$ 
for all unitaries $U_x$ and $\widetilde{U}_x$.  Therefore,
\begin{equation}
\label{sumequality1}
\sum_{i,j=0}^{d^2 - 1} K_{ij} \, \text{tr}(M^i \, U_x \, M^{j\, \dagger} \, U_x^\dagger) = \sum_{i,j=0}^{d^2 - 1} K_{ij} \, \text{tr}(M^i \, \widetilde{U}_x \, M^{j\,\dagger} \, \widetilde{U}_x^\dagger)
\end{equation}
for all $U_x, \widetilde{U}_x$.  In Eqn.~\eqref{sumequality1} above, the terms for which either $i$ or $j$ is zero have vanishing trace.  Also, the $i = j = 0$ term evaluates to one.  Then Eqn.~\eqref{sumequality1} simplifies to
\begin{equation}
\label{sumequality2}
\sum_{i,j=1}^{d^2 - 1} K_{ij} \, \text{tr}(M^i \, U_x \, M^{j\, \dagger} \, U_x^\dagger) = \sum_{i,j=1}^{d^2 - 1} K_{ij} \, \text{tr}(M^i \, \widetilde{U}_x \, M^{j\,\dagger} \, \widetilde{U}_x^\dagger)
\end{equation}
where the sums now run from $i,j = 1,....,d^2 - 1$.  Letting $\widetilde{U}_x = \textbf{1}$, we find that
\begin{equation}
\sum_{i,j=1}^{d^2 - 1} K_{ij} \, \text{tr}(M^i \, U_x \, M^{j\, \dagger} \, U_x^\dagger) = \sum_{i=1}^{d^2 - 1} K_{ii} = C
\end{equation}
for all $U_x$ and some constant $C$.  Using the Haar unitary integral
\begin{equation}
\int dU \, U_{nm}^*U_{k\ell} = \frac1{d}\,\delta_{nk}\delta_{m\ell}
\end{equation}
we find
\begin{equation}
\int dU_x \, \sum_{i,j=1}^{d^2 - 1} K_{ij} \, \text{tr}(M^i \, U_x \, M^{j\, \dagger} \, U_x^\dagger) = 0
\end{equation}
Therefore $C = 0$, implying that
\begin{equation}
\sum_{i,j=1}^{d^2 - 1} K_{ij} \, \text{tr}(M^i \, U_x \, M^{j\, \dagger} \, U_x^\dagger) = 0
\end{equation}
for all $U_x$.  Then we have
\begin{equation}
\left|\sum_{i,j=1}^{d^2 - 1} K_{ij} \, \text{tr}(M^i \, U_x \, M^{j\, \dagger} \, U_x^\dagger)\right|^2 = 0
\end{equation}
for all $U_x$. Using the Haar unitary integral
\begin{align}
\int dU\,{U_{n_1m_1}^*U_{n_2m_2}^*U_{k_1 \ell_1}U_{k_2 \ell_2}} &=
\frac1{d^2-1}\big[\delta_{n_1k_1}\delta_{m_1l_1}\delta_{n_2k_2}\delta_{m_2 \ell_2}+\delta_{n_1k_2}\delta_{m_1 \ell_2}\delta_{n_2k_1}\delta_{m_2 \ell_1} \nonumber \\
& \qquad \qquad \qquad -\frac1{d}\,\delta_{n_1k_1}\delta_{n_2k_2}\delta_{m_1 \ell_2}\delta_{m_2 \ell_1}-\frac1{d}\,\delta_{n_1k_2}\delta_{n_2k_1}\delta_{m_1 \ell_1}\delta_{m_2 \ell_2}\big]
\end{align}
we obtain
\begin{equation}
\int dU_x \, \left|\sum_{i,j=1}^{d^2 - 1} K_{ij} \, \text{tr}(M^i \, U_x \, M^{j\, \dagger} \, U_x^\dagger)\right|^2 = \frac{1}{d^2 - 1} \sum_{i,j=1}^{d^2 - 1} |K_{ij}|^2 = 0
\end{equation}
so that $K_{ij} = 0$ for $i,j=1,...,d^2 - 1$.

It follows that
\begin{equation}
\langle L| A_x \, B_y \, \rho_P \, B_y^\dagger \, A_x^\dagger | L \rangle = \frac{1}{d} \,K_{00} \, \text{tr}(\textbf{1} \, A_x \, \textbf{1} \, A_x^\dagger) + \sum_{i=0}^{d^2 - 1} \frac{K_{i0}}{\sqrt{d}} \, \text{tr}(M^i \, A_x \, \textbf{1} \, A_x^\dagger) + \sum_{j=0}^{d^2 - 1} \frac{K_{0j}}{\sqrt{d}} \, \text{tr}(\textbf{1} \, A_x \, M^{j\,\dagger} \, A_x^\dagger)
\end{equation}
which we can repackage into the desired equation
$$\langle L| A_x \, B_y \, \rho_P \, B_y^\dagger \, A_x^\dagger | L \rangle = \alpha \, \text{tr}(O_1 \, A_x \, A_x^\dagger) + \beta \, \text{tr}(A_x^\dagger \, A_x \, O_2)\,.$$

Conversely, if $\langle L| A_x \, B_y \, \rho_P \, B_y^\dagger \, A_x^\dagger | L \rangle = \alpha \, \text{tr}(O_1 \, A_x \, A_x^\dagger) + \beta \, \text{tr}(A_x^\dagger \, A_x \, O_2)$, then $\langle L| U_x \, B_y \, \rho_i \, B_y^\dagger \, U_x^\dagger |L\rangle$ is independent of unitaries $U_x$ which implies $\text{CI}(x \, : \, y) = 0$. $\square$ \\ \\

%
%
%
%

\section{Nonlocality of the quantum causal influence}
\label{sec:nonlocalityQCI}

Quantum causal influence captures the ability of one subsystem of a tensor network to affect another subsystem.  As remarked above, the quantum causal influence can behave in a peculiar way under the union of subsystems: in particular, we can have $\text{CI}(R:S_1) = \text{CI}(R: S_2) = 0$, whereas $\text{CI}(R : S_1 \cup S_2) > 0$.  In words, $R$ does not influence either $S_1$ or $S_2$ individually, but $R$ does influence their union $S_1 \cup S_2$.  More modest cases are also possible -- we may simply have that $\text{CI}(R:S_1)$, $\text{CI}(R: S_2)$ are close to zero whereas $\text{CI}(R : S_1 \cup S_2) > 0$ is significantly larger than zero.

How do we interpret the above cases, especially in the context of spacetime?  We will find that a core mechanism is the non-local encoding of information in spacetime.  For instance, in the spacetime setting, perturbations at $R$ can be non-locally encoded in the spacetime region $S_1 \cup S_2$, but not in the spacetime regions $S_1$ or $S_2$ alone.  We can find natural examples in which $S_1$ and $S_2$ can be vastly separated in both space and time.  Our analysis indicates that the non-local encoding of information in spacetime is a ubiquitous phenomenon.

A key tool for analyzing non-local quantum causal influence is the theory of quantum error correction codes.  We begin by discussing quantum error correction, and show how quantum error correction codes allow us to construct examples of non-local causal influence.  We then give a natural example of scrambling in a chaotic quantum many-body system.  Finally, we explore the causal structure of quantum teleportation.

\subsection{Quantum error correction codes}
\label{sec:QEC1}

Nonlocal features of quantum causal influence are intimately related to quantum error correction codes.  First, we briefly review quantum error correction codes, and quantum erasure codes in particular.  A nice overview written for high energy physicists is given in \cite{almheiri2015bulk}.

There are many equivalent definitions of quantum error correction codes, so we choose one which is most convenient for our analysis here.  Consider two Hilbert spaces $\mathcal{H}_A$, $\mathcal{H}_B$ with $\dim \mathcal{H}_A < \dim \mathcal{H}_B$.  We may think of $A$ as subsystem of $B$, so that $\mathcal{H}_B = \mathcal{H}_A \otimes \mathcal{H}_{\overline{A}}$.  Intuitively, imagine we have a noisy quantum system $B$, and that we want to construct a protocol which protects the state of some subsystem $A$ against our particular form of noise.  The idea is to redundantly \textit{encode} the state of the subsystem $A$ into a state of the larger system $B$, in such a way that the larger encoded state is robust to our form of noise.  Then we can subsequently \textit{decode} the larger encoded state to obtain the original state on $B$.

Now we formalize this intuition.  The space of density matrices on each Hilbert space $\mathcal{H}_A$, $\mathcal{H}_B$ are $\mathcal{S}(\mathcal{H}_A)$ and $\mathcal{S}(\mathcal{H}_B)$, respectively.  Suppose we have three quantum channels (i.e., completely positive trace-preserving (CPTP) maps):
\begin{align}
\mathcal{E} &: \mathcal{S}(\mathcal{H}_A) \longrightarrow \mathcal{S}(\mathcal{H}_B) \\ \nonumber \\
\mathcal{N} &: \mathcal{S}(\mathcal{H}_B) \longrightarrow \mathcal{S}(\mathcal{H}_B) \\ \nonumber \\
\mathcal{R} &: \mathcal{S}(\mathcal{H}_B) \longrightarrow \mathcal{S}(\mathcal{H}_A)\,.
\end{align} 
The channel $\mathcal{E}$ is the ``encoding'' channel, which maps density matrices on the subsystem $A$ to density matrices on the larger system $B$.  The channel $\mathcal{N}$ is the ``noise'' channel, which induces errors on density matrices on $B$.  Finally, the channel $\mathcal{R}$ is the ``recovery'' channel, which decodes density matrices on $B$ to density matrices on $A$.  Then we have a quantum error correction code if
\begin{equation}
(\mathcal{R} \circ \mathcal{N} \circ \mathcal{E})(\rho) = \rho\,, \quad \text{for all }\rho \in \mathcal{S}(\mathcal{H}_A)\,.
\end{equation} 
In words, the above equation means that for all states on the subsystem $A$, applying the encoding channel $\mathcal{E}$, the noise channel $\mathcal{N}$, and finally the recovery channel $\mathcal{R}$ gives back the state that we started with.

Notice that the description of a quantum error correction code depends on a specified form of noise, as provided by the given noise channel $\mathcal{N}$.  There are many kinds of quantum error correction codes which protect against varied forms of noise.  For our purposes, we will be most interested in noise which erases information.  The corresponding form of quantum error correction code which is robust to erasure errors is called a quantum erasure code.  These kinds of code are robust to an entire collection of noise channels $\{\mathcal{N}_S\}$, which we will define shortly.

To formally define a noise channel which causes erasure errors, consider again the Hilbert space $\mathcal{H}_B$, and let $S$ be a subsystem of $B$ with Hilbert space $\mathcal{H}_S$.  Then let $\mathcal{N}_S$ be a channel taking $\mathcal{S}(\mathcal{H}_B) \to \mathcal{S}(\mathcal{H}_B)$ which erases all information on the subsystem $S$.  The channel $\mathcal{N}_S$ is given by
\begin{equation}
\label{Noisy1}
\mathcal{N}_S(\rho) = \text{tr}_S(\rho) \otimes \frac{\textbf{1}_S}{\dim(\mathcal{H}_S)}
\end{equation}
where $\textbf{1}_S/\dim(\mathcal{H}_S)$ is the maximally mixed state on the subsystem $S$.

Now supposing that our system is a collection of qudits, let $|S|$ denote the number of qudits comprising the subsystem $S$.  Equivalently, $|S| = \log_d(\dim(\mathcal{H}_S))$.  Then a $k$--qudit quantum error correction code is given by quantum channels $\mathcal{E} : \mathcal{S}(\mathcal{H}_A) \to \mathcal{S}(\mathcal{H}_B)$, $\mathcal{R}_S : \mathcal{S}(\mathcal{H}_B) \to \mathcal{S}(\mathcal{H}_A)$ such that
\begin{equation}
(\mathcal{R}_S \circ \mathcal{N}_S \circ \mathcal{E})(\rho) = \rho\,, \quad \text{for all }S\text{ such that }|S| \leq k,\text{ and all }\rho \in \mathcal{S}(\mathcal{H}_A)\,.
\end{equation} 
In words, the $k$--qudit quantum error correction code can correct for the erasure of at most $k$ qudits of $B$.  Hence, the  $k$--qudit quantum error correction code corrects for the entire  collection of noise channels $\{\mathcal{N}_S\}_{|S| \leq k}$.  Notice that the recovery channel $\mathcal{R}_S$ depends on the choice of subsystem $S$ that is erased.

Now we provide an example of a $1$--qutrit\footnote{A qutrit is a three-level system, i.e. a qudit with $d=3$.} quantum erasure code, called the ``three qutrit code'' \cite{cleve1999share, beny2007generalization, beny2007quantum, almheiri2015bulk}.  This code protects against the erasure of a single qutrit, among three qutrits.  Let $\mathcal{H}_A = \text{span}\{|0\rangle, |1\rangle, |2\rangle\}$ be the space of a single qutrit (so that $\dim \mathcal{H}_A = 3$) and let $\mathcal{H}_B$ be the space of three qutrits (so that $\dim \mathcal{H}_B = 27$).  The encoding channel $\mathcal{E}$ is a unitary channel
\begin{equation}
\mathcal{E}(\rho) = U_{\text{encode}} \,\rho\, U_{\text{encode}}^\dagger
\end{equation}
where $U_{\text{encode}}$ acts by
\begin{align}
U_{\text{encode}}\sum_{i=0}^3 c_i \, |i\rangle = \sum_{i=0}^3 c_i \, |\widetilde{i}\rangle
\end{align}
and
\begin{align}
U_{\text{encode}}\left(|0\rangle \otimes |00\rangle\right) &= |\widetilde{0}\rangle = \frac{1}{\sqrt{3}}\left(|000\rangle + |111\rangle + |222\rangle \right) \\
U_{\text{encode}}\left(|1\rangle  \otimes |00\rangle \right) &= |\widetilde{1}\rangle = \frac{1}{\sqrt{3}}\left(|012\rangle + |120\rangle + |201\rangle \right) \\
U_{\text{encode}}\left(|2\rangle  \otimes |00\rangle \right) &= |\widetilde{2}\rangle = \frac{1}{\sqrt{3}}\left(|021\rangle + |102\rangle + |210\rangle \right)\,.
\end{align}
Then the noise channels $\mathcal{N}_S$ have the form of Eqn.~\eqref{Noisy1}, where $S$ is either $\{1\}$, $\{2\}$ or $\{3\}$, corresponding to erasing either the first, second or third qutrits.  Then the recovery maps $\mathcal{R}_S$ are
\begin{equation}
\mathcal{R}_S(\rho) = \text{tr}_{g(\overline{S}) \cup S}\big(\left(U_{\overline{S}} \otimes \textbf{1}_{S}\right) \, \rho \, \left(U_{\overline{S}}^\dagger \otimes \textbf{1}_{S}\right)\big)
\end{equation}
where $\overline{S}$ can be $\{1,2\}$, $\{2,3\}$ or $\{1,3\}$, and $g(\{1,2\}) = \{1\}$, $g(\{2,3\}) = \{2\}$, and $g(\{1,3\}) = \{3\}$. Here $U_{\overline{S}}$ is a unitary that takes
\begin{align}
U_{\overline{S}} |00\rangle &= |00\rangle\,, \quad U_{\overline{S}} |11\rangle = |01\rangle\,, \quad U_{\overline{S}} |22\rangle = |02\rangle\,, \\
U_{\overline{S}} |01\rangle &= |12\rangle\,, \quad U_{\overline{S}} |12\rangle = |10\rangle\,, \quad U_{\overline{S}} |20\rangle = |11\rangle\,, \\
U_{\overline{S}} |02\rangle &= |21\rangle\,, \quad U_{\overline{S}} |10\rangle = |22\rangle\,, \quad U_{\overline{S}} |21\rangle = |20\rangle\,.
\end{align}
This code has the property that for any operator $O$ on a qutrit state $|\psi\rangle$ in $\mathcal{H}_A$, we have the equivalences
\begin{equation}
\label{redundantencode1}
U_{\text{encode}} \, O |\psi\rangle = (\widetilde{O}_{12} \otimes \textbf{1}_3) |\widetilde{\psi}\rangle  = (\widetilde{O}_{23} \otimes \textbf{1}_1) |\widetilde{\psi}\rangle = (\widetilde{O}_{13} \otimes \textbf{1}_2) |\widetilde{\psi}\rangle
\end{equation}
for some operators $\widetilde{O}_{12}$, $\widetilde{O}_{23}$ and $\widetilde{O}_{13}$.  This result expresses that the effect of any operation on the original state can be expressed by an equivalent operator on any two of the three qutrits of the encoded state.

\begin{figure}[t]
\center
\includegraphics[width=3.5in]{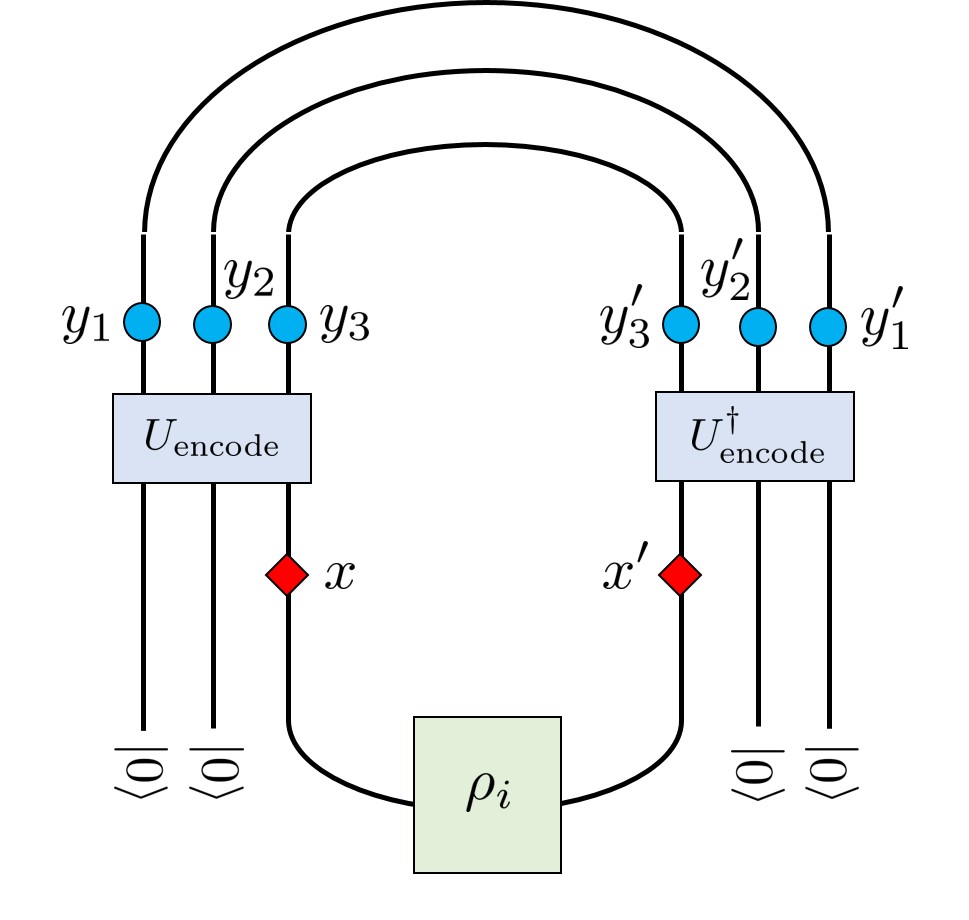}
\caption{The spacetime state for the three qutrit code.}
\label{fig:threequtrit}
\end{figure}

Now let us consider the three qutrit code in spacetime.  A diagram of a spacetime state which implements the three qutrit code is shown in Figure \ref{fig:threequtrit}.  The initial state of the qutrit we wish to encode is $\rho_i$, and the other two qutrits are initialized to $|0\rangle$.  From Eqn.~\eqref{redundantencode1}, it immediately follows that
\begin{equation}
\label{qutritcause1}
\text{CI}(x : y_1) = \text{CI}(x : y_2) = \text{CI}(x : y_3) = 0\,.
\end{equation}
However, we have
\begin{equation}
\label{qutritcause2}
\text{CI}(x : y_1 \, y_2) = \text{CI}(x : y_2 \, y_3) = \text{CI}(x : y_1 \, y_3) > 0
\end{equation}
and
\begin{equation}
\label{qutritcause3}
\text{CI}(x : y_1 \, y_2 \, y_3) > 0 \quad \text{is maximal.}
\end{equation}
By ``maximal,'' we mean that $\text{CI}(x : y_1 \, y_2 \, y_3)$ is as large as possible.  Taken together, Eqn's~\eqref{qutritcause1},~\eqref{qutritcause2} and~\eqref{qutritcause3} demonstrate how a peturbation at $x$ can be non-locally encoded in space so that in the future the perturbation can be detected by any two (or more) qutrits, but not any single qutrit.  More generally, all quantum erasure codes have non-local quantum causal influence between appropriate combinations of subsystems before and after the encoding.

\subsection{Scrambling}

While engineered quantum erasure codes provide examples of systems with nonlocal quantum causal influence, they are somewhat fine-tuned examples.  However, \textit{approximate} quantum error correction codes occur in various contexts in more natural systems.  The simplest example is that of a chaotic quantum many-body system which scrambles information.  The scrambling of information is ubiquitous in nature, since most all physical systems exhibit many-body chaos.  However, the most extreme examples of scrambling systems are black holes, which are the fastest scramblers in nature \cite{sekino2008fast, lashkari2013towards, shenker2014black, maldacena2016bound}.  It was in the context of black holes that scrambling was first explored.  We will not focus on any particular scrambling system, but instead use generic features of scrambling for our analysis.

There are many definitions of information scrambling in the literature. (See, for instance, \cite{hayden2007black, harrow2009random, brown2012scrambling, maldacena2016bound}.  For a short review of diagnostics of scrambling at infinite temperature, see Appendix A of \cite{cotler2017chaos}). Suppose we have a system with a large number $N$ of sites, and that the initial state of the system is $\rho_i$.  If the time evolution $U(t)$ of the system is chaotic, then the scrambling time $t_{\text{scr}}$ is the smallest time such that for any subsystem $a$ of $\mathcal{O}(1)$ size and any subsystem $B$ of size $N/2 + 1$, there exists a quantum channel $\mathcal{R}_{B \to a}$ such that
\begin{equation}
\mathcal{R}_{B \to a}\left[\text{tr}_{\overline{B}}\left(U(t_{\text{scr}}) \, \rho_i \,U^\dagger(t_{\text{scr}})\right)\right] \approx \text{tr}_{\overline{a}}(\rho_i)\,.
\end{equation}
In other words, any $\mathcal{O}(1)$--sized subsystem can be approximately recovered from just over half of the state after a scrambling time.  In this sense, unitary evolution for a scrambling time in a chaotic quantum system creates an (approximate) erasure code for initial subsystems of $\mathcal{O}(1)$ size.  The length of the scrambling time $t_{\text{scr}}$ depends on the types of interactions in the system, and typically scales with the number of degrees of freedom $N$ either polynomially in $N$ (if the interactions are geometrically local) or logarithmically in $N$ (for instance, if the interactions are $k$-local for $k \sim \mathcal{O}(1)$).

\begin{figure}[h]
\center
\includegraphics[width=6.5in]{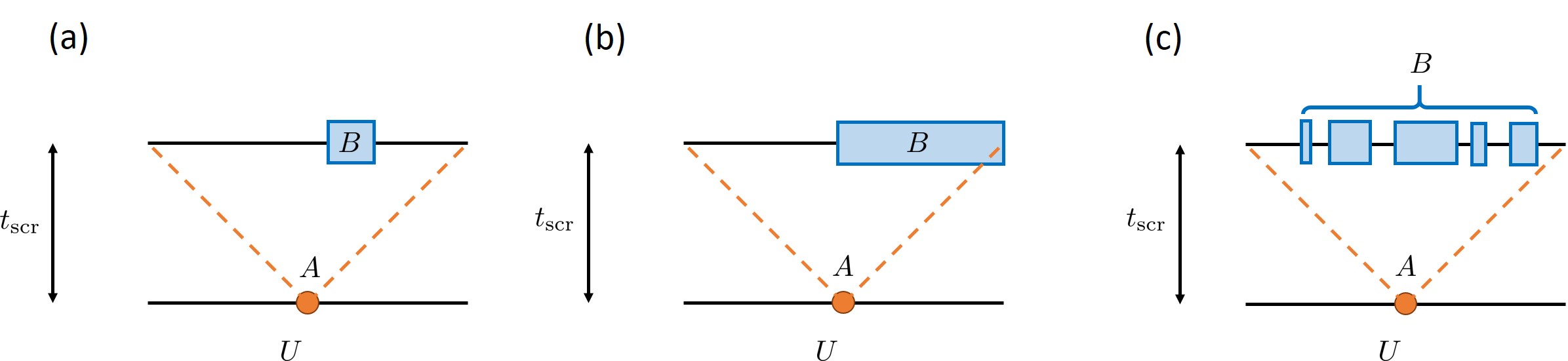}
\caption{A perturbation is made at some initial time, which then spreads out over a scrambling time $t_{\text{scr}}$, inside a cone (shown as dotted orange lines) bounded by the butterfly velocity $v_{B}$ \cite{shenker2014black, roberts2015localized, mezei2017entanglement}.  Here, time runs from bottom to top.  (a) A perturbation at $a$ barely causally influences the subregion $B$, since $B$ is less than half of the system size.  (b) and (c) A perturbation at $a$ strong causally influences $B$, if $B$ is greater than half of the system size.  The two figures illustrate the cases when region $B$ is a contiguous spatial region or the union of many contiguous regions. The conclusion applies to both cases. }
\label{fig:scramble}
\end{figure}

Now consider Figure \ref{fig:scramble} below, which shows a system scrambling (time goes from bottom to top).  In Figure \ref{fig:scramble}(a), we see that the causal influence $\text{CI}(a : B) \approx 0$ since $B$ is less than half of the system size.  However, in Figure \ref{fig:scramble}(b), the causal influence $\text{CI}(a : B)$ is sizeable, since $B$ is greater than half of the system size.  Finally, in Figure \ref{fig:scramble}(c), we have that $\text{CI}(a : B)$ is sizeable since $B$ is greater than half the system size, even though $B$ is not a spatially contiguous subregion.

We emphasize that any $\mathcal{O}(1)$--sized region at the initial time will have a negligible causal influence with any $\mathcal{O}(1)$--sized region in the future after the scrambling time, and conversely as well.  Relatedly, from the point of view of quantum causal influence, local subsystems in the present will appear approximately spacelike separated with local subsystems in the future after the entire system has thermalized.  Indeed, local notions of time disappear after a system thermalizes -- local properties of the past only weakly influence local properties of the far future.

\subsection{Quantum teleportation}

\begin{figure}[t]
\center
\includegraphics[width=5in]{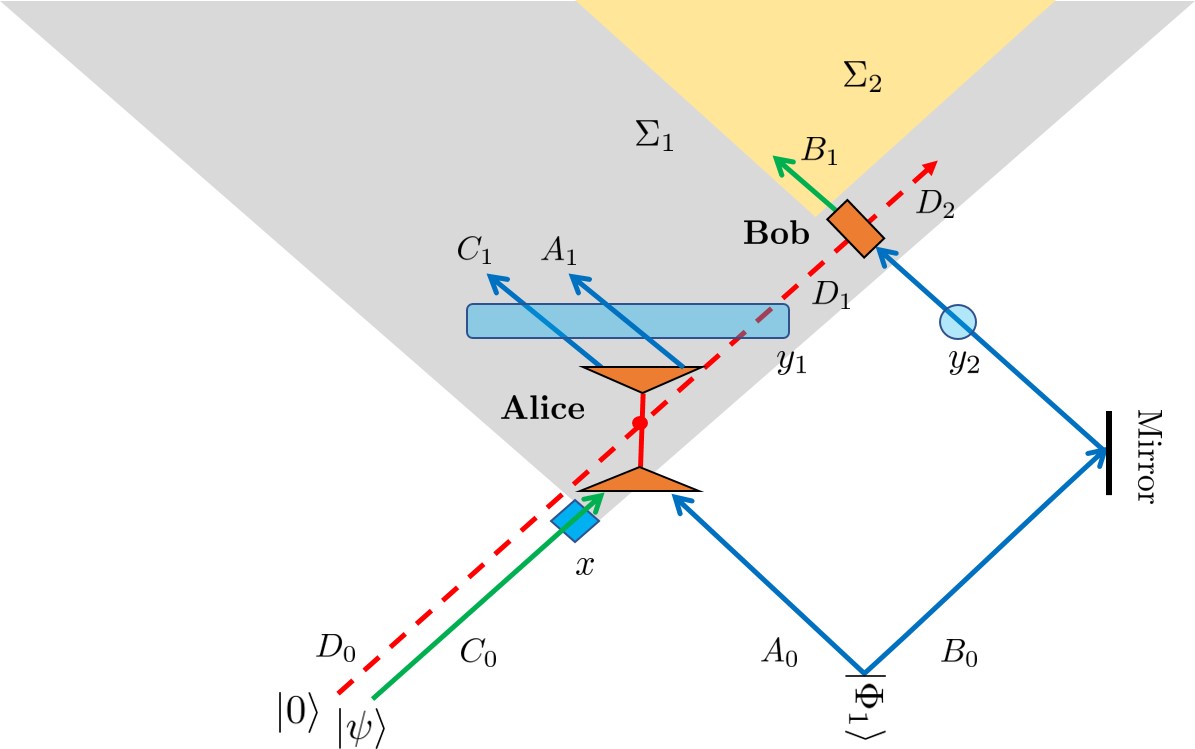}
\caption{A spacetime diagram of the quantum teleportation protocol.  Space runs horizontally, and time runs vertically from bottom to top.  The consequence of teleportation is that the future of $x$ (the teleportee $C_0$ right before teleportation happens) shrinks from the ordinary future light cone $\Sigma_1$ to a subset $\Sigma_2$ (the future of the point where teleportation is finished). Note that the EPR pair of $A_0,B_0$ is still outside the future of $x$, as is expected by microscopic causality.}
\label{fig:teleport1}
\end{figure}

Now we explore how quantum teleportation \cite{bennett1993teleporting} nonlocally encodes information in spacetime.  Quantum teleportation can be described by a tensor network, as shown in Figure \ref{fig:teleport1} below.  In the Figure, space runs horizontally, and time runs vertically from bottom to top.  Let us walk through the protocol step by step.

Consider the setup in Figure \ref{fig:teleport1}.  We suppose that all of the states involved are encoded into photons (say, in their polarization degrees of freedom), which have lightlike trajectories.  We start with a Bell state (i.e., an EPR pair of two qubits)
$$|\Phi_1\rangle_{AB} = \frac{1}{\sqrt{2}}\bigg(|0\rangle_{A_0} |0\rangle_{B_0} + |1\rangle_{A_0} |1\rangle_{B_0} \bigg)\,,$$
a state $|\psi\rangle_C$ that we wish to teleport (i.e., the teleportee), and an ancillary qubit $|0\rangle_{D_0}$.  One qubit of the Bell pair, as well as the joint state $|\psi\rangle_{C_0} \otimes |0\rangle_{D_0}$, are fed into a ``teleporter'' owned by Alice, denoted in the Figure by an orange triangle.  Letting
\begin{align*}
|\Phi_1\rangle &= \frac{1}{\sqrt{2}}\left(|0\rangle |0\rangle + |1\rangle |1\rangle \right) \\
|\Phi_2\rangle &= \frac{1}{\sqrt{2}}\left(|0\rangle |0\rangle - |1\rangle |1\rangle \right) \\
|\Phi_3\rangle &= \frac{1}{\sqrt{2}}\left(|0\rangle |1\rangle + |1\rangle |0\rangle \right) \\
|\Phi_4\rangle &= \frac{1}{\sqrt{2}}\left(|0\rangle |1\rangle - |1\rangle |0\rangle \right)
\end{align*}
denote the Bell states (which are the basis vectors of the Bell basis), the teleporter jointly measures $A_0 C_0$ in the Bell basis so that the output is $|\Phi_j\rangle_{A_1 C_1}$ for some $j$, and $|0\rangle_{D_0}$ is put into the state $|j\rangle_{D_1}$ to label the Bell state that has been measured.\footnote{The measurement can also be represented as a quantum channel, or equivalently by Stinespring dilation as a unitary with additional ancilla systems that are traced out.}  The $A_1$ and $C_1$ subsystems are discarded, while the $D_1$ subsystem goes on to Bob.  In the meantime, the $B$ subsystem of the Bell state is directed towards Bob with a mirror.  When Bob receives $B_0$ and $D_1$, he applies the unitary
\begin{equation}
U_{2,\,B_0 D_1} = \sum_{j=1}^4 U_{B_0,\,j} \otimes |j\rangle_{D_1} \langle j| 
\end{equation}
which is denoted by an orange box.  The unitary $U_{2,\,B_0 D_1}$ applies the unitary $U_{B_0,\,j}$ to the $B_0$ subsystem, controlled by the state of $D_1$.  The output of the $B_1$ subsystem will be the original state of $A_0$, namely $|\psi\rangle$, which has successfully been teleported to Bob.

Now we analyze the causal future of the initial state $|\psi\rangle_{C_0}$, denoted by the initial subsystem $C_0$. Apparently in the protocol, the future of $C_0$ is $B_1$.  In fact it can be checked that
\begin{equation}
\text{CI}(C_0 : B_1) > 0 \quad \text{is maximal}.
\end{equation}
(As before, ``maximal'' means that the quantum causal influence is as large as possible.)  However, denoting $y_1 = A_1 \cup C_1 \cup D_1$, we also have that
\begin{equation}
\text{CI}(C_0 : y_1) = 0
\end{equation}
and thus $C_0$ is spacelike separated from $A_1 \cup C_1 \cup D_1$ and any subset thereof.  We also have that
\begin{equation}
\text{CI}(C_0 : y_2) = 0
\end{equation}
which means that $C_0$ is spacelike separated from $B_0$.  This is consistent with the causal structure which Figure \ref{fig:teleport1} inherits from Minkowski space.

In summary, even though it appears that $C_0$ should be able to influence its whole future light cone $\Sigma_1$, it can only causally influence the subset $\Sigma_2$.  In words:
\begin{equation}
C_0\textit{ cannot influence any local region while it is being teleported.}
\end{equation}
Even though $\text{CI}(C_0:y_1) = 0$ and $\text{CI}(C_0:y_2) = 0$, we still have that
\begin{equation}
\text{CI}(C_0 : y_1 \cup y_2) > 0
\end{equation}
which is in fact maximal.  Thus, while the state of $C_0$ is not encoded in either $A_1 \cup C_1 \cup D_1$ alone or $B_0$ alone, $C_0$ \textit{is} encoded in $(A_1 \cup C_1 \cup D_1) \cup B_0$.

From another point of view, the example of quantum teleportation shows again that the causal structure depends on properties of the initial state, in this case the presence of the Bell state $|\Phi_1\rangle$.  Fine-tuning of the initial state can only reduce the size of the putative future of spatial subregions. Said simply, special initial states can {\it  remove} regions from the future. 



\section{Quantum gravity examples}
\label{sec:QGexamples}

In this section, we discuss several examples in holography as well as models of black holes for which quantum causal influence is a useful measure. In Section \ref{sec:holographicTN} we discuss holographic tensor networks and show how the causal influence correctly reproduces the bulk causal structure. In Section \ref{sec:finalstate} we discuss the causal structure in the Horowitz-Maldacena final state projection model of black hole.

\subsection{Holographic tensor networks}\label{sec:holographicTN}

\subsubsection{Holographic states}
\label{subsec:holoTNpart1}

An interesting instantiation of quantum error correction codes in high energy physics is in holographic systems, and specifically AdS-CFT \cite{maldacena1999large, witten1998anti}.  In AdS-CFT, there is a duality between a $(d+1)$--dimensional quantum gravity theory in AdS space (i.e., the bulk theory), and a $d$--dimensional conformal field theory which lives on a space isomorphic to the conformal boundary of AdS (i.e., the boundary theory).  There is necessarily an intricate relationship between  degrees of freedom in the bulk and the boundary, and in fact, low-energy degrees of freedom in the bulk are non-locally encoded in the boundary theory in the form of a quantum erasure code \cite{almheiri2015bulk}.  In particular, a local low energy operator acting in the bulk can be reconstructed from many distinct spatial regions in the boundary theory.

The quantum error correction property of AdS-CFT duality can be captured in toy models known as holographic tensor networks \cite{pastawski2015holographic,hayden2016holographic}.  We will consider quantum causal influence in holographic tensor networks, and study its relation to the bulk causal structure.

\begin{figure}[t]
\center
\includegraphics[width=3.5in]{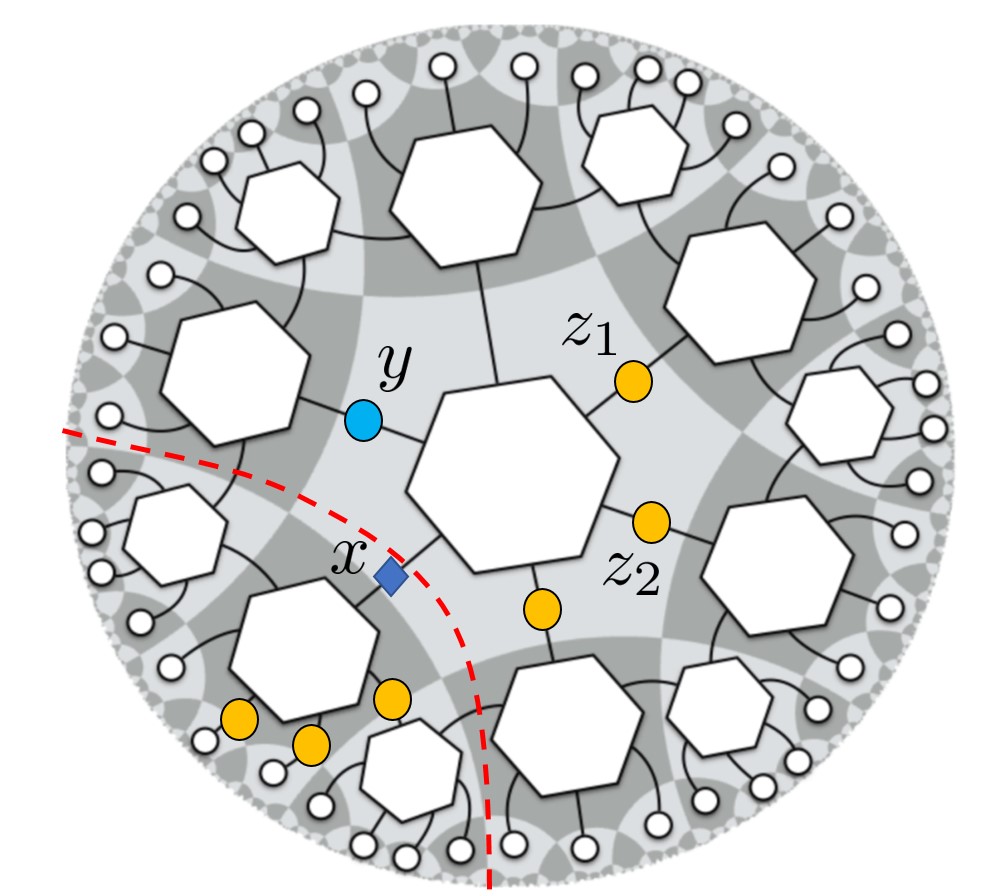}
\caption{A hyperbolic perfect tensor network in which $y$ is not in the future of $x$, as $U_x$ can be pushed to the boundary only using the circuit on the left-side of the geodesic (red dashed line). A multi-site region such as $z$ (6 yellow links) can be causally influenced by $x$ since there is no way to push $U_x$ to boundary without passing through $z$.  This type of non-local causal influence is characteristic of holographic systems, and does not occur, for example, on a fixed Cauchy slice for a quantum field theory.  This Figure is adapted from \cite{pastawski2015holographic}. }
\label{fig:holofig1}
\end{figure}

As an example, we consider the hyperbolic perfect tensor network state defined by the work of Pastawski et al. \cite{pastawski2015holographic}, shown in Figure \ref{fig:holofig1}.  (All the discussion in the following also applies to the random tensor networks in large bond dimension limit proposed in Ref. \cite{hayden2016holographic}.) A perfect tensor network state represents a many-body quantum state of the boundary legs, with its wavefunction defined by contracting perfect tensors.  Each perfect tensor is a rank $2n$ tensor $T_{a_1 \cdots a_{2n}}$ such that the bipartition of its indices into sets $A$ and $A^c$ with $|A| \leq |A^c|$ defines an isometry from $A$ to $A^c$ up to a normalization constant.  In Figure \ref{fig:holofig1}, we have considered the case $n=3$, and the only uncontracted legs of the tensor network state live near the boundary of a hyperbolic disk.\footnote{The hyperbolic disk has infinite area. We have imposed a radial cutoff so that it has finite area.  The uncontracted tensors live on the radial cutoff.}  Thus, the tensor network state in Figure \ref{fig:holofig1} forms a so-called ``holographic state'', which we denote by $|\Psi\rangle$.  The essential feature of this state is that if we break open any bulk leg (i.e., a non-boundary leg) of the tensor network state and stick in an operator, we can (non-uniquely) push it through the isometries out to the boundary, and so rewrite the operator as a ``boundary'' operator.  This mimics the AdS-CFT correspondence: operators inserted into the bulk can be rewritten non-uniquely as operators applied to some boundary state.

Suppose we break open two links $x$ and $y$ of $|\Psi\rangle$ to insert operators.  If we insert operators $A_x$ and $B_y$ into $x$ and $y$, respectively, we denote the resulting state by $|\Psi[A_x, B_y]\rangle$.  While we can express $\langle \Psi| \Psi\rangle$ as
\begin{equation}
\langle \Psi | \Psi\rangle = \langle L | \rho_P |L\rangle
\end{equation}
and similarly express $\langle \Psi[A_x, B_y]| \Psi[A_x, B_y]\rangle$ as
\begin{equation}
\langle \Psi[A_x, B_y]| \Psi[A_x, B_y]\rangle = \langle L| B_y^\dagger \, A_x^\dagger \, \rho_P \, A_x \, B_y |L\rangle
\end{equation}
As per our definition of GTN's, $\rho_P$ is the tensor product of vertex tensors (where we choose the boundary vertex tensors to be identity operators) and $|L\rangle$ is the link state comprised of EPR pairs.

We usually speak of the causal structure of a \textit{fully contracted} tensor network (such as the one which computes $\langle \Psi | \Psi\rangle$), but here is it convenient to speak of the causal structure of the \textit{state} $|\Psi\rangle$ (which has uncontracted legs).  This is purely for terminological convenience -- we always have in mind computing expectation values like $\langle \Psi[A_x, B_y]| \Psi[A_x, B_y]\rangle$.  So when we say ``the causal structure of $|\Psi\rangle$,'' we mean  ``the causal structure of $|\Psi\rangle$ contracted with itself.''

With our terminology defined, we now discuss quantum causal influence for the holographic state $|\Psi\rangle$ in Figure \ref{fig:holofig1}. For any two links $x$ and $y$, as long as they can be separated by a geodesic line on the hyperbolic disk, a unitary $U_x$ inserted at $x$ can be pushed to the boundary without using the $y$ link, so that $y$ is not in the causal future of $x$.  Examination of the holographic state reveals that any two links can be separated by a geodesic line on the holographic disk, and therefore
\begin{align}
&\langle \Psi[U_x, O_y]|\Psi[U_x, O_y]\rangle \text{  is independent of }U_x\,, \nonumber \\ \nonumber \\
&\langle \Psi[O_x, U_y]|\Psi[O_x, U_y]\rangle \text{  is independent of }U_y\,. \nonumber
\end{align}
It follows that $\text{CI}(x : y) = \text{CI}(y:x)=0$, so that any two links $x$ and $y$ in network are ``spacelike separated.''

Our operational definition of causal structure explains why perfect tensor network states should be understood as \textit{spatial} tensor network states even if their isometry conditions allow one to push operators around.  Indeed, the perfect tensor network state is an example where all small enough regions are spacelike separated, but larger size regions may be causally dependent (i.e., if such regions cannot be separated by a geodesic line on the hyperbolic disk). For example, in Figure \ref{fig:holofig1}, $x$ does not influence $y$, or any of the yellow points $z_1,~z_2,...$ individually.  Furthermore, $x$ does not influence the pair $z_1 \cup z_2$, since $x$ can be separated from $z_1 \cup z_2$ by a geodesic on the hyperbolic disk.  However, $x$ does causally influence the subregion that is the union of \textit{all} the yellow dots, since there is no way to push operators at $x$ to the boundary without overlapping with this subregion. 

\subsubsection{Exotic quantum Cauchy slicings of holographic states}
\label{subsec:exotic}
\begin{figure}[t]
\center
\includegraphics[width=3in]{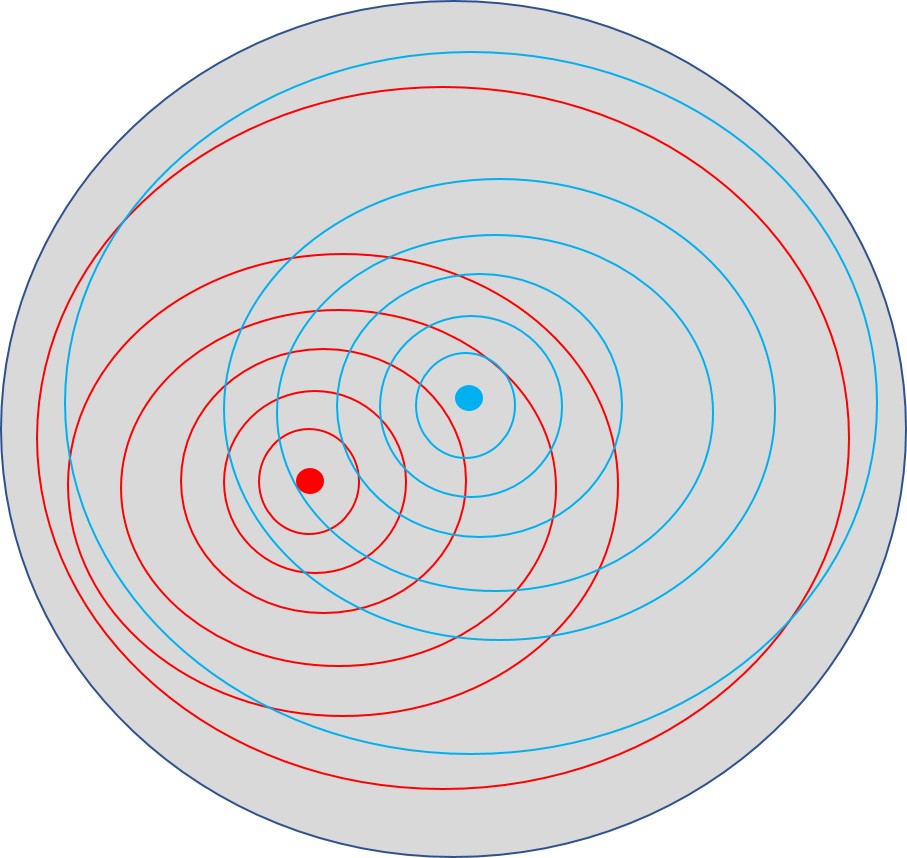}
\caption{Exotic quantum Cauchy slices of the HaPPY code holographic state.}
\label{fig:holofig2}
\end{figure}

In Figure \ref{fig:holofig2}, we provide some further illustration of the unconventional causal structure in the holographic tensor network state $|\Psi\rangle$.  In the Figure, the tensor network has been abstracted to a gray disk.  Consider a set of concentric rings on the hyperbolic disk (the red circles in Figure \ref{fig:holofig2}).  Each red ring defines a subsystem into which we can insert operators (i.e., corresponding to inserting operators into all links that the red ring cuts through).  Then we find that the subsystem corresponding to a red ring $R_1$ causally influences a subsystem corresponding to any bigger red ring $R_2$ that encloses $R_1$.  The influence is in fact maximal since there is an isometry from $R_1$ to $R_2$. Indeed, a pair of subsystems corresponding to a pair of concentric red rings has timelike separation with respect to the QCI.  Therefore, the concentric red rings are quantum analogs of Cauchy slicing of the holographic state.  We will not attempt to define quantum Cauchy slices in full generality, but will comment further in Section \ref{sec:conclusion}.  The concentric ring subsystems provide an exotic causal structure where the radial direction acts as time -- this is dramatically different from more familiar examples.  For instance, this exotic causal structure does not admit light cones.
 
There are many possible, incompatible Cauchy slicings of the holographic state, corresponds to different sets of concentric rings.  For instance, in Figure \ref{fig:holofig2}, the set of blue rings is another Cauchy slicing with the same property as the red rings. However, the red and blue Cauchy slicings are not compatible with each other, since the subsystem corresponding to some red ring may not be time-like separated with the subsystem corresponding to some blue ring. This situation never occurs with standard Cauchy slicings of a classical spacetime with Lorentzian signature.  The exotic Cauchy slicing found here is essential for bulk reconstruction to be consistent with the homogeneity of the bulk (i.e., there is no preferred point or preferred direction on the hyperbolic disk), which is the key difference between perfect tensor network states (as well as random tensor network states) and earlier proposals of MERA \cite{vidal2008class,swingle2012entanglement}.

In summary, the nonlocality of quantum causal influence characterizes how bulk locality is consistent with bulk reconstruction, as a consequence of the bulk's quantum error correction properties. The bulk contains a redundant encoding of boundary quantum information as is evident in the Cauchy surface structure, but this redundancy is invisible for local observers.

\subsubsection{Explicit time direction}

\begin{figure}[t]
\center
\includegraphics[width=5in]{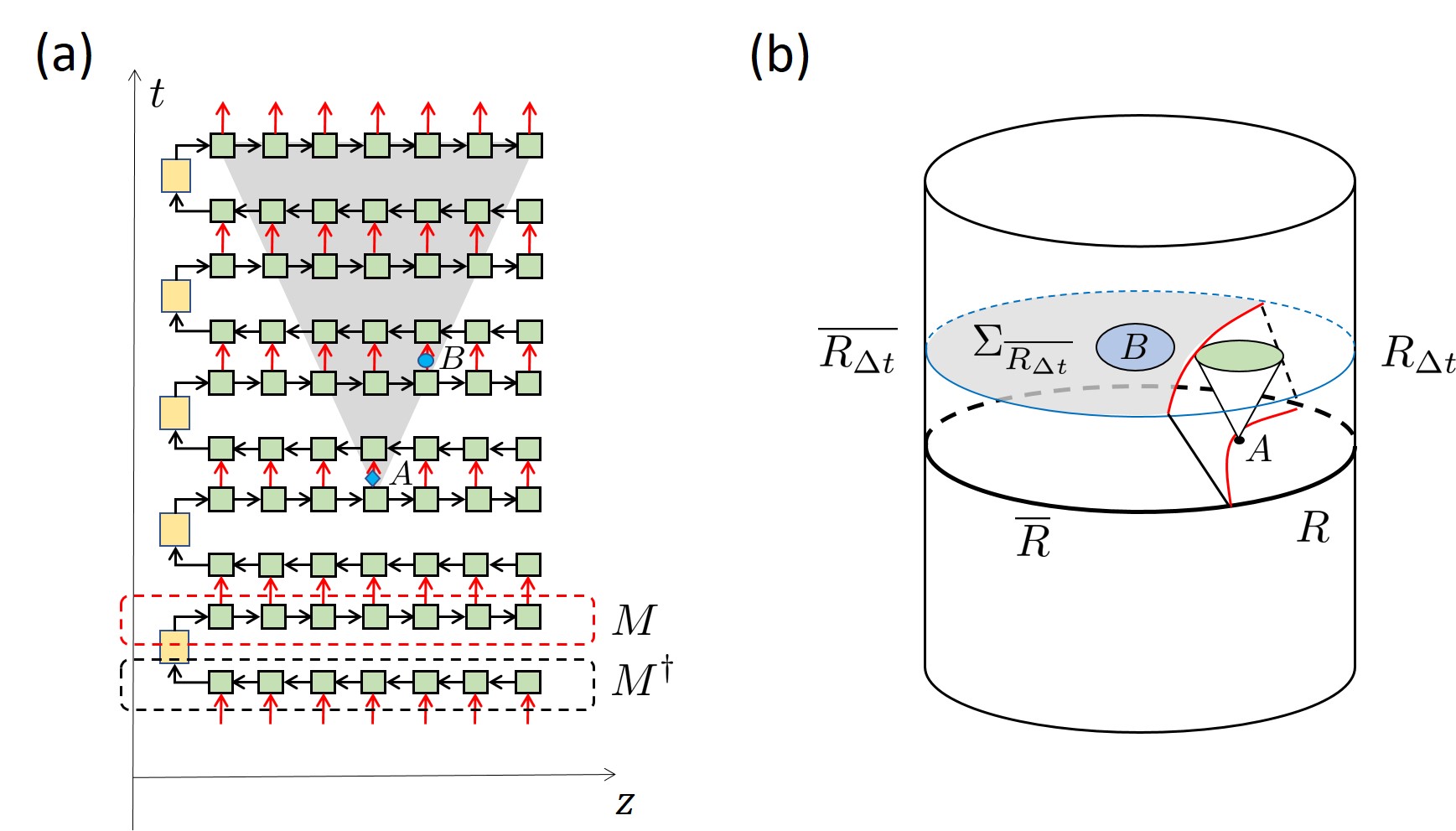}
\caption{Causal influence for two bulk regions at different times. (a) Bulk time evolution is defined by pulling back the boundary time evolution using a holographic tensor network (see text). (b) An illustration in a (2+1)d bulk.  An operator on a small region $A$ can be reconstructed in a boundary region $R$, which evolves into a slightly bigger region $R_{\Delta t}$ after a short time $\Delta t$. Therefore, all operators in the complement entanglement wedge $\Sigma_{\overline{R_{\Delta t}}}$ still commute with the operator at $A$, which proves that $A$ has no causal influence on any region $B\subset \Sigma_{\overline{R_{\Delta t}}}$. }
\label{fig:holofig3}
\end{figure}

The discussion above can be further generalized by considering an \textit{explicit} time direction via unitary evolution of the holographic state $|\Psi\rangle$. This section will be more technical, and we refer readers to \cite{hayden2016holographic} and \cite{qi2017butterfly} for details.  To describe the bulk dynamics of low-energy degrees of freedom, consider the holographic mapping (or holographic code) defined by a random tensor network with bulk and boundary indices.  Such a network defines a linear map
\begin{equation}
M : \mathcal{H}_{\text{bulk}} \longrightarrow \mathcal{H}_{\text{bdy}}
\end{equation}
from low-energy bulk degrees of freedom to the boundary.  The map is an isometry when the included bulk degrees of freedom have low enough dimension \cite{hayden2016holographic}.  We call the image of $\mathcal{H}_{\text{bulk}}$ under $M$ the ``code subspace'' of $\mathcal{H}_{\text{bdy}}$, which we denote by $\mathcal{H}_{\text{code}} := M(\mathcal{H}_{\text{bulk}})$.  Indeed, we have $\mathcal{H}_{\text{code}} \subset \mathcal{H}_{\text{bdy}}$.

In Figure \ref{fig:holofig3}, we illustrate such a mapping $M$ in the red dashed box. (The drawing is for a (1+1)d bulk for convenience, but the setup applies to arbitrary dimensions.) With this mapping $M$, boundary time evolution can be ``pulled back'' to the bulk and to define the bulk time evolution. With the boundary time evolution operator $e^{-iH\Delta t}$ for small $\Delta t$, the bulk time evolution is given by $U_{\rm bulk}=Me^{-iH\Delta t}M^\dagger$ (which is unitary in the code subspace if the boundary time evolution preserves the code subspace). Na\"{i}vely, this time evolution is very nonlocal in the bulk, since we have to map all operators to (non-local) operators on the boundary and then map them back after the time evolution. However, the quantum error correction properties and locality of boundary dynamics actually guarantees that the bulk evolution also has a local causal structure \cite{qi2017butterfly}.

The basic idea is illustrated in Figure \ref{fig:holofig3}(b). An operator $\phi_A$ in a small bulk region $A$ can be reconstructed in a boundary region $R$. Then due to boundary locality, the operator $\phi_A$ at a slightly later time $\Delta t$ will live in a slightly larger region $R_{\Delta t}$. Consequently, all bulk operators in the entanglement wedge $\Sigma_{\overline{R_{\Delta t}}}$ of the complement $\overline{R_{\Delta t}}$ still commute with the (slightly) Heisenberg-evolved operator $\phi_A$.  This implies that for any bulk region $B\in \Sigma_{\overline{R_{\Delta t}}}$, we have ${\rm CI}(A:B)=0$. Since the reconstruction can be done on different boundary regions $R$, the argument applies to each possible $R$. As long as $B$ is included in the complement of the entanglement wedge of some $R_{\Delta t}$, there will be no causal influence from $B$ to $R$ or $R_{\Delta t}$.


If we consider regions $B$ that are infinitesimal disks on the $\Delta t$ time slice, any $B$ that is outside the domain of support of $A$ at time $\Delta t$ is not influenced by $A$.  In Figure \ref{fig:holofig3}(b), we see that any small blue disk $B$ which does not intersect the green disc (which is the domain of support of $A$ at time $\Delta t$) is spacelike separated from the green disc.  Therefore, we recover the ordinary causal structure expected for the bulk theory. The boundary of the domain of support of $A$ at time $\Delta t$ (i.e., the green region in the Figure) defines an upper bound of the bulk speed of light \cite{qi2017butterfly}.

Now, if we consider more generic regions $B$ that are not small discs, the influence of $B$ with the domain of support of $A$ at time $\Delta t$ can be nontrivial even there is no intersection between these regions.  For example, if $B$ is a ring enclosing the domain of support of $A$ at time $\Delta t$, the causal influence will be nontrivial, since the reconstruction of operators in boundary region $R_{\Delta t}$ must use a bulk region that overlaps with $B$. This is similar to the exotic quantum Cauchy surfaces discussed above for the equal-time case.

\subsection{Black hole final state}\label{sec:finalstate}

\begin{figure}[t]
\center
\includegraphics[width=4.8in]{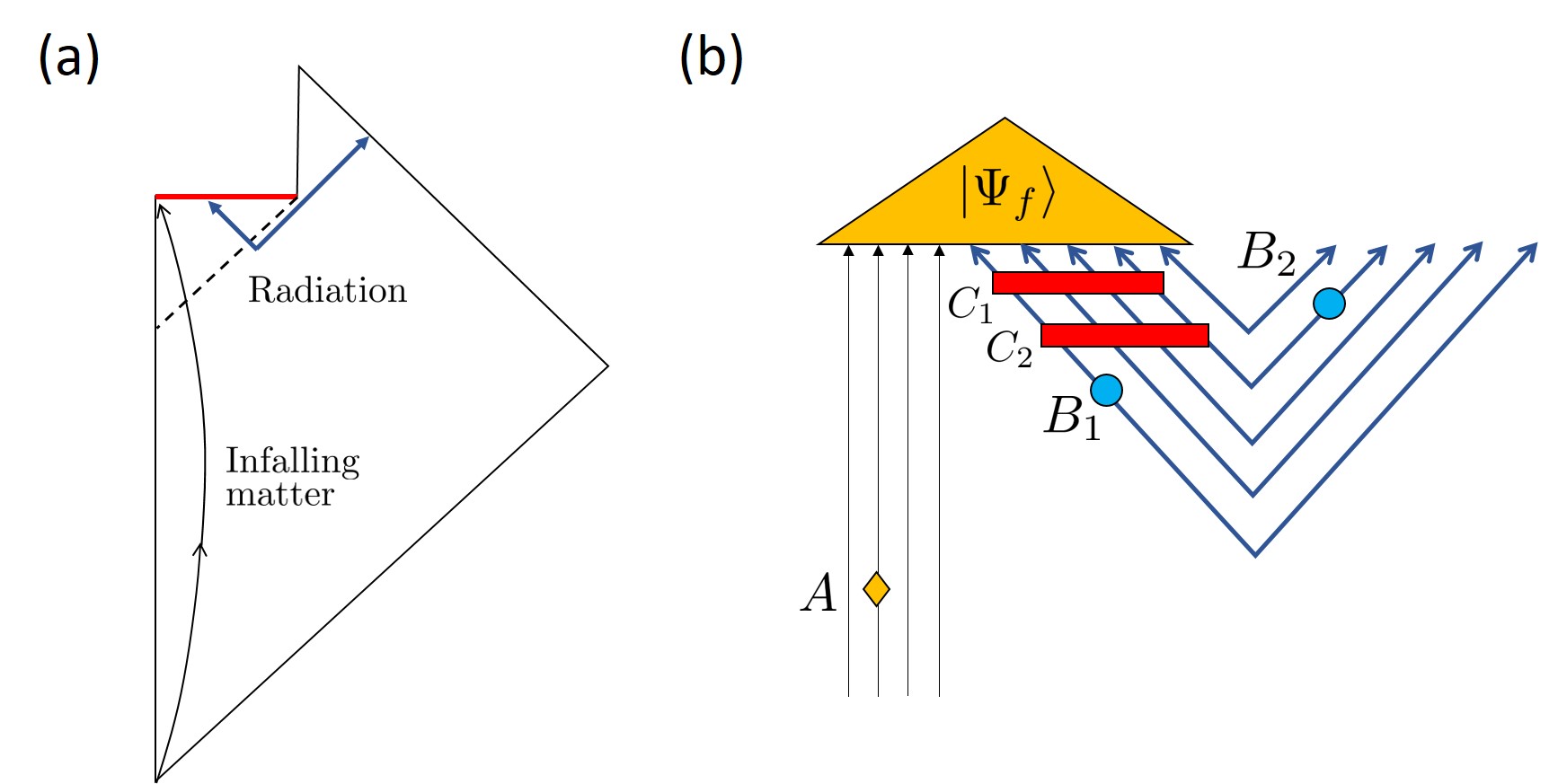}
\caption{(a) The Penrose diagram of a Schwarzchild black hole with infalling matter (black curve) and infalling and outgoing Hawking radiation (blue lines). The red line represents singularity. (b) The Horowitz-Maldacena final state projection model, with the infalling matter and infalling Hawking qubits projected to a pure state at the singularity. $A,B_1,B_2$ are small regions of infalling matter, infalling radiation and outgoing radiation, respectively. $C_1,C_2$ are bigger regions of the infalling radiation, for which artifacts of the final state projection are detectable.}
\label{fig:BHfinalstate}
\end{figure}

In Section \ref{sec:postselect} we discussed how for spacetime states, the causal influence depends in a similar manner on both the initial and final states.  The initial and final states act as boundary conditions for the spacetime state.  An interesting example of a nontrivial final state is the final state projection model of the black hole singularity, proposed by Horowitz and Maldacena \cite{horowitz2004black}. This model is illustrated in Figure \ref{fig:BHfinalstate}. There is infalling matter (the black curve), as well as infalling and outgoing radiation.  The outgoing radiation is Hawking radiation, and the infalling radiation can be thought of as the ``Hawking partner'' of the Hawking radiation \cite{harlow2016jerusalem}.  The outgoing and infalling radiation form a maximally entangled state.\footnote{The situation is more complicated when the entanglement is not maximal, but we will not discuss this here.}  The hypothesis is that there is a (post-selected) final state at the singularity, and that all matter and radiation falling into the singularity are projected onto that fixed final state. Such a projection will generically violate unitarity, but when the final state is chosen properly, the information content of infalling matter is mapped unitarily to outgoing radiation.  This is much like quantum teleportation: a desired state (infalling matter) and half of a maximally entangled state (infalling radiation) are jointly measured (projection onto black hole final state), and the desired state is teleported to the other half of the maximally entangled state (outgoing Hawking radiation).

For example, suppose the black hole final state $|\Psi_f\rangle$ is a Haar random state.  The state $|\Psi_f\rangle$ lives on the Hilbert space $\mathcal{H}_{M} \otimes \mathcal{H}_R$ where $\mathcal{H}_M$ is the Hilbert space of the infalling matter and $\mathcal{H}_R$ is the Hilbert space of the infalling radiation.  Then $|\Psi_f\rangle$ has the form
\begin{equation*}
|\Psi_f\rangle = \sum_{i,j} c_{ij} |i\rangle_M \otimes |j\rangle_R\,. 
\end{equation*}
By dualizing $\mathcal{H}_M$ to $\mathcal{H}_M^*$ (and thus $|i\rangle_M \to \langle i |_M$), we can re-express $|\Psi_f\rangle$ as a mapping $V_{\Psi_f} : \mathcal{H}_M \to \mathcal{H}_R$ from the infalling matter to the infalling radiation as
\begin{equation}
V_{\Psi_f} = \sum_{i,j} c_{ij} |j\rangle_R \langle i|_M\,.
\end{equation}
Indeed, if $|\Psi_f\rangle$ is Haar random and $\dim \mathcal{H}_M < \dim \mathcal{H}_I$ (i.e., the Hilbert space dimension of infalling matter is smaller than that of the infalling radiation), the mapping $V_{\Psi_f}$ is an isometry (up to exponentially small corrections in the number of degrees of freedom).  Since the infalling and outgoing radiation are maximally entangled, the net effect is that the information in the infalling matter is preserved in the outgoing Hawking radiation, and the unitarity of the quantum mechanics of the exterior region is restored (up to exponentially small corrections in the number of degrees of freedom) \cite{lloyd2006almost,verlinde2013black}. 

Since the final state plays the role of an (approximately) isometric mapping from the infalling matter to infalling radiation, unitary operations at $A$ have nontrivial causal influence on both the infalling and outgoing radiation. However, when $|\Psi_f\rangle$ is a Haar random state, its corresponding (approximately) isometric mapping $V_{\Psi_f}$ is a \textit{random} (approximate) isometry, and so the quantum causal influence of $A$ is highly nonlocal.  Accordingly, the quantum causal influence of $A$ on any small subsystem such as $B_1,B_2$ nearly vanishes. The influence due to $A$ is only nontrivial on large enough regions such as $C_1,C_2$.  This is the same phenomenon as the nonlocal causal influence we observed in quantum error correction codes (see \cite{verlinde2013black} for a related discussion).

The near vanishing of both $\text{CI}(A:B_1)$ and $\text{CI}(A:B_2)$ is consistent with the causal structure in the Penrose diagram in Figure \ref{fig:BHfinalstate}(a), since the Penrose diagram suggests that $A$ is spacelike separated from both $B_1$ and $B_2$. When we consider the quantum causal influence from $A$ to larger regions such as $C_1$ and $C_2$, we can observe abnormal causal structure that is at odds with the Penrose diagram. For example, we have $\text{CI}(A:C_1) \not = 0$ and $\text{CI}(A:C_2) \not = 0$.  Furthermore, the quantum causal influence between pairs of large regions also unveils abnormal quantum causal influence, for instance ${\rm CI}(C_1,C_2)\neq 0$ and ${\rm CI}(C_2,C_1)=0$, which means that the time ordering of big regions $C_1,C_2$ for infalling radiation has been reversed due to the final state projection. The reverse time ordering is consistent with the observation that measurements involving large regions can detect violations of standard (non-post-selected) quantum mechanics \cite{gottesman2004comment,bousso2014measurements}.

\section{Averaged quantum causal influence and spacetime quantum entropies}\label{sec:measures}

In this section, we perform a more quantitative analysis of the averaged quantum causal influence (aQCI) and discuss its relation to spacetime quantum entropies in the superdensity operator formalism.  We also use our results to analyze the quantum causal structure of evolving quantum spin chains as well as stabilizer tensor networks. 

\subsection{Relation to spacetime quantum R\'{e}nyi entropies}

\begin{figure}[t]
\center
\includegraphics[width=6.5in]{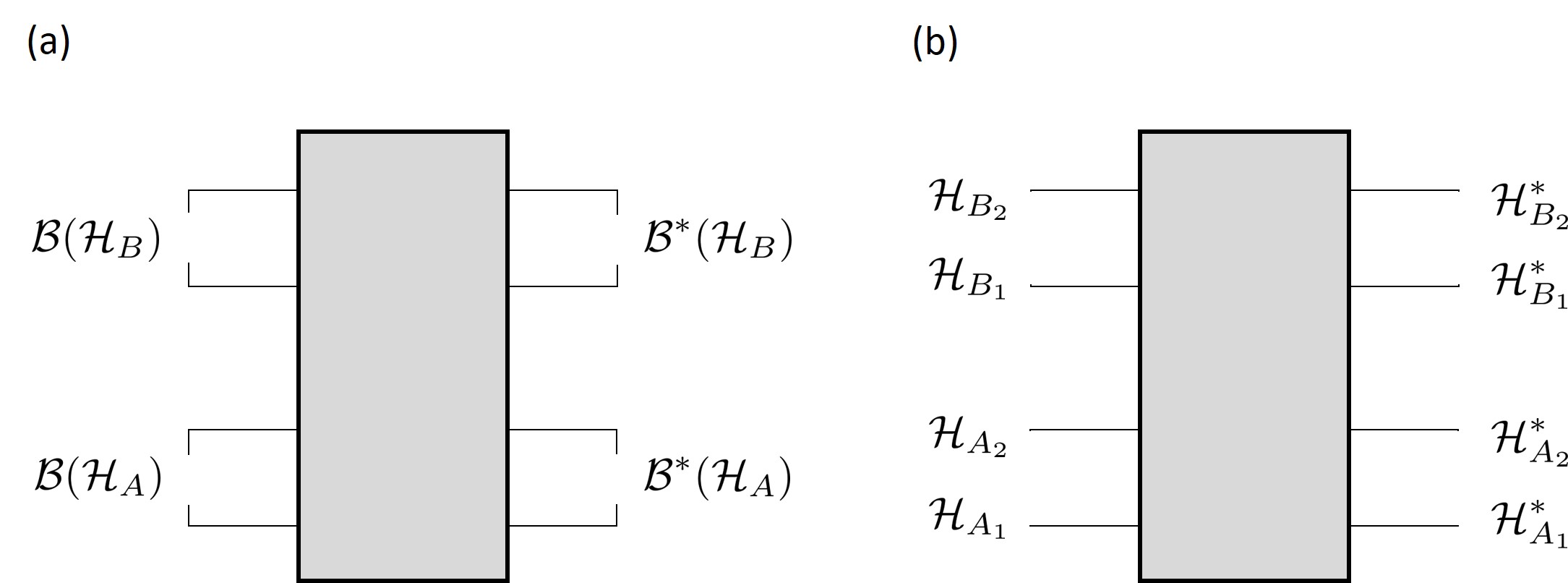}
\vskip.5cm
\caption{(a) A diagrammatic representation of the $R$ tensor, as per Eqn.~\eqref{Rundualized1}. (b) An equivalent diagrammatic representation of the $R$ tensor, where tensor legs have been relabelled by the isomorphism of Hilbert spaces as per Eqn.~\eqref{Rdualized1}.  \label{fig:Rop1}}
\end{figure}

In Section \ref{sec:setup} we presented two measures of quantum causal influence. The aQCI defined in Eqn.~\eqref{averagedinfluence} is easier to compute than the mQCI defined in Eqn.~\eqref{maximalinfluence}.  For the aQCI, we can in fact explicitly carry out the average over $U_A$ and $O_B$. The aQCI can be written as
\begin{align}
\overline{\text{CI}}(A:B)&=\int_{\|O_B\|_2^2 = 1} dO_B\int dU_A \, \left|M(U_A:O_B)\right|^2-\int_{\|O_B\|_2^2 = 1} dO_B\left|\int dU_A \, M(U_A:O_B)\right|^2
\end{align}
To obtain a more explicit expression of $\overline{\text{CI}}(A:B)$, we define an orthonormal basis $\{\ket{n_A}\}$ of $\mathcal{H}_A$, and similarly $\{\ket{n_B}\}$ of $\mathcal{H}_B$. Since $M(U_A : O_B)$ is quadratic in $U_A$ and in $O_B$, we can define a tensor $R_{nm\alpha\beta}^{k\ell\gamma\delta}$, such that
\begin{eqnarray}
M(U_A : O_B)=U_A^{nm*}O_B^{\alpha\beta*}R_{nm\alpha\beta}^{k\ell\gamma\delta}U_A^{k\ell}O_B^{\gamma\delta}\label{superdensitydef}
\end{eqnarray}
Here, $R$ can be thought of as a positive semidefinite operator mapping 
\begin{equation}\\
\label{Rundualized1}
R : \mathcal{B}(\mathcal{H}_A) \otimes \mathcal{B}^*(\mathcal{H}_{A}) \otimes \mathcal{B}(\mathcal{H}_B) \otimes \mathcal{B}^*(\mathcal{H}_{B}) \longrightarrow \mathbb{C}
\end{equation}
which is depicted in Figure \ref{fig:Rop1}(a).  Since $\mathcal{B}(\mathcal{H}_A) \simeq \mathcal{H}_{A_1} \otimes \mathcal{H}_{A_2}$ where $\mathcal{H}_{A_1} \simeq \mathcal{H}_A$ and $\mathcal{H}_{A_2} \simeq \mathcal{H}_A$ (and similarly for $\mathcal{B}^*(\mathcal{H}_A)$, $\mathcal{B}(\mathcal{H}_B)$, $\mathcal{B}^*(\mathcal{H}_B)$), we can treat $R$ as a mapping
\begin{equation}
\label{Rdualized1}
R : (\mathcal{H}_{A_1} \otimes \mathcal{H}_{A_2}) \otimes (\mathcal{H}_{A_1}^* \otimes \mathcal{H}_{A_2}^*) \otimes (\mathcal{H}_{B_1} \otimes \mathcal{H}_{B_2}) \otimes (\mathcal{H}_{B_1}^* \otimes \mathcal{H}_{B_2}^*) \longrightarrow \mathbb{C}
\end{equation}
which is depicted in Figure \ref{fig:Rop1}(b).  If $A$ and $B$ are each unitary regions, with proper normalization, $R$ for a spacetime tensor network is an example of a superdensity operator \cite{cotler2017superdensity}. The Haar average of $U_A$ and $O_B$ can be carried out with the following identities:
\begin{eqnarray}
\int dU \, U_{nm}^*U_{k\ell}&=&\frac1{d_A}\,\delta_{nk}\delta_{m\ell}\nonumber\\
\int_{\|O\|_2^2 = 1} dO \, O_{\alpha \beta}^*O_{\gamma\delta}&=&\frac1{d_B^2}\,\delta_{\alpha\gamma}\delta_{\beta\delta}\nonumber\\ \nonumber\\
\int dU\,{U_{n_1m_1}^*U_{n_2m_2}^*U_{k_1 \ell_1}U_{k_2 \ell_2}}&=&
\frac1{d_A^2-1}\left[\delta_{n_1k_1}\delta_{m_1l_1}\delta_{n_2k_2}\delta_{m_2 \ell_2}+\delta_{n_1k_2}\delta_{m_1 \ell_2}\delta_{n_2k_1}\delta_{m_2 \ell_1}\right.\nonumber\\ \nonumber\\
& &\left.-\frac1{d_A}\,\delta_{n_1k_1}\delta_{n_2k_2}\delta_{m_1 \ell_2}\delta_{m_2 \ell_1}-\frac1{d_A}\,\delta_{n_1k_2}\delta_{n_2k_1}\delta_{m_1 \ell_1}\delta_{m_2 \ell_2}\right]\nonumber\\ \nonumber\\
\int_{\|O\|_2^2 = 1} dO \, O_{\alpha_1\beta_1}^*O_{\alpha_2\beta_2}^*O_{\gamma_1\delta_1}O_{\gamma_2\delta_2}&=&\frac1{d_B^4+d_B^2}\left[\delta_{\alpha_1\gamma_1}\delta_{\alpha_2\gamma_2}\delta_{\beta_1\delta_1}\delta_{\beta_2\delta_2}+\delta_{\alpha_1\gamma_2}\delta_{\alpha_2\gamma_1}\delta_{\beta_1\delta_2}\delta_{\beta_2\delta_1}\right] \nonumber
\end{eqnarray}
Using these identities, $\overline{\text{CI}}(A:B)$ can be written as
\begin{align}
\overline{\text{CI}}(A:B)&=\frac1{\left(d_A^2-1\right)\left(d_B^4+d_B^2\right)}\,{\rm tr}\bigg[\left(X_{A_1}-\frac{\textbf{1}_{A_1}}{d_{A}} \otimes \frac{\textbf{1}_{A_1}}{d_A}\right)\left(X_{A_2}-\frac{\textbf{1}_{A_2}}{d_{A}} \otimes \frac{\textbf{1}_{A_2}}{d_{A}}\right) \nonumber \\
& \qquad \qquad \qquad \qquad \qquad \qquad \qquad \quad \times \left((\textbf{1}_{B_1} \otimes \textbf{1}_{B_1}) \otimes (\textbf{1}_{B_2} \otimes \textbf{1}_{B_2})+X_{B_1} \otimes X_{B_2}\right)
\, R^{\otimes 2}\bigg]\nonumber\\\label{sigmaAB}
\end{align}
where $X_{A_1}$ is the swap operator $\left[X_{A_1}\right]_{n_1n_2,k_1k_2}=\delta_{n_1k_2}\delta_{n_2k_1}$ on $\mathcal{H}_{A_1} \otimes \mathcal{H}_{A_1}$ (and so swaps the $A_1$ subsystem of the first copy of $R$ with the $A_1$ subsystem of the second copy of $R$), and $X_{A_2}$, $X_{B_1}$, $X_{B_2}$ are defined similarly. 

If $A$ and $B$ are mutually unitary regions, we can relate $R$ to the superdensity operator $\varrho$ for operator insertions on the regions $A$ and $B$.  (For a review of superdensity operators, see Appendix B.)  In this case, if we multiplicatively normalize the tensor network so that
\begin{eqnarray}
M(\textbf{1}_A : \textbf{1}_B)=1\,,
\end{eqnarray}
then by the definition of mutually unitary regions in Section \ref{sec:setup} and Eqn.~\eqref{unitaryregion2}, we have
\begin{eqnarray}
1=\int dU_A \, dU_B \, M\left(U_A : U_B\right)=\frac1{d_A d_B}\,{\rm tr}(R)\,,
\end{eqnarray}
and thus
\begin{eqnarray}
\varrho = \frac1{d_A d_B}\, R
\end{eqnarray}
is a superdensity operator.  As per Eqn.~\eqref{Rdualized1}, we can treat $\varrho$ as a density operator on $\mathcal{H}_{A_1} \otimes \mathcal{H}_{A_2} \otimes \mathcal{H}_{B_1} \otimes \mathcal{H}_{B_2}$.  Interestingly, we can write $\overline{\text{CI}}(A:B)$ in terms of R\'{e}nyi-2 entropies $S^{(2)}$ of $\varrho$ as
\begin{align}
\label{aQCIrenyi1}
\overline{\text{CI}}(A:B) &=\frac{d_A^2}{\left(d_B^2+1\right)\left(d_A^2-1\right)}\left[e^{-S^{(2)}_{A_1}}+e^{-S^{(2)}_{A_1 A_2 B_1 B_2}}-\frac1{d_A}\left(e^{-S^{(2)}_{A_2}}+e^{-S^{(2)}_{A_2 B_1 B_2}}\right)\right. \nonumber\\
& \qquad \qquad \qquad \qquad \qquad \qquad \qquad -\frac1{d_A}\left(e^{-S^{(2)}_{A_1}}+e^{-S^{(2)}_{A_1 B_1 B_2}}\right)\left.+\frac1{d_A^2}\left(1+e^{-S^{(2)}_{B_1 B_2}}\right)\right]\,.
\end{align}
In the above equation, we have, for instance
$$S^{(2)}_{A_1} := - \log \text{tr}(\varrho_{A_1}^2)$$
where $\varrho_{A_1} = \text{tr}_{A_2 B_1 B_2}(\varrho)$.  The R\'{e}nyi-2 entropies of other combinations of subsystems are defined similarly.  Note that Eqn.~\eqref{aQCIrenyi1} is particularly interesting since it relates causality to \textit{spacetime} entropies.


\subsection{Spin chain examples}

\begin{figure}[h]
\center
\includegraphics[width=11.5cm]{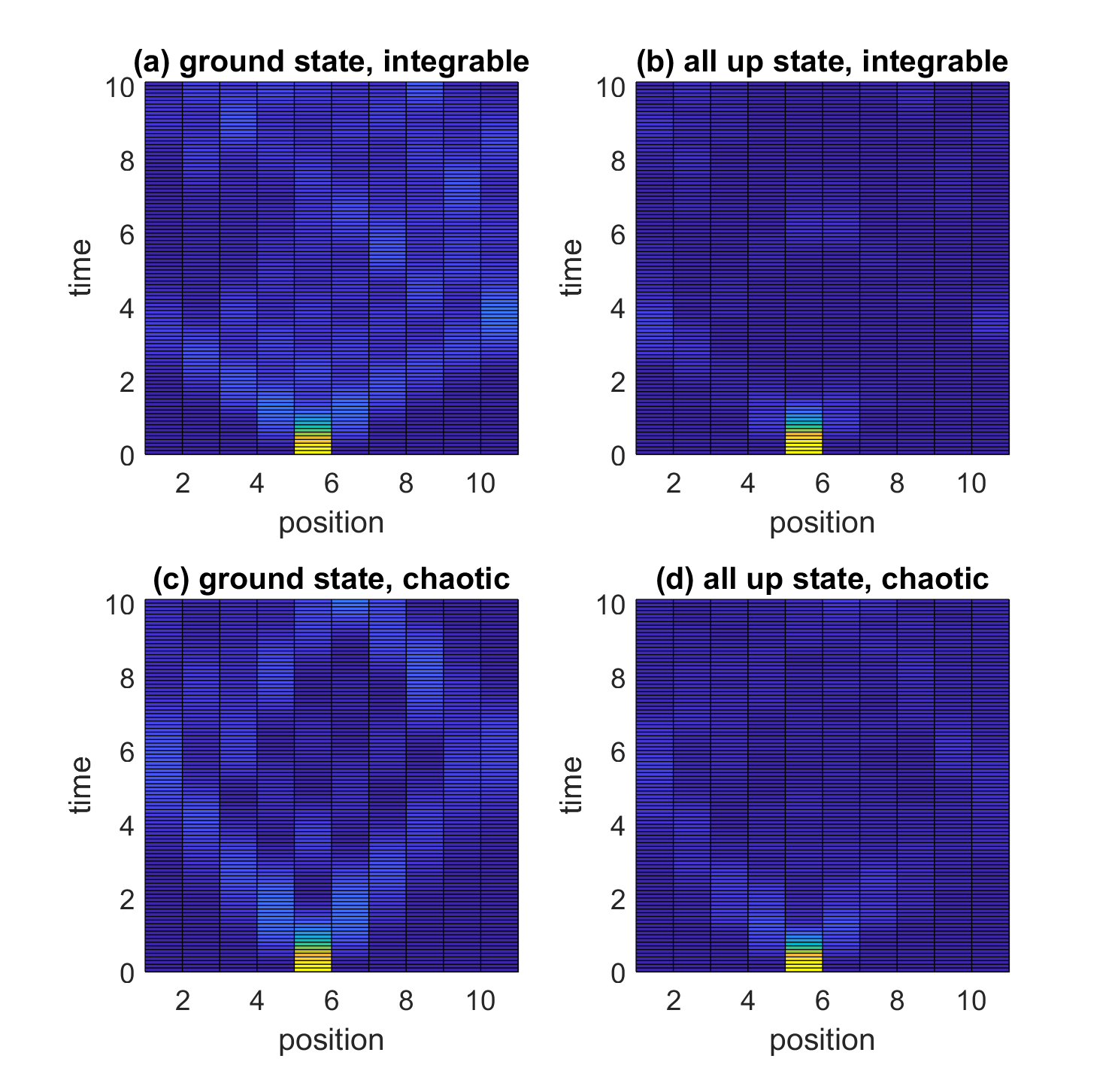}
\caption{The averaged quantum causal influence $\overline{\text{CI}}(A:B)$ for a quantum spin chain for length-$1$ regions $A$ and $B$. Region $A$ is at site $5$ in the middle of the chain at time $t=0$. The heat maps depict $\overline{\text{CI}}(A:B)$ as a function of the position and time of $B$. Results are obtained for two different initial states: the ground state and the ``all-up'' state. The calculation is done for a quantum Ising model with $10$ sites.  The Hamiltonian has nearest neighbor $ZZ$ interactions with coupling $J=1$, a transverse field, and open boundary conditions.  The coupling for the transverse field is $\vec{h}=(1,0,0)$ for the integrable model (see (a) and (b)) and $\vec{h}=(1.48,0,-0.7)$ for the chaotic model (see (c) and (d)).} 
\label{fig:spinchain}
\end{figure} 

The aQCI, $\overline{\text{CI}}(A:B)$, serves as an unbiased measure of causal influence, which only depends on the $A,B$ regions and the tensor network.  To obtain more intuition about its behavior, we study $\overline{\text{CI}}(A:B)$ in an example system. Consider a spin chain with continuous time evolution. Here, $A$ and $B$ are single-site subsystems at two different times $t_1,t_2$, as is illustrated earlier in Figure \ref{fig:1d}. It should be noted that the tensor network description and the definition of causal influence apply to continuous time evolution, since we can treat a time evolution operator such as $U(t_2,t_1)=e^{-iH(t_2-t_1)}$ as a big tensor with $2L$ legs (i.e., $L$ input legs and $L$ output legs), when the spin chain has $L$ sites. Our numerical results for $\overline{\text{CI}}(A:B)$ are shown in Figure \ref{fig:spinchain}. We studied the dependence of $\overline{\text{CI}}(A:B)$ on initial states and the Hamiltonian. The model we consider is an Ising model with a generic magnetic field:
\begin{eqnarray}
H=J\sum_{n=1}^L\sigma^z_{n}\sigma^z_{n+1}+\sum_{\alpha=x,y,z}h_\alpha \sum_{n=1}^L\sigma_n^\alpha
\end{eqnarray}
The model is integrable if the magnetic field $\vec{h}$ is in the $xy$--plane, and the model is chaotic otherwise.

As seen in Figure \ref{fig:spinchain}, the aQCI is strong and long-lasting if the system is integrable and the initial state is the ground state. If the system is chaotic and the initial state is the ground state, the aQCI is a bit weaker, but still lasts for long times.  In contrast, if the system is integrable and the initial state is a finite energy density state (here we use the ``all-up'' state as an example), the causal influence has some revivals but otherwise decays.  Finally, if the system is chaotic and the initial state is a finite energy density state, the causal influence decays uniformly with time.
 
\begin{figure}[h]
\center
\includegraphics[width=12cm]{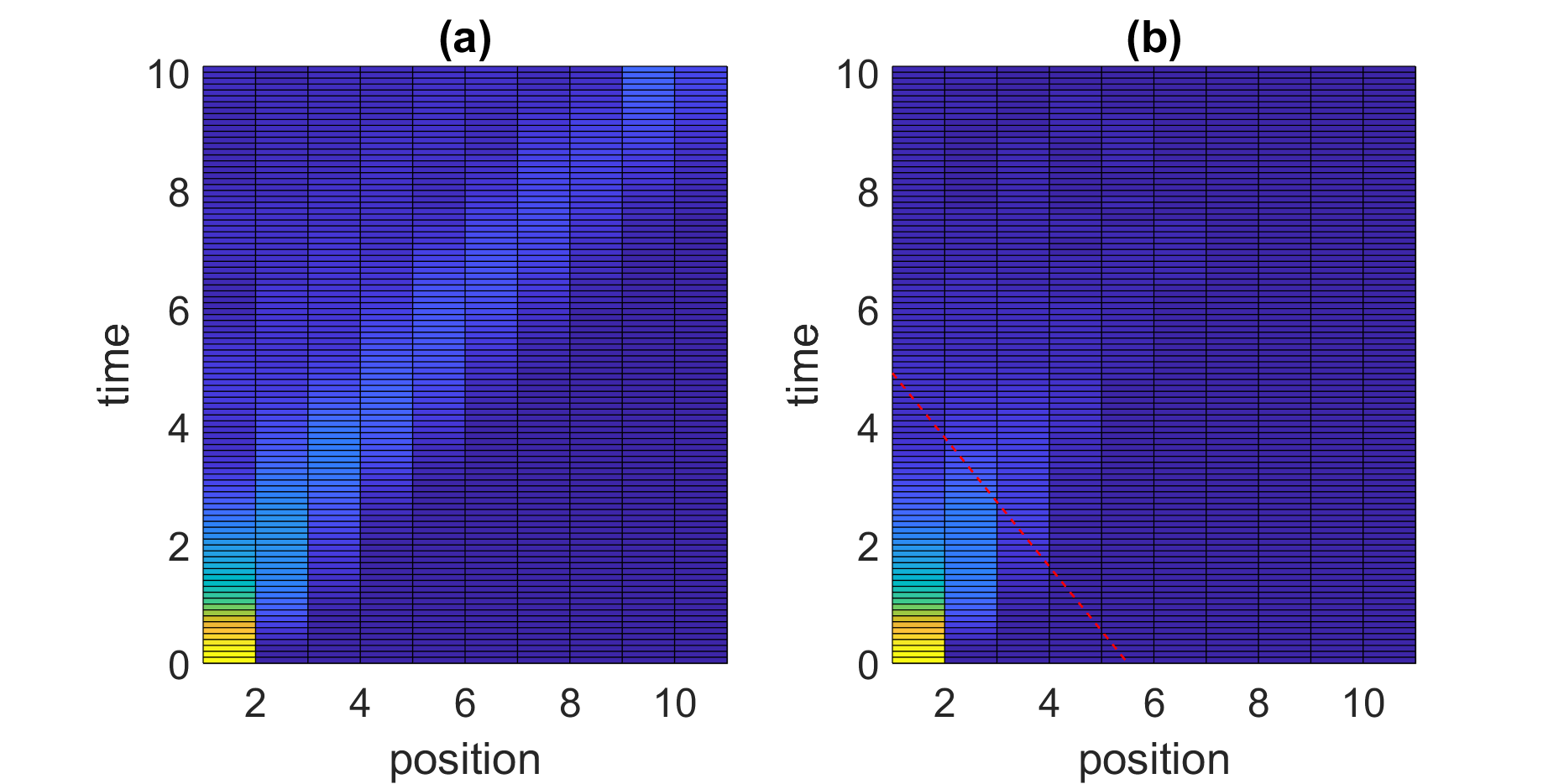}
\caption{Initial state dependence of the aQCI. (a) The aQCI, $\overline{CI}(A:B)$, of the quantum Ising model with region $A$ being site $1$ at $t=0$, as a function of the position and time of the single site region $B$.  The initial state is the ground state $|G\rangle$. (b) The same quantity with the initial state $U_R|G\rangle$, where $U_R$ is a Haar random unitary operator acting on the right half of the system. The red dashed line is a visual guide of the ``heat wavefront.'' The calculation is performed for the quantum Ising model with $J=1,~\vec{h}=[1.48,0,0.70]$, with open boundary conditions.}
\label{fig:CIfirewall}
\end{figure}

To further investigate the initial state dependence of quantum causal influence, we start from the ground state $|G\rangle$ of the spin chain and apply a Haar random unitary $U_R$ to the right half of the system (see Figure \ref{fig:CIfirewall}). The resulting state $U_R|G\rangle$ has a high energy density in its right half (sites $6$ through $10$ in the Figure) and the ground state energy density in its left half (sites $1-5$).  Evolving the system in time, energy propagates into the left half and ultimately heats up the whole system.  Consequently, the quantum causal influence of a region in the left half, such as site $1$ at $t=0$, behaves like quantum causal influence in the ground state until the ``heat wave'' arrives. This is consistent with the numerical results in Figure \ref{fig:CIfirewall} (b).

\subsection{Stabilizer tensor network examples}
\label{sec:stabsection}

\begin{figure}[t]
\center
\includegraphics[width=0.48\textwidth]{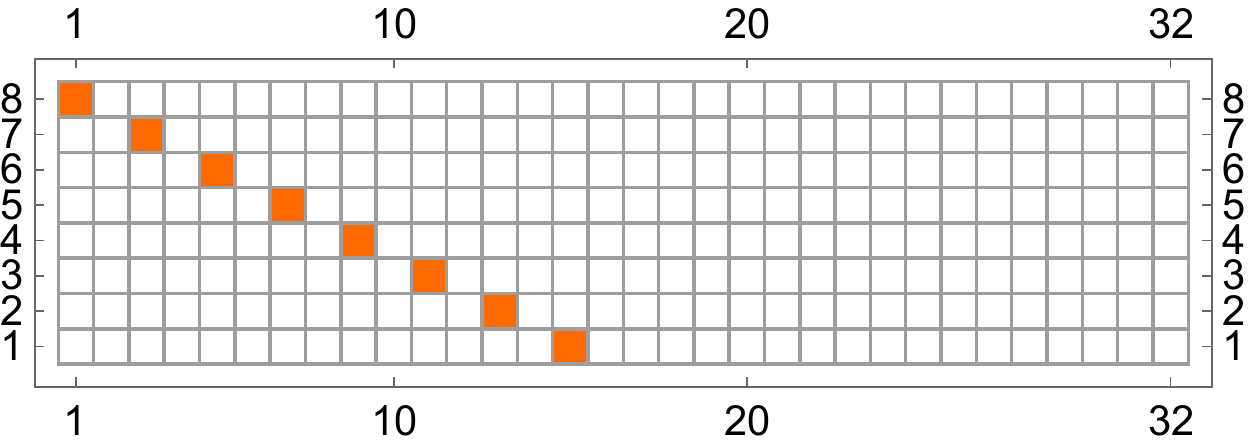}
\includegraphics[width=0.48\textwidth]{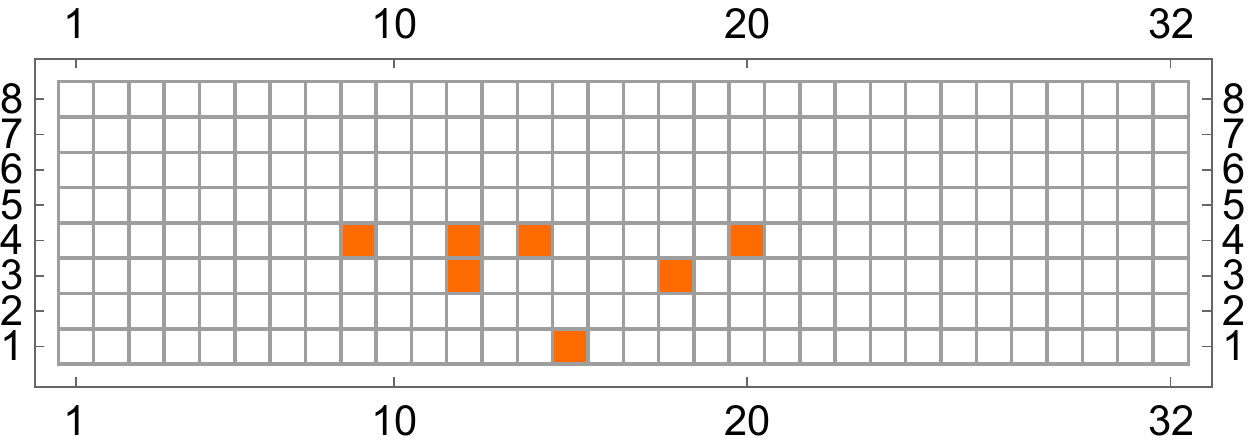}
\includegraphics[width=0.48\textwidth]{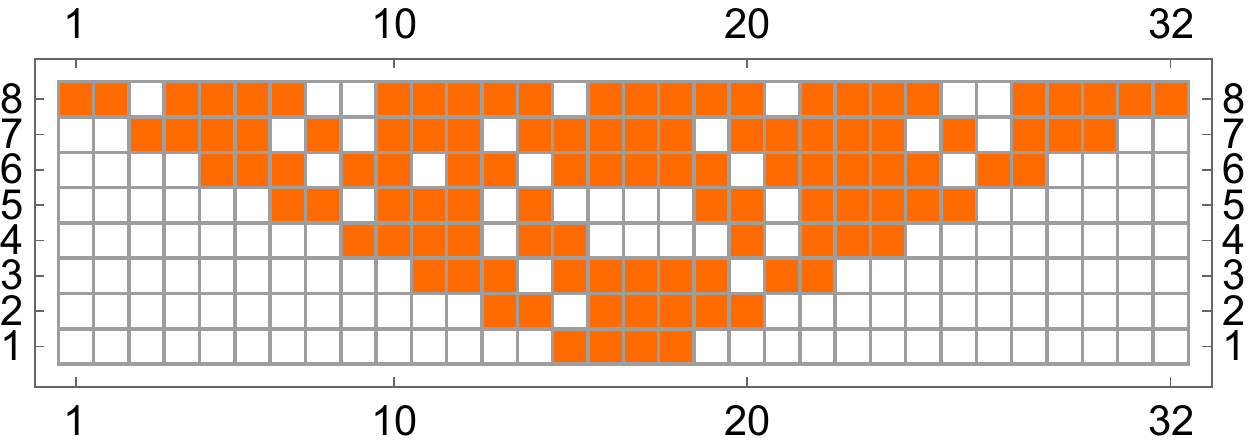}
\includegraphics[width=0.48\textwidth]{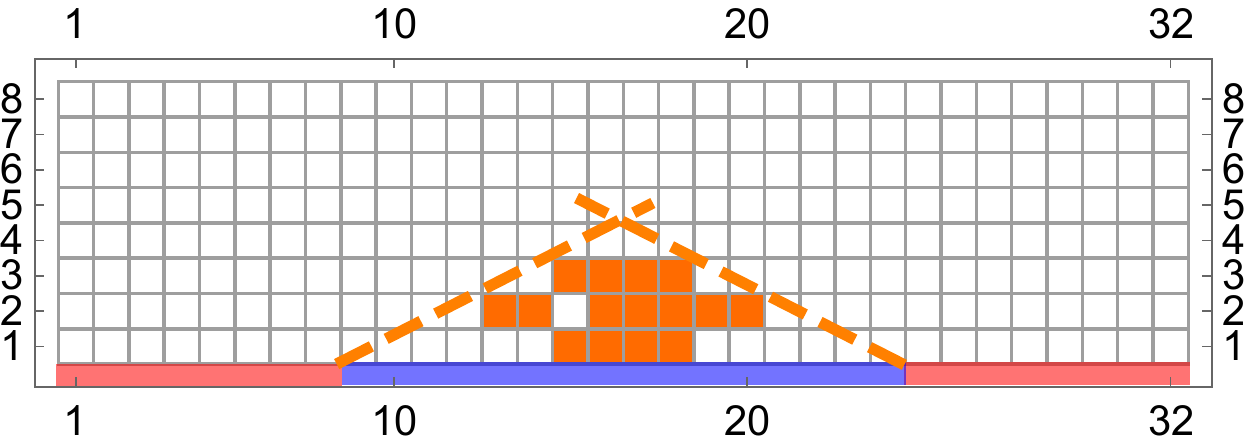}
\caption{The causal future of a link at $t = 0$ (pointing to the upper left in the center of the lowest layer) is colored orange.  In particular, the orange points are individually causally influenced by the link at the initial time.  The vertical axis is time and the horizontal axis is space (links) with periodic boundaries. Top-left: the integrable swap code with a random stabilizer initial state; top-right: the perfect [[4,0,3]] qutrit code with the same initial state; bottom-left: the same perfect code with initial state $\bigotimes |0\rangle \langle 0|$; bottom-right: the same perfect code with an infinite temperature initial state in the region marked red and $\bigotimes |0\rangle \langle 0|$ marked blue. Dashed lines are visual guides for the light cones of the red regions.}
\label{fig:stab}
\end{figure}

Here we apply our formula for the aQCI to stabilizer tensor networks \cite{gottesman1997stabilizer}, which provide a numerically tractable toy model for Trotterized Hamiltonian evolution.  Stabilizer tensor networks are reviewed in Appendix D.  In such networks, the entanglement entropy of any subsystem, as well as reduced density matrices of small subsystems, can be evaluated exactly in polynomial time in the network size \cite{fattal2004entanglement}. Our chosen geometry is shown in Figure \ref{fig:Trotter1}, where every vertex tensor is a stabilizer code. The horizontal direction is viewed as space (with periodic boundary conditions) and the vertical direction is viewed as time.  As the network structure is periodic with respect to pairs of layers of tensors, the time is set to increase by one for every two layers.  Furthermore, links in each layer are positioned at 1, 2, \ldots so that the speed of light in Figure \ref{fig:Trotter1} is $c=2$.

In the following, we will consider two examples of qutrit stabilizer tensor networks (i.e., there is a three-dimensional Hilbert space assigned to each link in the network) with stabilizer initial states $\rho_i$. For clarity, details of the stabilizers and algorithms are recapitulated in Appendix D and only physically relevant features of these codes will be discussed here. In the first example, all tensors are chosen to be the swap code; as a unitary two-to-two gate each tensor is written as $|i\rangle |j \rangle \mapsto |j \rangle |i \rangle$ where $i, j \in \mathbb{F}_3$. This may serve as a toy model for integrable systems where particles propagate ballistically without scattering.

In the second example, all tensors are chosen to be the perfect [[4, 0, 3]] code, $|i\rangle |j \rangle \mapsto |\frac{i + j}{2} \rangle |\frac{i - j}{2} \rangle$ where division by two is evaluated in $\mathbb{F}_3$. It is straightforward to verify that the tensor, viewed as a gate from any two of the four links to the other two, is unitary (such tensors are called perfect, as mentioned in Section \ref{subsec:holoTNpart1}).
Interestingly, the Heisenberg evolution of operators in such networks exhibits the growth of operator length (linearly in time), which captures some salient physics of scrambling in systems with spatial locality. 

For a fixed $U_x$ insertion at time $t = 0$, all positions $y$ for which $\text{CI}(x : y) > 0$ are colored orange in Figure \ref{fig:stab}. In the case of swap codes, the information from the $U_x$ insertion propagates ballistically and the causal future coincides with the future light cone of $x$. The specific direction of information propagation in the Figure depends on which link (left- or right-moving) $U_x$ acts on.

Results for the perfect code are remarkably different. For a generic initial state, as shown in the top-right panel of Figure \ref{fig:stab}, the causal influence of a point $x$ at $t=0$ on local regions in the future is small and vanishes for late times, which shows that information at $x$ spreads into nonlocal degrees of freedom. However, for the special initial state $\rho_i = \bigotimes |0 \rangle \langle 0 |$, the causal future of $x$ (with respect to local subregions) is the filled future light cone. Although there is not a sharp notion of thermal initial states in stabilizer tensor networks, such a causal influence structure suggests that $\rho_i$ is similar to a ``cold'' low-energy state of a local Hamiltonian (although energy is not well-defined in this Trotterized tensor network) because the causal influence does not decay substantially in the future (and hence does not quickly ``thermalize'').  Previously, we saw that low energy states of a quantum Ising model exhibit similar behavior, justifying our use of ``cold'' and ``low-energy'' in describing $\bigotimes |0\rangle \langle 0|$ for our stabilizer tensor network.

In Figure \ref{fig:stab} we have implemented an initial state $\rho_i = \bigotimes_{\text{hot}} \frac{1}{3} I \otimes \bigotimes_{\text{cold}} |0 \rangle \langle 0 |$ where in ``hot'' regions the initial state is at infinite temperature and in ``cold'' regions it is the product state. The causal future of $x$ terminates when it is engulfed by heatwaves from the infinite temperature subsystem.  The initial state dependence of quantum causal influence is manifest in these examples.

\subsection{An upper bound by spacetime quantum mutual information}
\label{subsec:mutualbound}

Recall that the quantum mutual information provides a bound on spacelike connected correlation functions \cite{wolf2008area}.  An analogous bound on \textit{spacetime} correlation functions was given in terms of superdensity operators in \cite{cotler2017superdensity} (a short discussion of this can be found in Appendix A).  It is natural that the causal influence between two regions is bounded by the spacetime mutual information of a corresponding superdensity operator.  Here, we will prove such an inequality:  \\ \\
\textbf{Bound on causal influence by spacetime quantum mutual information:} Consider two spacetime subregions $A$ and $B$ corresponding to Hilbert spaces $\mathcal{H}_A$ and $\mathcal{H}_B$, and a corresponding superdensity operator $\varrho_{AB}$. 
If $A,B$ are mutually unitary regions, we have
\begin{equation}
\text{CI}(A : B)^2 \leq 2 \, d_A^2 \, I_{\varrho_{AB}}(A : B)
\end{equation}
where $I_{\varrho_{AB}}(A : B)$ is the superdensity quantum mutual information between $A$ and $B$. \\ \\
\noindent \textit{Proof.} The proof of the inequality is easiest to understand diagrammatically.  First we write $\text{CI}(A:B)^2$ as
\begin{equation}
\text{CI}(A:B)^2 = \sup_{U_A, O_B} \frac{1}{\|O_B\|_2^4} \left|M(U_A : O_B) - \int dU_A \, M(U_A : O_B) \right|^2
\end{equation}
which can be expressed diagrammatically in superdensity operator notation as
\begin{equation}
\includegraphics[width=5.5in]{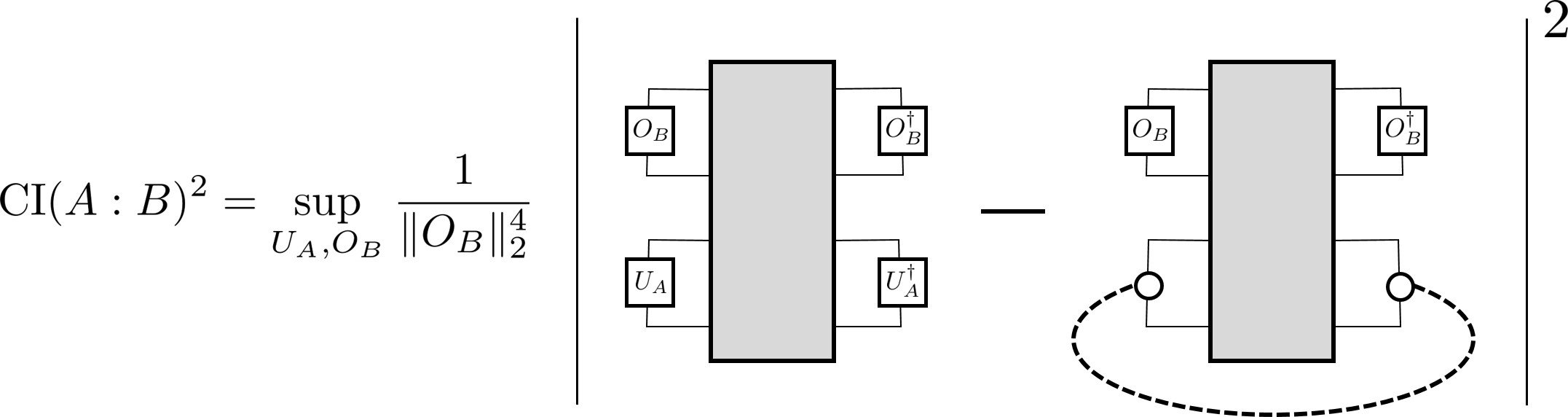}
\end{equation}
The dotted lines denote the $\int dU_A$ integration.  The identity $\int dU \, U_{ij} U_{k\ell}^\dagger = \frac{1}{d} \, \delta_{i\ell} \delta_{jk}$ is depicted by
\begin{equation}
\includegraphics[width=4in]{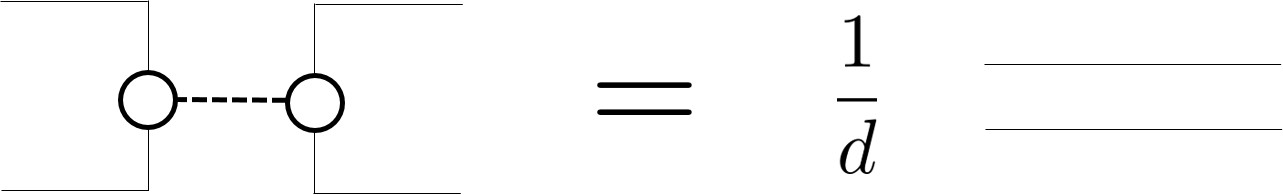}
\end{equation}
and so our diagram for $\text{CI}(A:B)$ becomes
\begin{equation}
\includegraphics[width=5.5in]{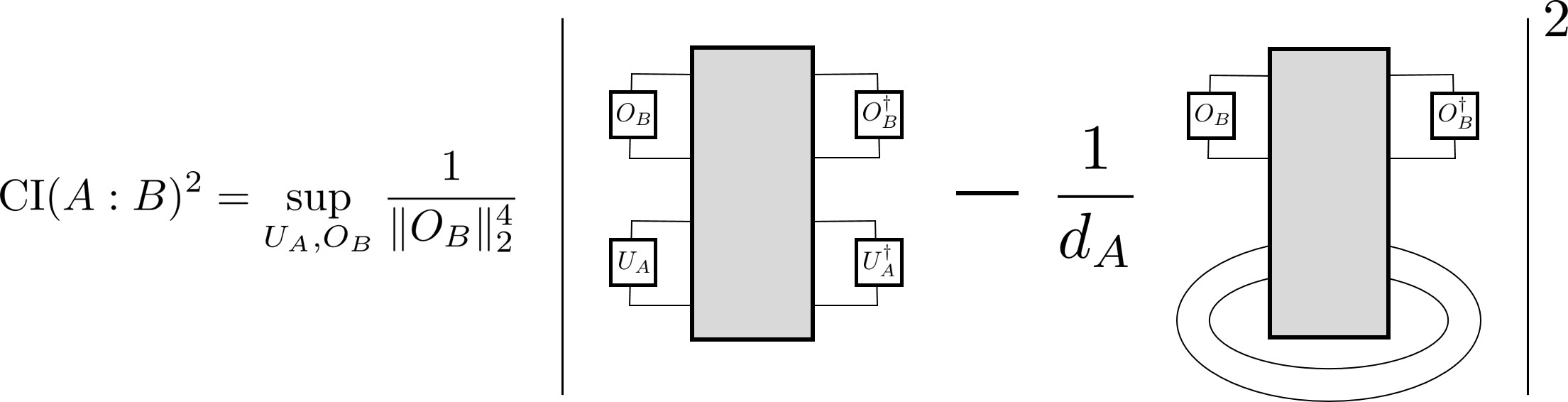}
\end{equation}
Now consider the identity
\begin{equation}
\includegraphics[width=5.5in]{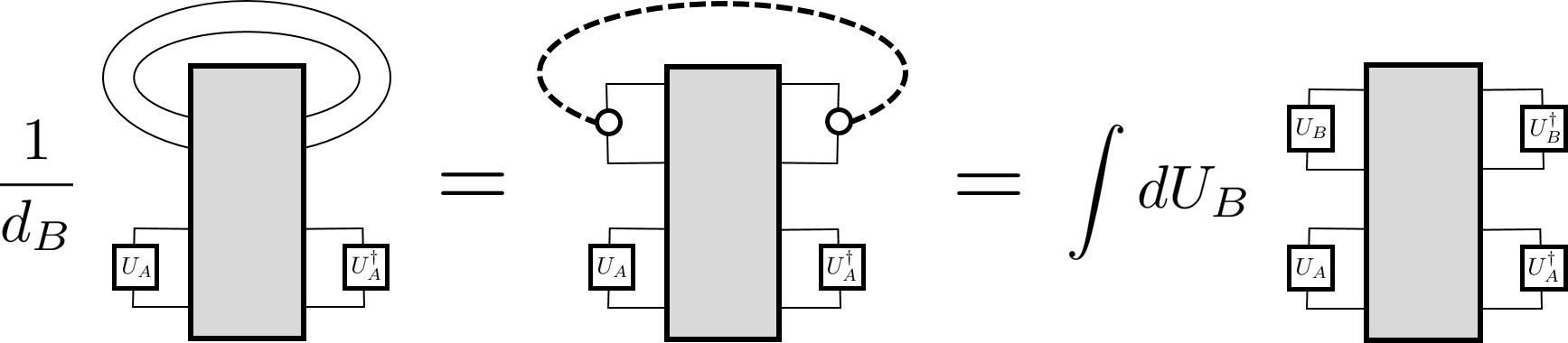}
\end{equation}
The term inside the integral on the right-hand side is actually independent from $U_A$ and $U_B$ if $A$, $B$ are mutually unitary regions.  Then we can replace the $U_A$ contractions by an average over $U_A$,
\begin{equation}
\includegraphics[width=4.5in]{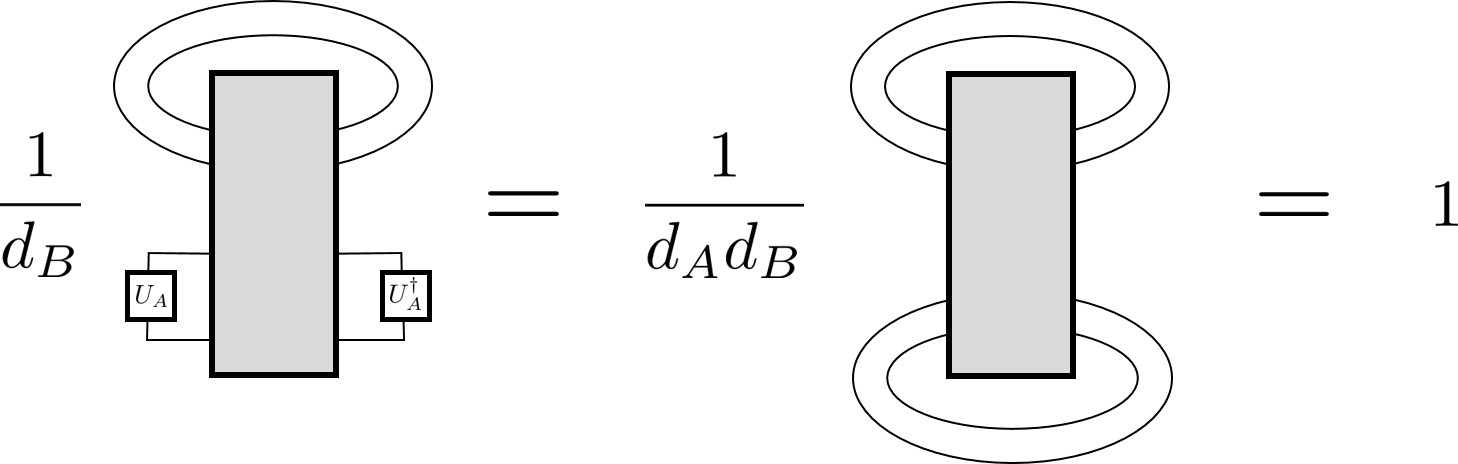}
\end{equation}
where the last equivalence is just the statement $\text{tr}(\varrho_{AB}) = 1$.  Then we can insert this factor of unity into our expression for $\text{CI}(A:B)^2$ to obtain
\begin{equation}
\includegraphics[width=6in]{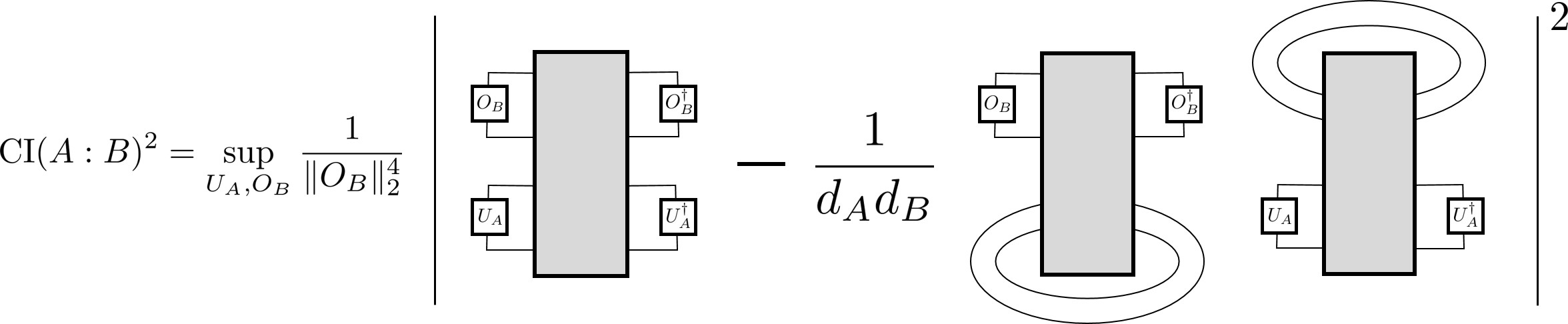}
\end{equation}
The term inside the absolute value bars is a connected correlation function with respect to the superdensity operator $\varrho_{AB}$.  Thus, we can use the superdensity operator quantum mutual information bound on connected correlation functions (see \cite{cotler2017superdensity} and Appendix B for a review), which gives us
\begin{equation}
\includegraphics[width=6in]{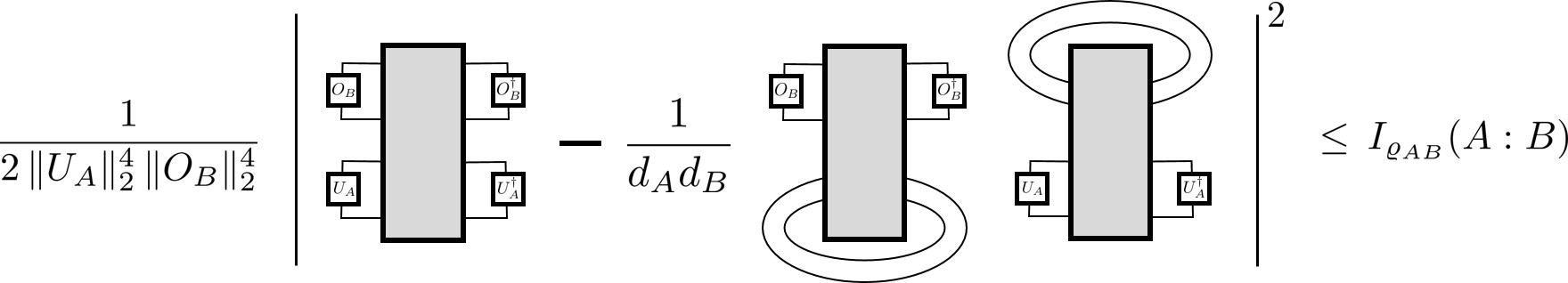}
\end{equation}
Comparing to our expression for $\text{CI}(A:B)^2$, we obtain the desired inequality $\text{CI}(A:B)^2 \leq 2 \, d_A^2 I_{\varrho_{AB}}(A:B)$. $\square$ \\ \\

\section{Conclusion and further discussion}
\label{sec:conclusion}

In this paper, we have proposed a new measure of causal structure, the quantum causal influence, in quantum many-body systems. We used the framework of general tensor networks to describe quantum many-body systems without a pre-fixed causal structure. In this framework, we showed how the causal influence between two spacetime regions $A,B$ can be probed by the effect of unitary operations in region $A$ on observables in region $B$. Unitarity plays an essential role in the asymmetry of the causal influence between two regions. Accordingly, the entanglement inherent in a general tensor network can be seen as building up space, time, and the causal relationships between local and collective spacetime degrees of freedom.  Our definition of quantum causal influence provides a new unified perspective on many seemingly disconnected phenomena.

Through examples and more abstract results, we have shown that the quantum causal influence, and therefore the direction of ``time's arrow,'' depends on the initial state and final state of the time evolution.  In particular, a maximally mixed subregion of either the initial or final state cannot causally influence other regions.  It would be interesting to understand in detail what happens when the initial or final states have subsystems that merely have high entropy (instead of having maximal entropy by virtue of being maximally mixed).

An important feature of the quantum causal influence is its nonlocality: a region $A$ can have trivial influence on regions $B,C$ while having nontrivial influence on their union $B\cup C$. Quantum error correction and quantum teleportation are both examples of such non-local causal influence. The non-locality of causal influence plays an essential role in holographic duality, where small disk-shape regions in the bulk have ordinary causal structure as prescribed by general relativity, while nonlocal regions have a different (and more exotic) causal structure required by the holographic principle.  Specifically, any given bulk operator can be reconstructed on a big enough region of the boundary, which means (using our definition) the quantum causal influence of a bulk point on the boundary is nontrivial, even if the point is spacelike separated from the boundary from a Riemannian geometry point of view.

We also discussed how unconventional causal structures appear in the Horowitz-Maldacena final state proposal of the black hole singularity, where again the non-locality of quantum causal influence plays an essential role in reconciling the ordinary causal structure of the black hole geometry (between small disks) and the unitarity of time evolution.  Additionally, we studied multiple probes of quantum causal influence, and discussed their relation to other quantum information quantities such as the quantum mutual information and R\'{e}nyi entropies.


There are many open questions that can be studied with the quantum causal influence.  For instance, it is interesting to ask whether there is a precise generalization of Cauchy surfaces defined in terms of the QCI.  For instance, such a plausible quantum generalization of Cauchy surfaces is a foliation of a general tensor network into disjoint subsystems $C_1,C_2,...,C_N$ such that $C_i$ only has nontrivial causal influence with $C_j$ if  $j>i$.  In addition, one should require that for each $C_i$, all of its disjoint subregions are spacelike separated from one another other.  In \ref{subsec:exotic} we discussed an example of such quantum Cauchy surfaces in holographic tensor networks. In general systems, can Cauchy surfaces always be found? When Cauchy surfaces are defined, is it always possible to define a ``quantum state'' on each surface, as in the (semi-)classical setting?

Another open question is how to generalize the quantum causal influence to measure (the quantum generalization) of spacetime geometry. In a similar vein, there have previously been proposals relating spatial distances between local subsystems to their quantum mutual information \cite{van2010building,qi2013exact}. It would be interesting to investigate whether a combination of these ideas can lead to a generalization of quantum causal influence which probes a (quantum generalization of a) spacetime metric. An even more general question concerns whether quantum causal influence can be applied to spacetime tensor networks with fluctuating geometries, such as those proposed in \cite{qi2017holographic}.
\\ \\
\noindent{\bf Acknowledgements.} We would like to thank C. Brukner, W. Donnelly, P. Hayden, D. Huse, A. May, J. Preskill, D. Ranard, R. Spekkens, and F. Wilczek for helpful discussions. JC is supported by the Fannie and John Hertz Foundation and the Stanford Graduate Fellowship program.  XH is supported by the Stanford Graduate Fellowship program.  XLQ is supported by the National Science Foundation under grant No. 1720504 and the David and Lucile Packard Foundation. ZY is supported by the David and Lucile Packard Foundation.  This work is supported in part by the U. S. Department of Energy award DE-SC0019380.

\newpage
\appendix

\section{Quantum causal influence for non-unitary regions}


Suppose we have a general tensor network given by $\{\{\mathcal{H}_i\}, |L\rangle, \rho_P\}$, and that $R_1$ is a subregion which is \textit{not} a unitary region.  This means that
\begin{equation}
\langle L| U_{R_1} \, \rho_P \, U_{R_1}^\dagger |L\rangle \not = \langle L | \rho_P |L\rangle
\end{equation}
for some unitary $U_{R_1}$.  This situation can occur even in some more modest examples, such as systems with post-selection.

In this context, it is natural to define quantum causal influence for non-unitary regions.  We let
\begin{equation}
M'(U_{R_1} : O_{R_2}) := \frac{\langle L| (U_{R_1} \otimes O_{R_2}) \rho_P \, (U_{R_1}^\dagger \otimes O_{R_2}^\dagger)|L\rangle}{\langle L | U_{R_1} \, \rho_P \, U_{R_1}^\dagger |L\rangle}
\end{equation}
where $R_1$ is not a unitary region.  Here, $M$ has been furnished with a prime $'$ to distinguish it from the usual $M(U_{R_1} : O_{R_2})$.  Then the corresponding mQCI for non-unitary regions is
\begin{equation}
\text{CI}\,'(R_1:R_2)=\sup_{U_{R_1}, O_{R_2}}\frac1{||O_{R_2}||_2^2}\left|M'(U_{R_1}:O_{R_2})-\int dU_{R_1} \, M'(U_{R_1}:O_{R_2})\right|
\end{equation}
and similarly, the corresponding aQCI for non-unitary regions is
\begin{equation}
\overline{\text{CI}}\,'(R_1:R_2)=\int dU_R\int_{||O_{R_2}||_2^2=1} dO_{R_2} \,\left|M'(U_{R_1}:O_{R_2})-\int dU_{R_1} \, M'(U_{R_1}:O_{R_2})\right|^2\,.
\end{equation}
Notice that modified mQCI and the modified aQCI are also furnished with primes $'$ to distinguish them for their unmodified counterparts.

Note that if $R_1$ \textit{is} a unitary region, then
\begin{align}
\text{CI}\,'(R_1:R_2) &= \frac{1}{\langle L | \rho_P |L\rangle} \, \text{CI}(R_1:R_2) \\
\overline{\text{CI}}\,'(R_1:R_2) &= \frac{1}{\langle L | \rho_P |L\rangle} \, \overline{\text{CI}}(R_1:R_2)\,,
\end{align}
meaning the modified and unmodified mQCI and aQCI are related by a multiplicative constant in this case.  Of course, if $\langle L| U_{R_1} \, \rho_P \, U_{R_1}^\dagger |L\rangle = 1$ for all $U_{R_1}$, then the multiplicative constant becomes one.

\section{Review of the superdensity operator formalism}
\label{sec:SuperAppendix}
Throughout the paper, we make use of the superdensity operator formalism to analyze spacetime correlation functions.  We review superdensity operators here, and a full exposition can be found in \cite{cotler2017superdensity}.

A superdensity operator is a spacetime analog of a density operator, so first we begin by examining density operators.  Consider a Hilbert space $\mathcal{H}$ of dimension $d$ so that the space of density operators on $\mathcal{H}$ is denote by $\mathcal{S}(\mathcal{H})$.  A density operator is denoted by $\rho$ and is defined by: \\ \\
\noindent \textbf{Definition (density operator):} \textit{A density operator $\rho$ is a bilinear form}
\begin{equation*}
\rho : \mathcal{H}^* \otimes \mathcal{H} \longrightarrow \mathbb{C}
\end{equation*}
\textit{satisfying the conditions:} \\ \\
\indent 1. $\rho^\dagger = \rho$ \qquad \qquad \qquad \qquad \qquad \qquad \qquad \qquad \quad (\textit{Hermitian}) \\ \\
\indent 2. $\rho \succeq 0 $\,, \textit{meaning $\langle \phi|\rho|\phi \rangle \geq 0$ for all} $|\phi\rangle$ \qquad \quad  \!(\textit{positive semi-definite}) \\ \\
\indent 3. $\tr(\rho) = 1$ \qquad \qquad \qquad \qquad \qquad \qquad \qquad \quad \,\,\,\,\,(\textit{unit trace}) \\ \\ \\
\noindent 
Since $\rho : \mathcal{H}^* \otimes \mathcal{H} \longrightarrow \mathbb{C}$, we can represent $\rho$ by the tensor diagram
\vskip.3cm
\begin{center}
\includegraphics[width=1.5in]{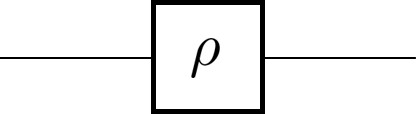}
\end{center}
where
\begin{center}
\includegraphics[width=4in]{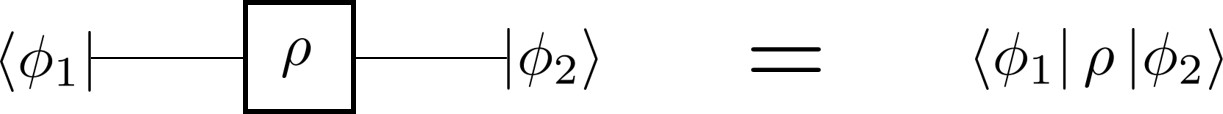}
\end{center}
Equivalently, we can think of $\rho$ as a map from operators in $\mathcal{B}(\mathcal{H})$ to correlation functions (i.e., a map from $\mathcal{B}(\mathcal{H}) \to \mathbb{C}$) by re-writing the tensor diagram as
\begin{center}
\includegraphics[width=1.5in]{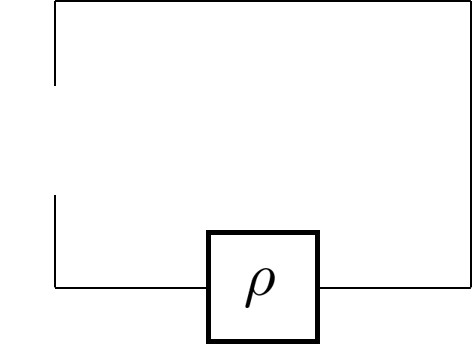}
\end{center}
where similarly \\
\begin{center}
\includegraphics[width=3.5in]{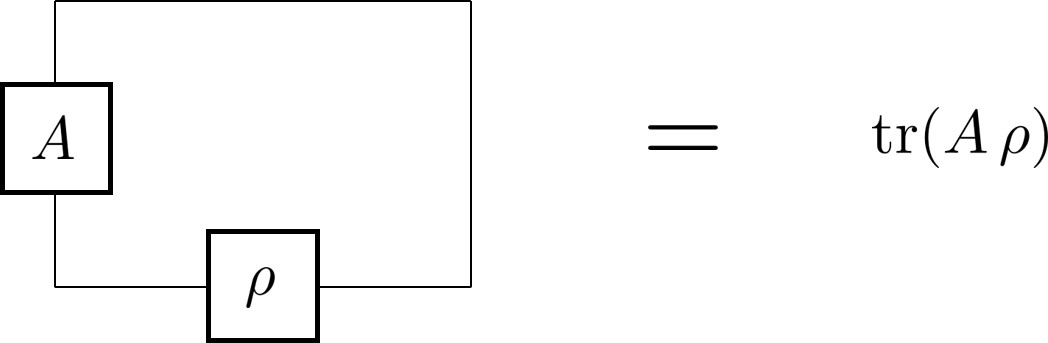}
\end{center}
\indent Now we introduce a new object which may at first appear peculiar, but will later appear natural.  It is given diagramatically by
\begin{center}
\includegraphics[width=2in]{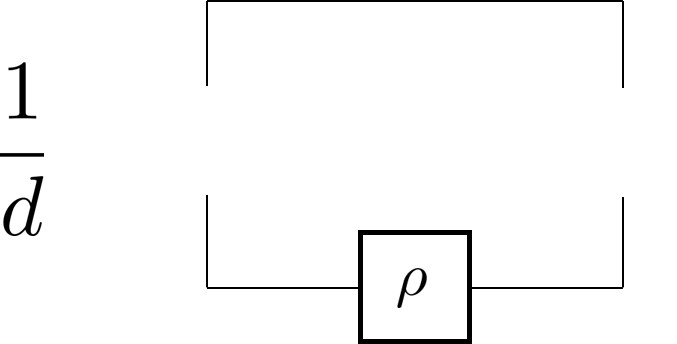}
\end{center}
This object satisfies \\
\begin{center}
\includegraphics[width=4in]{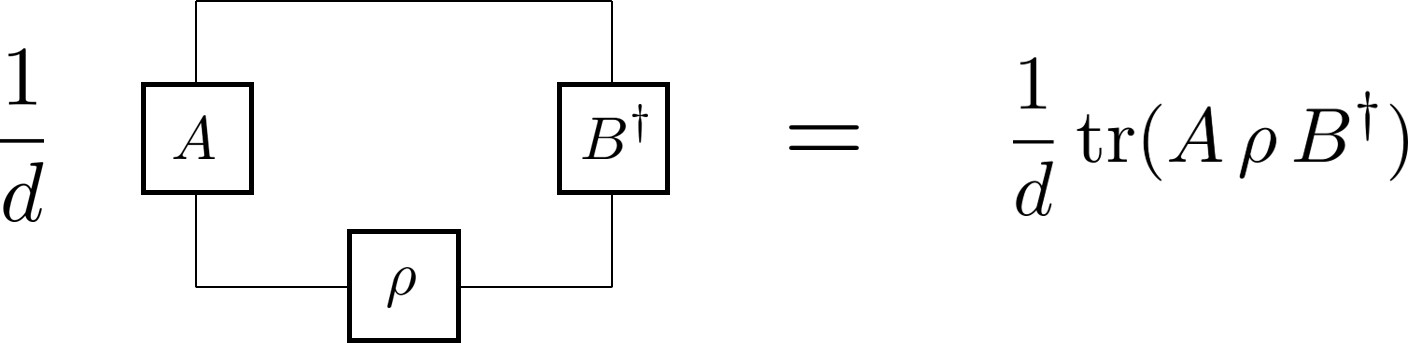}
\end{center}
and so is a bilinear form from $\mathcal{B}^*(\mathcal{H}) \otimes \mathcal{B}(\mathcal{H}) \to \mathbb{C}$.

This new object is clearly a repackaging of $\rho$, since it contains all of the same data.  Now let us write this new object in non-diagrammatic notation, and call it $\varrho_{\text{super}}$.  Consider the space of operators on $\mathcal{H}$, denoted by $\mathcal{B}(\mathcal{H})$.  Let $\{X_i\}_{i=1}^{d^2}$ be an orthonormal basis of operators for $\mathcal{B}(\mathcal{H})$, so that $\text{tr}(X_i^\dagger X_j) = \delta_{ij}$.  Since $\mathcal{B}(\mathcal{H})$ is itself a Hilbert space, we can write its basis in bra-ket notation as $\{|X_i\rangle\}_{i=1}^{d^2}$ where $\langle X_i | X_j\rangle := \text{tr}(X_i^\dagger X_j) = \delta_{ij}$.  Then we can write $\varrho_{\text{super}}$ in this basis as
\begin{equation}
\varrho_{\text{super}} = \frac{1}{d}\sum_{i,j=1}^{d^2} \text{tr}(X_i \, \rho \, X_j^\dagger) \, |X_i\rangle \langle X_j|\,.
\end{equation}
Then we have
\begin{equation}
\langle A | \, \varrho_{\text{super}} \, |B\rangle = \frac{1}{d} \, \text{tr}(A \, \rho \, B^\dagger)
\end{equation}
which matches the diagram above.

Several comments are in order.  The object $\varrho_{\text{super}}$ is our first example of a superdensity operator, which we will define shortly.  While a standard density operator $\rho$ is a map $\rho : \mathcal{H}^* \otimes \mathcal{H} \to \mathbb{C}$, the object $\varrho_{\text{super}}$ is a map $\varrho_{\text{super}} : \mathcal{B}^*(\mathcal{H}) \otimes \mathcal{B}(\mathcal{H}) \to \mathbb{C}$.  In fact, it is easy to check that $\varrho_{\text{super}}$ is Hermitian, positive semi-definite, and has unit trace.  Therefore, just as $\rho$ is a density operator on $\mathcal{H}$, we have that $\varrho_{\text{super}}$ is a density operator on $\mathcal{B}(\mathcal{H})$ (and hence a \textit{super}density operator).

So far, we have merely repackaged $\rho$ as the superdensity operator $\varrho_{\text{super}}$.  Both objects capture the data of correlation functions of a system at a single time.  But now suppose we want to capture the data of the correlation functions of a system at two times.  Letting $U$ be the unitary evolution between these two times, we can write down the new superdensity operator $\sigma_{\text{super}}$, namely
\begin{center}
\includegraphics[width=2in]{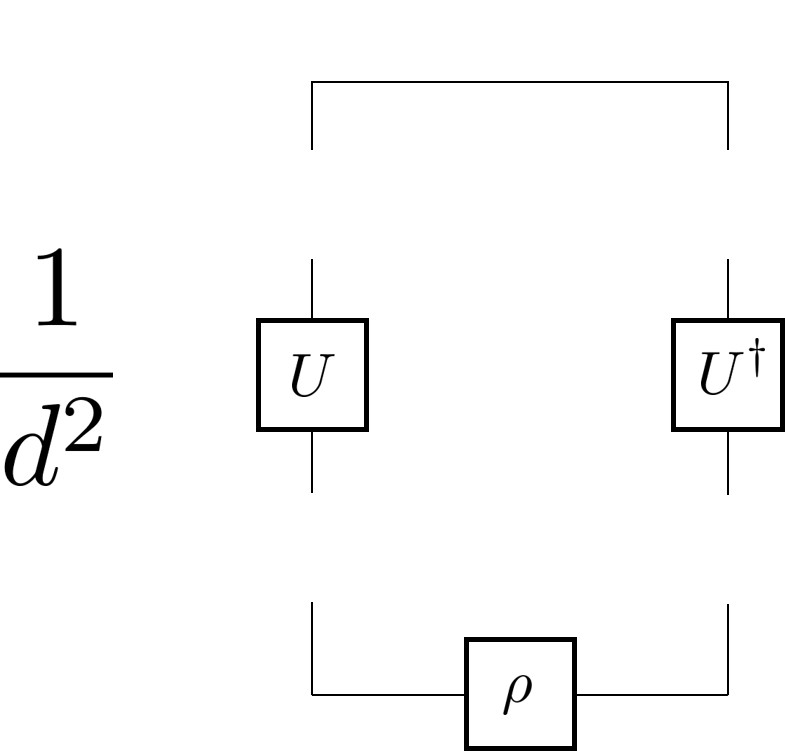}
\end{center}
which satisfies
\begin{center}
\includegraphics[width=5in]{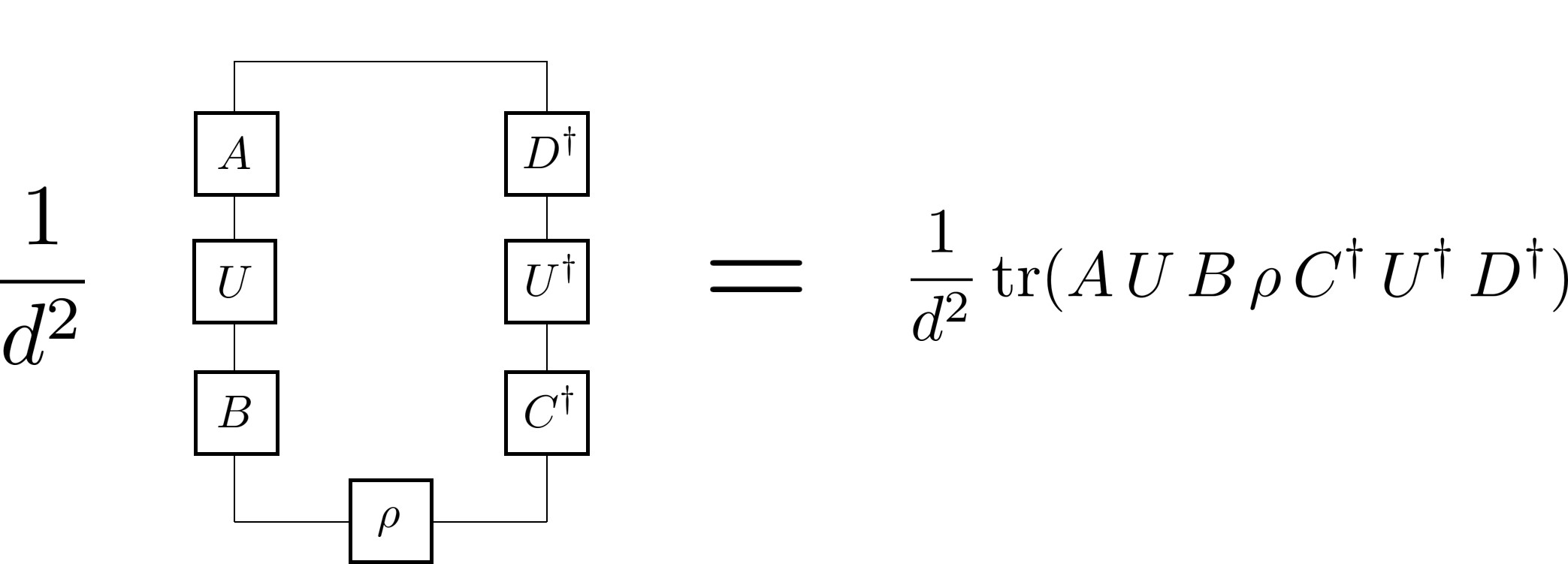}
\end{center}
and can be written non-diagrammatically as
\begin{equation}
\sigma_{\text{super}} = \frac{1}{d^2}\sum_{i,j,k,\ell=1}^{d^2} \text{tr}(X_i \, U \, X_j \, \rho \, X_k^\dagger \, U^\dagger \, X_\ell^\dagger) \, |X_j\rangle \langle X_k| \otimes |X_i \rangle \langle X_\ell|\,.
\end{equation}
Here, $\sigma_{\text{super}}$ maps operators at an initial time $t_1$ and operators at a final time $t_2$ to a correlation function.  We can write this map as
\begin{equation}
\sigma_{\text{super}} : \big(\mathcal{B}^*(\mathcal{H}_{t_1}) \otimes \mathcal{B}(\mathcal{H}_{t_1})\big) \otimes \big(\mathcal{B}^*(\mathcal{H}_{t_2}) \otimes \mathcal{B}(\mathcal{H}_{t_2})\big) \longrightarrow \mathbb{C}\,,
\end{equation}
or isomorphically
\begin{equation}
\sigma_{\text{super}} : \mathcal{B}^*(\mathcal{H}_{t_1} \otimes \mathcal{H}_{t_2}) \otimes \mathcal{B}(\mathcal{H}_{t_1} \otimes \mathcal{H}_{t_2}) \longrightarrow \mathbb{C} \,.
\end{equation}
Indeed, $\sigma_{\text{super}}$ is Hermitian, positive semi-definite, and has unit trace.  Therefore, $\sigma_{\text{super}}$ is a density operator on the operator space $\mathcal{B}(\mathcal{H}_{t_1} \otimes \mathcal{H}_{t_2})$.  We refer to Hilbert spaces of the form $\bigotimes_t \mathcal{H}_t$, such as $\mathcal{H}_{t_1} \otimes \mathcal{H}_{t_2}$, as ``history Hilbert spaces.''

As illustrated above, $\sigma_{\text{super}}$ contains the data of two-time correlation functions of a system, all packaged into a density operator on an appropriate operator space (for instance, $\mathcal{B}(\mathcal{H}_{t_1} \otimes \mathcal{H}_{t_2})$).  The reason we package this data into a density operator is because we can immediately use many of the tools and techniques of quantum information theory, which are designed for generic density operators (although they are typically applied only to standard density operators).  For instance, one can compute spacetime quantum entropies, spacetime quantum mutual information, and so on, and the results are physically and mathematically meaningful (see \cite{cotler2017superdensity} for an in-depth discussion of these points).  We will remark on the quantum mutual information below.

Of course, our construction above naturally generalizes to any number of times $t_1, t_2,..., t_n$.  The construction also generalizes to subsystems of the Hilbert space in the following way.  Consider a Hilbert space $\mathcal{H}$ which has (possibly overlapping) subsystems $\mathcal{H}_A$ and $\mathcal{H}_B$ with dimensions $d_A$ and $d_B$, respectively.  We will consider, for concreteness, a two-time superdensity operator $\chi_{\text{super}}$, given diagrammatically by
\begin{center}
\includegraphics[width=3in]{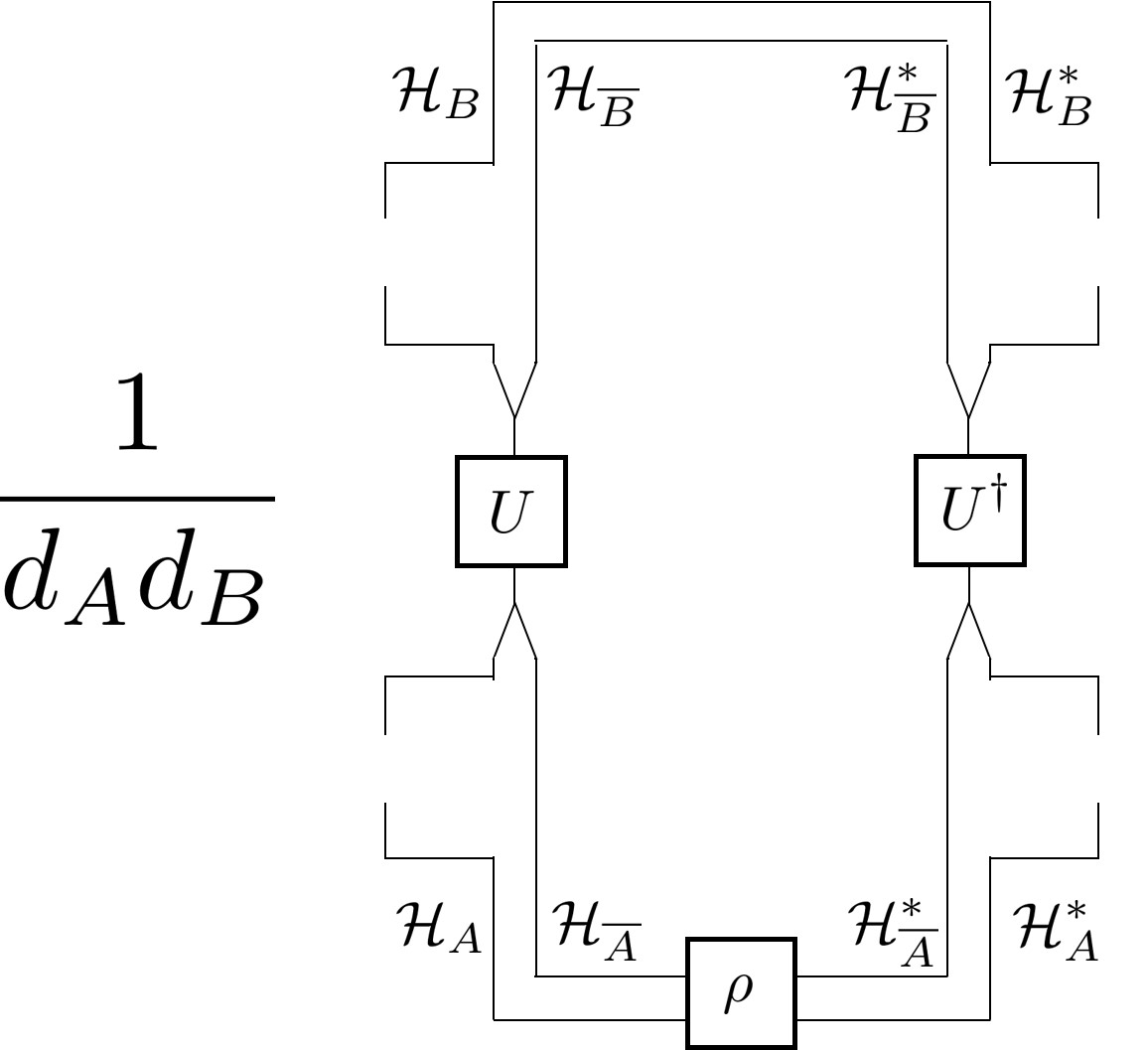}
\end{center}
satisfying
\begin{center}
\includegraphics[width=6in]{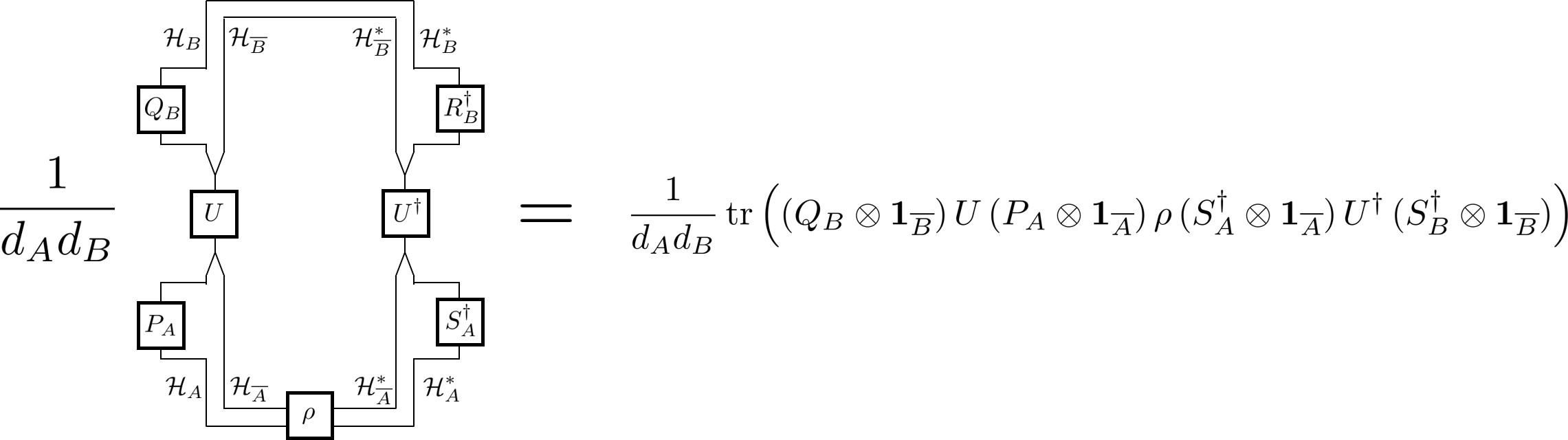}
\end{center}
and written in non-diagrammatic notation as
\begin{align}
\chi_{\text{super}} &= \frac{1}{d_A d_B}\sum_{i,j=1}^{d_A^2} \sum_{k,\ell=1}^{d_B^2} \text{tr}\left( \left(X_k^B \otimes \textbf{1}_{\overline{B}}\right) \, U \, \left(X_i^A \otimes \textbf{1}_{\overline{A}} \right) \, \rho \, \left(X_j^{\dagger \, A} \otimes \textbf{1}_{\overline{A}} \right) \, U^\dagger \, \left(X_\ell^{\dagger \, B} \otimes \textbf{1}_{\overline{B}}\right) \right) \nonumber \\
& \qquad \qquad \qquad \qquad \qquad \qquad \qquad \qquad \qquad \qquad \qquad \qquad \,\,\,\times  |X_i^A\rangle \langle X_j^A| \otimes |X_k^B \rangle \langle X_\ell^B|\,.
\end{align}
In this case, $\chi_{\text{super}}$ is a map from 
\begin{equation}
\chi_{\text{super}} : \mathcal{B}^*(\mathcal{H}_{A,\,t_1} \otimes \mathcal{H}_{B,\,t_2}) \otimes \mathcal{B}(\mathcal{H}_{A,\,t_1} \otimes \mathcal{H}_{B,\,t_2}) \longrightarrow \mathbb{C} \,,
\end{equation}
and is likewise Hermitian, positive semi-definite, and has unit trace.  Then $\chi_{\text{super}}$ captures the data of two-time correlation functions with operators on the subsystem $A$ at time $t_1$ and operators on the subsystem $B$ at time $t_2$.  This construction generalizes naturally to many times $t_1, t_2,..., t_n$ and arbitrary subsystems at each time.

The superdensity operators we have considered so far have a particular form: an initial state followed by slots for operator insertions, followed by unitary evolution, followed by more slots for operators insertions, and so on until a final trace is taken.  These kinds of superdensity operators can also be thought of as the quantum state of ancillary apparatus which couples to an evolving system in a certain manner (see \cite{cotler2017superdensity} for details).

More generally, we might be agnostic to the internal structure of a superdensity operator $\varrho$, and notate it as
\begin{center}
\includegraphics[width=2in]{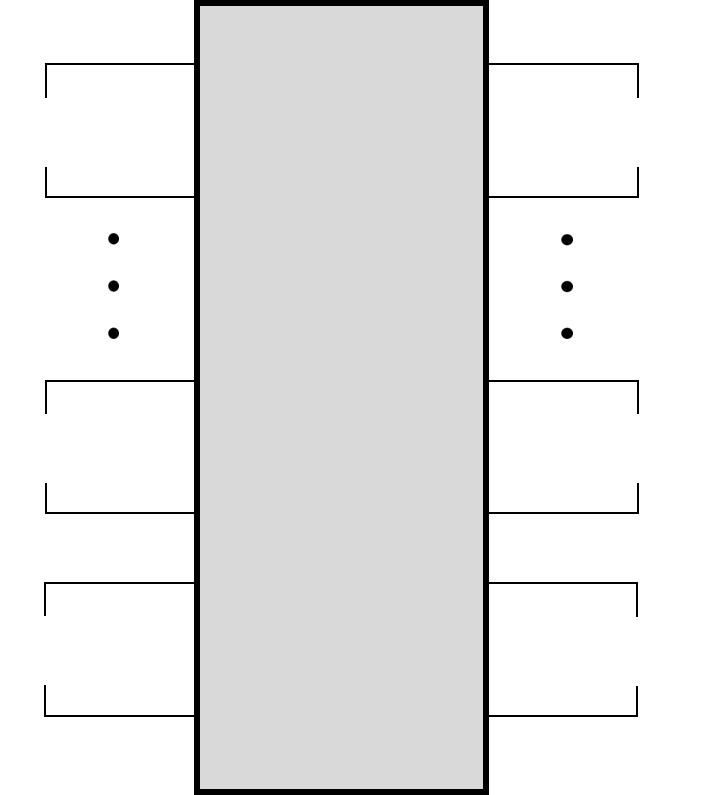}
\end{center}
which is a bilinear map
\begin{equation}
\varrho : \mathcal{B}^*(\mathcal{H}_{\text{hist.}}) \otimes \mathcal{B}(\mathcal{H}_{\text{hist.}}) \longrightarrow \mathbb{C}
\end{equation}
for some Hilbert space $\mathcal{H}_{\text{hist.}}$ that we designate as the history Hilbert space (in keeping with our previous terminology).  We may require that $\varrho$ is Hermitian, positive semi-definite, and has unit trace, so that it is formally a density operator (albeit on an operator space $\mathcal{B}(\mathcal{H}_{\text{hist.}})$).  This brings us to the definition: \\ \\
\noindent \textbf{Definition (superdensity operator):} \textit{A superdensity operator $\varrho$ is a bilinear form}
\begin{equation*}
\varrho : \mathcal{B}^*(\mathcal{H}_{\text{hist.}}) \otimes \mathcal{B}(\mathcal{H}_{\text{hist.}}) \longrightarrow \mathbb{C}
\end{equation*}
\textit{satisfying the conditions:} \\ \\
\indent 1. $\varrho^\dagger = \varrho$ \qquad \qquad \qquad \qquad \qquad \qquad \qquad \qquad \quad (\textit{Hermitian}) \\ \\
\indent 2. $\varrho \succeq 0 $\,, \textit{meaning $\langle W|\varrho|W \rangle \geq 0$ for all} $|W\rangle$ \quad \,\,\,\!(\textit{positive semi-definite}) \\ \\
\indent 3. $\tr(\varrho) = 1$ \qquad \qquad \qquad \qquad \qquad \qquad \qquad \quad \,\,\,\,\,(\textit{unit trace}) \\ \\ \\
As mentioned above, measures of quantum information of density operators can be upgraded to be measures of spacetime quantum information of superdensity operators.  These upgraded measures are meaningful \cite{cotler2017superdensity}.  For instance, recall the quantum mutual information bound \cite{wolf2008area}
\begin{equation}
\frac{1}{2 \,\|P_A\|_1^2 \, \|Q_B\|_1^2} \, \bigg| \text{tr}\left((P_A \otimes Q_B) \, \rho \right) - \text{tr}\left(P_A \, \rho \right) \text{tr}\left(Q_B \, \rho \right) \bigg|^2 \leq I_\rho(A : B)
\end{equation}
where $\mathcal{H} = \mathcal{H}_A \otimes \mathcal{H}_B \otimes \cdots$ and $I_\rho(A : B)$ is the quantum mutual information between $A$ and $B$ with respect to $\rho$.  Here, $A$ and $B$ are arbitrary disjoint spatial subregions.

One can straightforwardly show \cite{cotler2017superdensity} that the superdensity analog is
\begin{align}
&\frac{1}{2 \,\|P_A\|_2^2 \, \|Q_B\|_2^2 \, \|R_B\|_2^2 \,\|S_A\|_2^2 } \nonumber \\
& \quad \times \bigg| \big(\langle P_A| \otimes \langle Q_B|\big) \, \varrho_{\text{super}}^{AB} \, \big(|S_A\rangle \otimes |R_B\rangle\big) -  \langle P_A| \, \text{tr}_{\mathcal{B}(\mathcal{H}_B)}\left( \varrho_{\text{super}}^{AB} \right) \, |S_A\rangle \, \langle Q_B| \, \text{tr}_{\mathcal{B}(\mathcal{H}_A)}\left( \varrho_{\text{super}}^{AB} \right) \, |R_B\rangle\bigg|^2 \nonumber \\
& \qquad \qquad \qquad \qquad \qquad \qquad \qquad \qquad \qquad \qquad \qquad \qquad \qquad \qquad \qquad \qquad \qquad \qquad \leq I_{\varrho_{\text{super}}^{AB}}(A : B) \nonumber \\
\end{align}
where $\mathcal{B}(\mathcal{H}_{\text{hist.}}) = \mathcal{B}(\mathcal{H}_A) \otimes \mathcal{B}(\mathcal{H}_B) \otimes \cdots$ and $I_{\varrho_{\text{super}}^{AB}}(A : B)$ is the (spacetime) quantum mutual information between $A$ and $B$ with respect to $\varrho_{\text{super}}^{AB}$ which can be depicted by \\
\begin{center} 
\includegraphics[width=3in]{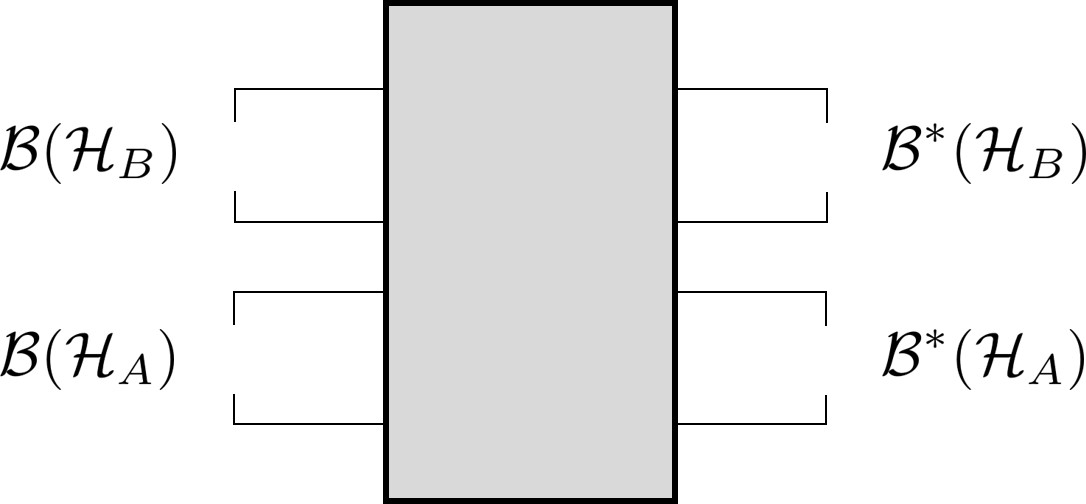}
\end{center}
Here, by contrast, $A$ and $B$ are arbitrary disjoint spacetime subregions.  The spacetime quantum mutual information bound can be depicted diagrammatically by
\begin{center}
\includegraphics[width=6.5in]{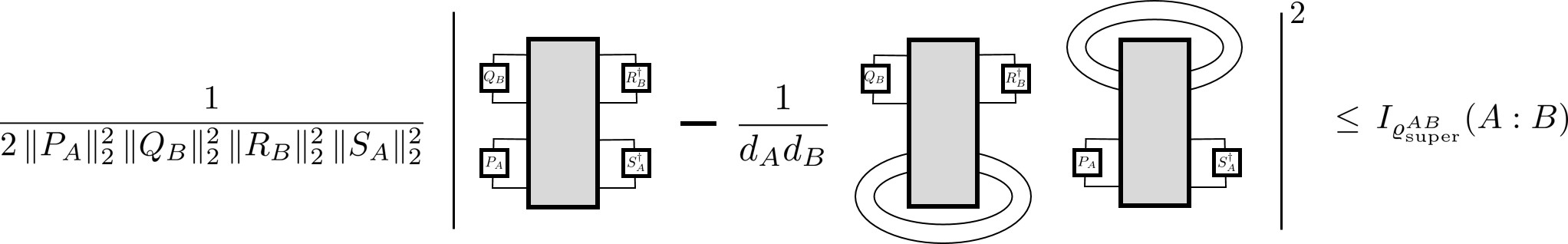}
\end{center}
which we utilize in Section \ref{subsec:mutualbound}.

\newpage
\section{Classical analog of non-local causal influence}
In this paper, we have been primarily focused on causal influence in quantum systems.  Here, we will explore features of causal influence in classical systems, and in particular focus on non-local aspects of causal influence.  We will compare and contrast with the quantum case, and find key differences.

In order to adapt our framework to the classical setting, we find it convenient to embed a classical system into a quantum system, and continue to use bra-ket notation and the operator formalism.  First, we establish how to present a classical system in this notation.  Suppose we have $n$ qubits, and consider the canonical basis $\{|i_1 \cdots i_n\rangle\}_{i_1,...,i_n = 0}^1$ which picks out the $z$-basis for every qubit.  We will refer to this basis as the classical basis, and write it more compactly using multi-index notation as $\{|I\rangle\}_{I \in \{0,1\}^n}$.  We require that a classical density operator $\rho_{\text{classical}}$ is a convex combination of projectors onto classical basis elements, namely of the form
\begin{equation}
\rho_{\text{classical}} =\sum_{I \in \{0,1\}^n} p_I \, |I\rangle \langle I |\,, \qquad \sum_{I\in \{0,1\}^n} p_I = 1\,, \qquad p_I \geq 0 \text{ for all }I.
\end{equation}
In words, a classical density operator is a probabilistic (incoherent) mixture of classical states in which each qubit has a definite $z$-direction.

Now we construct operators which act on classical states.  An arbitrary operator $A$ has the form
\begin{equation}
\label{classicalop1}
A = \sum_{I \in \{0,1\}^n} a_I \, |f(I)\rangle \langle I|\,,
\end{equation}
where $f$ is an arbitrary function $f : \{0,1\}^n \to \{0,1\}^n$ and the $a_I$'s are complex numbers.  Notice that this operator maps pure classical states to pure classical states (up to a complex scalar prefactor) since $O |J\rangle = c_J |f(J)\rangle$.  We can specialize to Hermitian operators $B$ which have the form
\begin{equation}
B = \sum_{I \in \{0,1\}^n} b_I \, |f(I)\rangle \langle I|\,,\qquad f\circ f = \text{Identity}\,, \qquad b_I = b_{f(I)}^*\,.
\end{equation}
Here, we see that $f : \{0,1\}^n \to \{0,1\}^n$ is its own inverse, meaning that $f\circ f$ is the identity map.

Now we turn to observables.  In the classical context, observables $C$ are Hermitian operators that satisfy the superselection rule $\langle I | O |J\rangle = 0$ if $I \not = J$, so that the eigenvectors cannot be superpositions of classical states.  Thus, observables have the form
\begin{equation}
O = \sum_{I \in \{0,1\}^n} c_I \, |I \rangle \langle I|
\end{equation} 
where the $c_I$'s are real numbers.

Finally, the classical analog of unitary operators are invertible operators satisfying $P^\dagger P = P P^\dagger = \textbf{1}$.  Comparing to Eqn.~\eqref{classicalop1}, we see that such a $P$ must have the form
\begin{equation}
P = \sum_{I \in \{0,1\}^n} |f(I)\rangle \langle I|\,, \qquad f\text{ invertible.}
\end{equation}
This means that $P$ is a permutation operator on the classical basis elements.  This is intuitive: the classical analog of unitary evolution can only interchange classical states.

Now we define the classical analog of causal influence for our $n$-qubit system.  Analogous to Eqn.~\eqref{maximalinfluence}, we define the classical maximal influence by 
\begin{equation}
\label{classicalmaximalinfluence}
\text{CI}_{\text{classical}}(A : B) :=\sup_{\substack{P_A \in \,\text{permutations on }A \\ O_B \in \,\text{classical operators on }B}}\frac1{||O_B||_2^2}\left|M\left(P_A:O_B\right)- \frac{1}{n!} \sum_{P_A \in \text{Perms}} M\left(P_A:O_B\right)\right|\,.
\end{equation}

Having set up classical causal influence, we turn to an example.\footnote{We thank Robert Spekkens for suggesting this example.}  We will consider a hallmark of classical cryptography: the one-time pad.  Suppose we have two parties Alice and Bob, and that Alice has a secret message that she wishes to share with Bob.  For concreteness, suppose that this secret message $M$ comprises of an $n$-bit string.  In the one-time pad protocol, Alice and Bob share in advance a secret key $K$, called the one-time pad, which is likewise an $n$-bit string that is unknown to anyone else.  This secret key $K$ has been sampled from a uniform distribution on all $n$-bit strings and must be discarded the protocol is completed (i.e., only used ``one time'').  Suppose Alice's messages is $(x_1, x_2,..., x_n)$ with $x_i \in \{0,1\}$, and the secret key is $(y_1, y_2,..., y_n)$ with $y_i \in \{0,1\}$.  Then Alice produces an encrypted message $E$, whose $i$th bit is the sum, modulo $2$, of the $i$th bits of $M$ and $K$.  The encrypted message $E$ would be
\begin{equation}
\label{classicalencoding1}
\left((x_1 \oplus y_1), (x_2 \oplus y_2),...,(x_n \oplus y_n)\right)\,,
\end{equation}
where here $\oplus$ denotes summation modulo $2$.  This encrypted message is then sent to Bob.  Bob decodes the message by taking its $i$th bit, and adding it modulo $2$ to the $i$th bit of the secret key.  The result is
\begin{align}
&\left((x_1 \oplus y_1 \oplus y_1), (x_2 \oplus y_2 \oplus y_2),...,(x_n \oplus y_n \oplus y_n)\right) \nonumber \\
= \, & (x_1, x_2,..., x_n)\,,
\end{align}
which is exactly Alice's original message $M$.  The secret key $K$ (i.e., the one-time pad) cannot be used in subsequent instantiations of the protocol since an eavesdropper can glean information about encrypted messages by looking for patterns, although we will not discuss this in detail here.

Let us express the encoding step of this protocol in terms of a superdensity operator.  Consider the diagram in Figure \ref{fig:classicalfig} below.  Let $\rho_{\text{message}} = |M\rangle \langle M|$, which is a classical state corresponding to the secret message.  Let $\sigma_{\text{key}}$ be the uniform distribution over classical states, namely the maximally mixed state $\sigma_{\text{key}} = \frac{1}{2^n}\sum_{J \in \{0,1\}^n} |J\rangle \langle J| = \frac{1}{2^n}\textbf{1}$.  We also let $P$ map
\begin{equation}
\label{Pmap1}
P |I\rangle \otimes |J\rangle = |I \oplus J\rangle \otimes |J\rangle\,, 
\end{equation}
where $I \oplus J$ represents bitwise addition modulo $2$ as per Eqn.~\eqref{classicalop1}.
\begin{figure}[t]
\center
\includegraphics[width=3in]{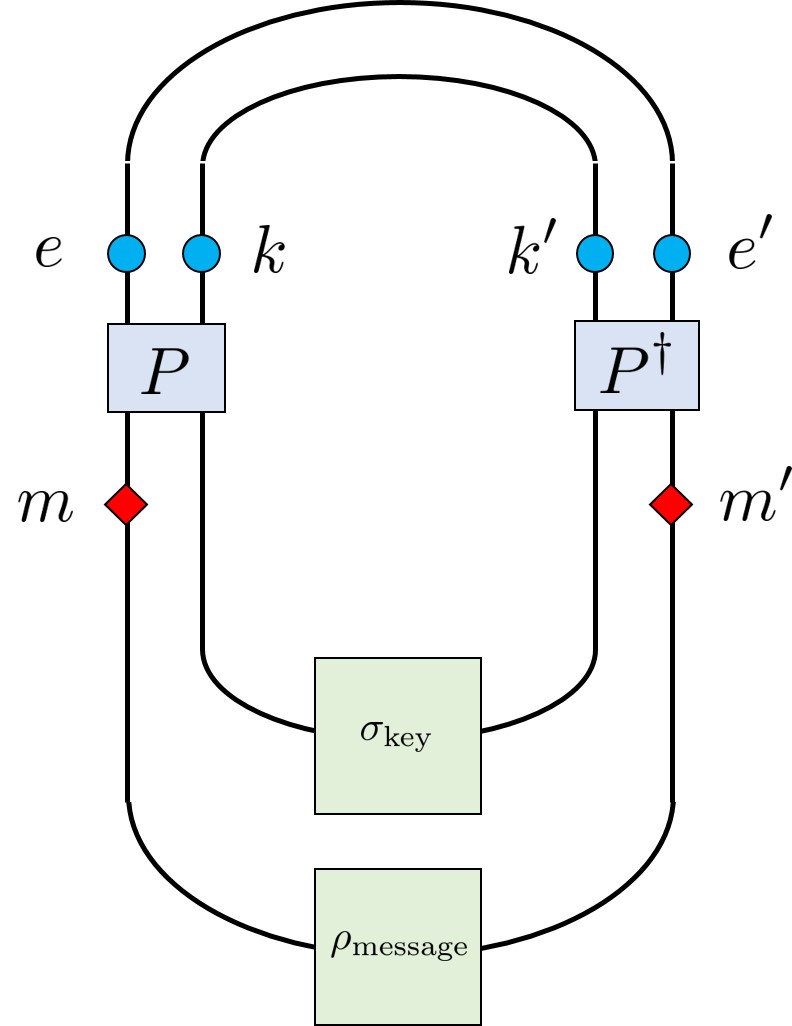}
\caption{A diagram for the one-time pad.  Here, $\rho_{\text{message}}$ is the state of the message, $\sigma_{\text{key}}$ is the state of the secret key, and $P$ is encrypts the message using the secret key, as described in Eqn.~\eqref{Pmap1} and the surrounding text. \label{fig:classicalfig}}
\end{figure}
Then we have
\begin{equation}
P (\rho_{\text{message}} \otimes \sigma_{\text{key}}) P^\dagger = \frac{1}{2^n} \sum_{J \in \{0,1\}^n} |M \oplus J\rangle \langle M \oplus J| \otimes |J\rangle \langle J|\,.
\end{equation}
Now let us consider the classical causal influence between $m$ (a place where an operator insertion affects the message) and $e$ (a place where an operator insertion probes the encrypted message).  Since
\begin{equation}
\text{tr}_{\text{key}}\left(P (\rho_{\text{message}} \otimes \sigma_{\text{key}}) P^\dagger\right) = \frac{1}{2^n} \, \textbf{1}\,,
\end{equation}
it follows that
\begin{equation}
\text{CI}_{\text{classical}}(m : e) = 0\,.
\end{equation}
This is intuitive -- it means that manipulating the message at $m$ does not affect the encrypted message at $e$, and hence no information from the message is contained in $e$ alone.  Thus, if an eavesdropper was positioned at $e$ and could tamper with the encrypted message, the secret message could not be discovered.

Similarly, we can consider the classical causal influence between $m$ and $k$ (a place where an operator insertion probes the secret key).  Since
\begin{equation}
\text{tr}_{\text{encrypted message}}\left(P (\rho_{\text{message}} \otimes \sigma_{\text{key}}) P^\dagger\right) = \frac{1}{2^n} \, \textbf{1}\,,
\end{equation}
we find
\begin{equation}
\text{CI}_{\text{classical}}(m : k) = 0\,.
\end{equation}
This is not surprising at all, since the initial message is not correlated with the secret key.

However, if we consider the classical causal influence between $m$ and $e \cup k$, we find
\begin{equation}
\text{CI}_{\text{classical}}(m : e \cup k) > 0\,.
\end{equation}
The result again is intuitive, since given access to both the encrypted message and the secret key, one can recover the initial message. This is an example of classical non-local causal influence: even though $m$ does not influence either $e$ and $k$, it influences $e \cup k$.

This example appears superficially similar to examples of non-local causal influence earlier in the paper, such as the quantum erasure code example in Section \ref{sec:QEC1} above.  However, there are key differences.  In our classical example, we treated the state of the key as a uniform distribution over all $n$-bit strings.  But in an actual instantiation of the protocol, a \textit{particular} key $K$ is chosen, and so $\sigma_{\text{key}} = |K\rangle \langle K|$ would be a pure state.  In this case, we would find $\text{CI}_{\text{classical}}(m : k) = 0$, $\text{CI}_{\text{classical}}(m : e) > 0$ and $\text{CI}_{\text{classical}}(m : e \cup k) > 0$, which is \textit{not} an example of non-local causal influence.

So why did we choose $\sigma_{\text{key}} = \textbf{1}/2^n$?  We did this because in the context of the protocol, a putative eavesdropper has a uniform prior on the state of the key, and so to her it is \textit{as if} the key was in a maximally mixed state.  But this is a reflection of the eavesdropper's particular knowledge, and not the state of the universe in which she lives.

If the classical universe of the protocol starts in a pure state, it will remain in a pure state for all time, and so it would instead be correct to use $\sigma_{\text{key}} = |K\rangle \langle K|$ for some particular $K$.  In such a universe, there can be no non-local causal influence.
If the universe was, in fact, at least partially in a mixed state, then we could harness some of the randomness to produce something like $\sigma_{\text{key}} = \textbf{1}/2^n$.

Now we summarize the key point.  In the classical setting, if the global state of the system is pure (i.e., not a probabilistic mixture), then the state of any subsystem is likewise pure.  This is emphatically not the case for a quantum system due to entanglement, and so subsystems of a pure quantum state are often mixed states.  If a classical universe starts in a pure classical state which remains pure and classical for all time, then there cannot be non-local causal influence with respect to subsystems.  However, if a quantum universe starts in a pure quantum state which is pure for all time, then there can be non-local causal influence with respect to subsystems.


\section{Numerical recipe for stabilizer tensor networks}
Here we review stabilizer tensor networks, and explain how we implement numerical calculations of these networks as discussed in Section \ref{sec:measures}.

\begin{figure}[t]
\center
\includegraphics[width=2.5in]{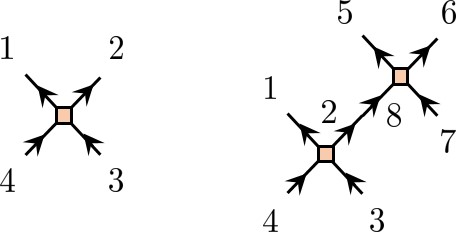}
\caption{Ordering of links in two simple geometries: one rank-four tensor and two rank-four tensors with a pair of links contracted. \label{fig:tensorindex}}
\end{figure}

To begin, stabilizer tensor networks are tensor networks comprised of connected unit stabilizer codes. Each unit stabilizer code is a tensor defined as the state fixed by a set of operators (stabilizers). Pictorially, a tensor can be represented as a vertex, and there is a Hilbert space on each link. The basic units we consider here are rank-four qutrit codes, i.e., there is a three-dimensional Hilbert space associated with each link and each vertex is degree four. The space of operators on each three-dimensional Hilbert space has a complex basis $X^{n} Z^m$ where $n, m = 0, 1, 2,$ and
\begin{equation}
X = \left(
\begin{array}{c c c}
0 & 1 & 0 \\
0 & 0 & 1 \\
1 & 0 & 0
\end{array}
\right), \quad
Z = \left(
\begin{array}{c c c}
1 & 0 & 0 \\
0 & e^{i 2 \pi / 3} & 0 \\
0 & 0 & e^{i 4 \pi / 3}
\end{array}
\right),
\end{equation}
in a preferred basis $\{|0\rangle, |1\rangle, |2\rangle\}$ of the Hilbert space. 
Note that $X Z = \exp (i 2 \pi / 3) Z X$, and so the basis operators $X^n Z^m$ all commute up to phases. Stabilizer operators are products of such basis operators, for example, $X \otimes I \otimes X \otimes I$, where operators on different links are separated by $\otimes$ and links are ordered as in Figure \ref{fig:tensorindex}. 

\newcommand{\bx}{\overline{X}}
\newcommand{\bz}{\overline{Z}}

A more convenient notation for stabilizer operators would be vectors with elements in $\mathbb{F}_3$, i.e., the field of three elements. For example, stabilizer operators for the rank-four swap code can be written as (denote $\bx = X^2 = X^{-1}$ and $\bz = Z^2 = Z^{-1}$)
\begin{equation}
\label{stabrow1}
 \left(
\begin{array}{c | c c c c c c c c}
0 & 1 & 0 & 0 & 0 & 1 & 0 & 0 & 0 \\
0 & 0 & 0 & 1 & 0 & 0 & 0 & 1 & 0  \\
0 & 0 & 1 & 0 & 0 & 0 & 2 & 0 & 0 \\
0 & 0 & 0 & 0 & 1 & 0 & 0 & 0 & 2
\end{array}
\right) \Leftrightarrow
\left(
\begin{array}{c}
X \otimes I \otimes X \otimes I \\
I \otimes X \otimes I \otimes X \\
Z \otimes I \otimes \bz \otimes I \\
I \otimes Z \otimes I \otimes \bz
\end{array}
\right),
\end{equation} 
that is,
\begin{align}
&(
\begin{array}{c | c c c c c c c c}
k & n_1 & m_1 & n_2 & m_2 & n_3 & m_3 & n_4 & m_4
\end{array}
) \Leftrightarrow e^{i 2 \pi k / 3} X^{n_1} Z^{m_1} \otimes X^{n_2} Z^{m_2} \otimes  X^{n_3} Z^{m_3} \otimes X^{n_4} Z^{m_4}.
\end{align}
Indeed, it is easy to verify that the code 
\begin{equation}
	\sum_{i, j \in \mathbb{F}_3} |i\rangle \otimes |j\rangle \otimes |i \rangle \otimes |j \rangle
\end{equation}
is (up to a multiplicative constant) the only state fixed by these four stabilizers given by the rows of Eqn.~\eqref{stabrow1}. If we regard this state as a unitary gate from links 3, 4 to 1, 2, it merely transports the state from link 3 to 1, and from link 4 to 2, hence is called a ``swap'' gate.  The dynamics of multiple catenated and layered swap gates simply propagates qutrits along diagonal lines in the stabilizer tensor network, and so clearly corresponds to integrable time evolution.

Another code that we use is the $[[4,0,3]]$ perfect code where the state is (note that division is in $\mathbb{F}_3$)
\begin{equation}
	\sum_{i, j \in \mathbb{F}_3} |i \rangle \otimes |j\rangle \otimes |(i - j) / 2 \rangle \otimes |(i + j) / 2 \rangle,
\end{equation}
corresponding to a set of stabilizers 
\begin{equation}
\left(
\begin{array}{c | c c c c c c c c}
0 & 0 & 1 & 0 & 1 & 0 & 0 & 0 & 1 \\
0 & 0 & 1 & 0 & 2 & 0 & 1 & 0 & 0  \\
0 & 1 & 0 & 1 & 0 & 0 & 0 & 1 & 0 \\
0 & 1 & 0 & 2 & 0 & 1 & 0 & 0 & 0
\end{array}
\right) \Leftrightarrow
\left(
\begin{array}{c}
Z \otimes Z \otimes I \otimes Z \\
Z \otimes \bz \otimes Z \otimes I \\
X \otimes X \otimes I \otimes X \\
X \otimes \bx \otimes X \otimes I
\end{array}
\right).
\end{equation}
Of course the full set of stabilizer operators of this code should contain products of these operators as well, so the choice of four generating operators is not unique. 

Now we proceed to finding stabilizers for networks composed of simple rank-four tensors. As an example, consider contracting two swap codes (identifying links 2 and 8 as in Figure \ref{fig:tensorindex}). Taking the product of operators corresponds to addition in the vector notation, so a general stabilizer (up to phase factors) of two swap codes takes the form
\begin{equation}
\left(
\begin{array}{c c c c c c c c}
a_1 & a_2 & a_3 & a_4 & b_1 & b_2 & b_3 & b_4
\end{array}
\right)
 \left(
\begin{array}{c c c c c c c c c c c c c c c c}
1 & 0 & 0 & 0 & 1 & 0 & 0 & 0 & 0 & 0 & 0 & 0 & 0 & 0 & 0 & 0 \\
0 & 0 & 1 & 0 & 0 & 0 & 1 & 0 & 0 & 0 & 0 & 0 & 0 & 0 & 0 & 0 \\
0 & 1 & 0 & 0 & 0 & 2 & 0 & 0 & 0 & 0 & 0 & 0 & 0 & 0 & 0 & 0 \\
0 & 0 & 0 & 1 & 0 & 0 & 0 & 2 & 0 & 0 & 0 & 0 & 0 & 0 & 0 & 0 \\
0 & 0 & 0 & 0 & 0 & 0 & 0 & 0 & 1 & 0 & 0 & 0 & 1 & 0 & 0 & 0\\
0 & 0 & 0 & 0 & 0 & 0 & 0 & 0 & 0 & 0 & 1 & 0 & 0 & 0 & 1 & 0\\
0 & 0 & 0 & 0 & 0 & 0 & 0 & 0 & 0 & 1 & 0 & 0 & 0 & 2 & 0 & 0\\
0 & 0 & 0 & 0 & 0 & 0 & 0 & 0 & 0 & 0 & 0 & 1 & 0 & 0 & 0 & 2
\end{array}
\right),
\label{mat:gensol}
\end{equation}
where we have temporarily suppressed the prefactor column for simplicity. The stabilizers on contracted links should cancel to give an operator acting on the remaining links only. Specifically, if the stabilizer on link 2 is $X^{n} Z^{m}$, then the stabilizer on link 8 must be $X^{n} Z^{- m}$. To find such solutions, only columns 3, 4 (link 2) and columns 15, 16 (link 8) in the matrix are relevant. The algebraic equation in $\mathbb{F}_3$ is thus
\begin{equation}
\left(
\begin{array}{c c c c c c c c}
a_1 & a_2 & a_3 & a_4 & b_1 & b_2 & b_3 & b_4
\end{array}
\right)
 \left(
\begin{array}{c c c c}
0 & 0 & 0 & 0 \\
1 & 0 & 0 & 0 \\
0 & 0 & 0 & 0 \\
0 & 1 & 0 & 0 \\
0 & 0 & 0 & 0\\
0 & 0 & 1 & 0\\
0 & 0 & 0 & 0\\
0 & 0 & 0 & 2
\end{array}
\right) 
\left(
\begin{array}{c c}
1 & 0\\
0 & 1\\
-1 & 0 \\
0 &  1
\end{array}
\right) = 0
,
\end{equation}
and the solution is $a_2 = b_2$, $a_4 = b_4$ and $a_i, b_i \in \mathbb{F}_3$ for $i = 1, 2, 3, 4$, i.e., the row space of
\begin{equation}
\left(
\begin{array}{c c c c c c c c}
1 & 0 & 0 & 0 & 0 & 0 & 0 & 0 \\
0 & 1 & 0 & 0 & 0 & 1 & 0 & 0 \\
0 & 0 & 1 & 0 & 0 & 0 & 0 & 0 \\
0 & 0 & 0 & 1 & 0 & 0 & 0 & 1 \\
0 & 0 & 0 & 0 & 1 & 0 & 0 & 0 \\
0 & 0 & 0 & 0 & 0 & 0 & 1 & 0
\end{array}
\right).
\label{mat:sol}
\end{equation}
Hence a generating set of stabilizers is the product of (\ref{mat:sol}) with (\ref{mat:gensol}) (with columns corresponding to contracted links dropped)
\begin{equation}
\left(
\begin{array}{c c c c c c c c c c c c c c c c}
1 & 0 & 1 & 0 & 0 & 0 & 0 & 0 & 0 & 0 & 0 & 0 \\
0 & 0 & 0 & 0 & 1 & 0 & 0 & 0 & 1 & 0 & 0 & 0 \\
0 & 1 & 0 & 2 & 0 & 0 & 0 & 0 & 0 & 0 & 0 & 0 \\
0 & 0 & 0 & 0 & 0 & 2 & 0 & 0 & 0 & 1 & 0 & 0 \\
0 & 0 & 0 & 0 & 0 & 0 & 1 & 0 & 0 & 0 & 1 & 0 \\
0 & 0 & 0 & 0 & 0 & 0 & 0 & 1 & 0 & 0 & 0 & 2
\end{array}
\right)
\Leftrightarrow
\left(
\begin{array}{c}
X \otimes X \otimes I \otimes I \otimes I \otimes I \\
I \otimes I \otimes X \otimes I \otimes X \otimes I \\
Z \otimes \bz \otimes I \otimes I \otimes I \otimes I \\
I \otimes I \otimes \bz \otimes I \otimes Z \otimes I \\
I \otimes I \otimes I \otimes X \otimes I \otimes X \\
I \otimes I \otimes I \otimes Z \otimes I \otimes \bz
\end{array}
\right),
\label{mat:contracted}
\end{equation}
which are indeed stabilizers for
\begin{equation}
\sum_{i, j, k \in \mathbb{F}_3} |i \rangle \otimes |i \rangle \otimes |j \rangle \otimes |k \rangle \otimes |j \rangle \otimes |k \rangle.
\end{equation}
Intuitively, this code transports states from link 3 to 1, 4 to 6 and 7 to 5. 

For general codes, phase factors must be taken into account when multiplying operators. Addition rules for phases are modified due to the non-commutativity of $X$ and $Z$ operators. For each link, 
\begin{equation}
X^n Z^m \times X^{n'} Z^{m'} = e^{- i 2 \pi m n' / 3} X^{n + n'} Z^{m + m'},
\end{equation}
that is,
\begin{equation}
\left(
\begin{array}{c | c c}
k & n & m
\end{array}
\right)
+ 
\left(
\begin{array}{c | c c}
k' & n' & m'
\end{array}
\right)
= 
\left(
\begin{array}{c | c c}
k + k' - m n' & n + n' & m + m'
\end{array}
\right).
\end{equation}
And the total phase is a sum of contributions from each link $i$:
\begin{equation}
\left(
\begin{array}{c | c c}
k & n_i & m_i
\end{array}
\right)
+ 
\left(
\begin{array}{c | c c}
k' & n'_i & m'_i
\end{array}
\right)
= 
\left(
\begin{array}{c | c c}
k + k' - \sum_i m_i n'_i & n_i + n_i' & m_i + m_i'
\end{array}
\right).
\end{equation}

Then determining stabilizers of the network is reduced to a linear algebra problem that can be solved in time polynomial in the network size. More specifically, the algorithm consists of three steps:
\begin{enumerate}
\item List the stabilizers of all constituent tensors;
\item Solve the linear equations imposed by requiring that operators on contracted links cancel;
\item Use the solution to the linear equations to find combinations of the stabilizers in step 1 that are the identity on the contracted links (taking into account the phase additions).
\end{enumerate}

Given stabilizers $O_1, \ldots, O_n$, the state fixed by all stabilizers is then the eigenstate of $O_1 + \cdots + O_n$ with eigenvalue $n$ because the spectrum of each operator $O_i$ only contains values $\exp (i 2\pi k / 3)$, $k = 0, 1, 2$. The superdensity operator of stabilizer tensor networks with few-vertex insertions (as shown in Figure \ref{fig:TrotterSuper}) is then itself a stabilizer state which can be computed up to a prefactor in polynomial time. The prefactor can be fixed by requiring the trace of the superdensity operator to be one. Causal influence is evaluated according to Eqn.~\eqref{sigmaAB} using the superdensity operator, which produces Figure \ref{fig:stab}. 

%
%
%
%
%
%
%
%

\bibliographystyle{JHEP}
\bibliography{refs}

\providecommand{\href}[2]{#2}\begingroup\raggedright\begin{thebibliography}{10}

\bibitem{lieb1972finite}
E.~H. Lieb and D.~W. Robinson, \emph{The finite group velocity of quantum spin
  systems},  in \emph{Statistical Mechanics}, pp.~425--431.
\newblock Springer, 1972.

\bibitem{oreshkov2012quantum}
O.~Oreshkov, F.~Costa and {\v{C}}.~Brukner, \emph{Quantum correlations with no
  causal order}, {\emph{Nature communications} {\bfseries 3} (2012) 1092}.

\bibitem{brukner2014quantum}
{\v{C}}.~Brukner, \emph{Quantum causality}, {\emph{Nature Physics} {\bfseries
  10} (2014) 259}.

\bibitem{aharonov2014each}
Y.~Aharonov, S.~Popescu and J.~Tollaksen, \emph{Each instant of time a new
  universe},  in \emph{Quantum theory: a two-time success story}, pp.~21--36.
\newblock Springer, 2014.

\bibitem{fitzsimons2015quantum}
J.~F. Fitzsimons, J.~A. Jones and V.~Vedral, \emph{Quantum correlations which
  imply causation}, {\emph{Scientific reports} {\bfseries 5} (2015) 18281}.

\bibitem{ried2015quantum}
K.~Ried, M.~Agnew, L.~Vermeyden, D.~Janzing, R.~W. Spekkens and K.~J. Resch,
  \emph{A quantum advantage for inferring causal structure}, {\emph{Nature
  Physics} {\bfseries 11} (2015) 414}.

\bibitem{pienaar2015graph}
J.~Pienaar and {\v{C}}.~Brukner, \emph{A graph-separation theorem for quantum
  causal models}, {\emph{New Journal of Physics} {\bfseries 17} (2015) 073020}.

\bibitem{brukner2015bounding}
{\v{C}}.~Brukner, \emph{Bounding quantum correlations with indefinite causal
  order}, {\emph{New Journal of Physics} {\bfseries 17} (2015) 083034}.

\bibitem{costa2016quantum}
F.~Costa and S.~Shrapnel, \emph{Quantum causal modelling}, {\emph{New Journal
  of Physics} {\bfseries 18} (2016) 063032}.

\bibitem{oreshkov2016causal}
O.~Oreshkov and C.~Giarmatzi, \emph{Causal and causally separable processes},
  {\emph{New Journal of Physics} {\bfseries 18} (2016) 093020}.

\bibitem{ringbauer2016experimental}
M.~Ringbauer, C.~Giarmatzi, R.~Chaves, F.~Costa, A.~G. White and A.~Fedrizzi,
  \emph{Experimental test of nonlocal causality}, {\emph{Science advances}
  {\bfseries 2} (2016) e1600162}.

\bibitem{allen2017quantum}
J.-M.~A. Allen, J.~Barrett, D.~C. Horsman, C.~M. Lee and R.~W. Spekkens,
  \emph{Quantum common causes and quantum causal models}, {\emph{Physical
  Review X} {\bfseries 7} (2017) 031021}.

\bibitem{maclean2017quantum}
J.-P.~W. MacLean, K.~Ried, R.~W. Spekkens and K.~J. Resch,
  \emph{Quantum-coherent mixtures of causal relations}, {\emph{Nature
  communications} {\bfseries 8} (2017) 15149}.

\bibitem{castro2018dynamics}
E.~Castro-Ruiz, F.~Giacomini and {\v{C}}.~Brukner, \emph{Dynamics of quantum
  causal structures}, {\emph{Physical Review X} {\bfseries 8} (2018) 011047}.

\bibitem{oreshkov2015operational}
O.~Oreshkov and N.~J. Cerf, \emph{Operational formulation of time reversal in
  quantum theory}, {\emph{Nature Physics} {\bfseries 11} (2015) 853}.

\bibitem{oreshkov2016operational}
O.~Oreshkov and N.~J. Cerf, \emph{Operational quantum theory without predefined
  time}, {\emph{New Journal of Physics} {\bfseries 18} (2016) 073037}.

\bibitem{hardy2012operator}
L.~Hardy, \emph{The operator tensor formulation of quantum theory},
  {\emph{Phil. Trans. R. Soc. A} {\bfseries 370} (2012) 3385}.

\bibitem{hardy2016operational}
L.~Hardy, \emph{Operational general relativity: Possibilistic, probabilistic,
  and quantum}, {\emph{arXiv preprint arXiv:1608.06940} (2016) }.

\bibitem{jia2017generalizing}
D.~Jia et~al., \emph{Generalizing entanglement}, {\emph{Physical Review A}
  {\bfseries 96} (2017) 062132}.

\bibitem{jia2018tensor}
D.~Jia, N.~Sakharwade et~al., \emph{Tensor products of process matrices with
  indefinite causal structure}, {\emph{Physical Review A} {\bfseries 97} (2018)
  032110}.

\bibitem{jia2018quantum}
D.~Jia et~al., \emph{Quantum theories from principles without assuming a
  definite causal structure}, {\emph{Physical Review A} {\bfseries 98} (2018)
  032112}.

\bibitem{pastawski2015holographic}
F.~Pastawski, B.~Yoshida, D.~Harlow and J.~Preskill, \emph{Holographic quantum
  error-correcting codes: Toy models for the bulk/boundary correspondence},
  {\emph{Journal of High Energy Physics} {\bfseries 2015} (2015) 149}.

\bibitem{hayden2016holographic}
P.~Hayden, S.~Nezami, X.-L. Qi, N.~Thomas, M.~Walter and Z.~Yang,
  \emph{Holographic duality from random tensor networks}, {\emph{Journal of
  High Energy Physics} {\bfseries 2016} (2016) 9}.

\bibitem{horowitz2004black}
G.~T. Horowitz and J.~Maldacena, \emph{The black hole final state},
  {\emph{Journal of High Energy Physics} {\bfseries 2004} (2004) 008}.

\bibitem{cotler2017superdensity}
J.~Cotler, C.-M. Jian, X.-L. Qi and F.~Wilczek, \emph{Superdensity operators
  for spacetime quantum mechanics}, {\emph{arXiv:1711.03119} (2017) }.

\bibitem{vidal2003efficient}
G.~Vidal, \emph{Efficient classical simulation of slightly entangled quantum
  computations}, {\emph{Physical review letters} {\bfseries 91} (2003) 147902}.

\bibitem{verstraete2004renormalization}
F.~Verstraete and J.~I. Cirac, \emph{Renormalization algorithms for
  quantum-many body systems in two and higher dimensions}, {\emph{arXiv
  preprint cond-mat/0407066} (2004) }.

\bibitem{levin2007tensor}
M.~Levin and C.~P. Nave, \emph{Tensor renormalization group approach to
  two-dimensional classical lattice models}, {\emph{Physical review letters}
  {\bfseries 99} (2007) 120601}.

\bibitem{qi2018space}
X.-L. Qi and Z.~Yang, \emph{Space-time random tensor networks and holographic
  duality}, {\emph{arXiv preprint arXiv:1801.05289} (2018) }.

\bibitem{almheiri2015bulk}
A.~Almheiri, X.~Dong and D.~Harlow, \emph{Bulk locality and quantum error
  correction in ads/cft}, {\emph{Journal of High Energy Physics} {\bfseries
  2015} (2015) 163}.

\bibitem{cleve1999share}
R.~Cleve, D.~Gottesman and H.-K. Lo, \emph{How to share a quantum secret},
  {\emph{Physical Review Letters} {\bfseries 83} (1999) 648}.

\bibitem{beny2007generalization}
C.~B{\'e}ny, A.~Kempf and D.~W. Kribs, \emph{Generalization of quantum error
  correction via the heisenberg picture}, {\emph{Physical review letters}
  {\bfseries 98} (2007) 100502}.

\bibitem{beny2007quantum}
C.~B{\'e}ny, A.~Kempf and D.~W. Kribs, \emph{Quantum error correction of
  observables}, {\emph{Physical Review A} {\bfseries 76} (2007) 042303}.

\bibitem{sekino2008fast}
Y.~Sekino and L.~Susskind, \emph{Fast scramblers}, {\emph{Journal of High
  Energy Physics} {\bfseries 2008} (2008) 065}.

\bibitem{lashkari2013towards}
N.~Lashkari, D.~Stanford, M.~Hastings, T.~Osborne and P.~Hayden, \emph{Towards
  the fast scrambling conjecture}, {\emph{Journal of High Energy Physics}
  {\bfseries 2013} (2013) 22}.

\bibitem{shenker2014black}
S.~H. Shenker and D.~Stanford, \emph{Black holes and the butterfly effect},
  {\emph{Journal of High Energy Physics} {\bfseries 2014} (2014) 67}.

\bibitem{maldacena2016bound}
J.~Maldacena, S.~H. Shenker and D.~Stanford, \emph{A bound on chaos},
  {\emph{Journal of High Energy Physics} {\bfseries 2016} (2016) 106}.

\bibitem{hayden2007black}
P.~Hayden and J.~Preskill, \emph{Black holes as mirrors: quantum information in
  random subsystems}, {\emph{Journal of High Energy Physics} {\bfseries 2007}
  (2007) 120}.

\bibitem{harrow2009random}
A.~W. Harrow and R.~A. Low, \emph{Random quantum circuits are approximate
  2-designs}, {\emph{Communications in Mathematical Physics} {\bfseries 291}
  (2009) 257}.

\bibitem{brown2012scrambling}
W.~Brown and O.~Fawzi, \emph{Scrambling speed of random quantum circuits},
  {\emph{arXiv preprint arXiv:1210.6644} (2012) }.

\bibitem{cotler2017chaos}
J.~Cotler, N.~Hunter-Jones, J.~Liu and B.~Yoshida, \emph{Chaos, complexity, and
  random matrices}, {\emph{Journal of High Energy Physics} {\bfseries 2017}
  (2017) 48}.

\bibitem{roberts2015localized}
D.~A. Roberts, D.~Stanford and L.~Susskind, \emph{Localized shocks},
  {\emph{Journal of High Energy Physics} {\bfseries 2015} (2015) 51}.

\bibitem{mezei2017entanglement}
M.~Mezei and D.~Stanford, \emph{On entanglement spreading in chaotic systems},
  {\emph{Journal of High Energy Physics} {\bfseries 2017} (2017) 65}.

\bibitem{bennett1993teleporting}
C.~H. Bennett, G.~Brassard, C.~Cr{\'e}peau, R.~Jozsa, A.~Peres and W.~K.
  Wootters, \emph{Teleporting an unknown quantum state via dual classical and
  einstein-podolsky-rosen channels}, {\emph{Physical review letters} {\bfseries
  70} (1993) 1895}.

\bibitem{maldacena1999large}
J.~Maldacena, \emph{The large-n limit of superconformal field theories and
  supergravity}, {\emph{International journal of theoretical physics}
  {\bfseries 38} (1999) 1113}.

\bibitem{witten1998anti}
E.~Witten, \emph{Anti de sitter space and holography}, {\emph{arXiv preprint
  hep-th/9802150} (1998) }.

\bibitem{vidal2008class}
G.~Vidal, \emph{Class of quantum many-body states that can be efficiently
  simulated}, {\emph{Physical review letters} {\bfseries 101} (2008) 110501}.

\bibitem{swingle2012entanglement}
B.~Swingle, \emph{Entanglement renormalization and holography}, {\emph{Physical
  Review D} {\bfseries 86} (2012) 065007}.

\bibitem{qi2017butterfly}
X.-L. Qi and Z.~Yang, \emph{Butterfly velocity and bulk causal structure},
  {\emph{arXiv:1705.01728} (2017) }.

\bibitem{harlow2016jerusalem}
D.~Harlow, \emph{Jerusalem lectures on black holes and quantum information},
  {\emph{Reviews of Modern Physics} {\bfseries 88} (2016) 015002}.

\bibitem{lloyd2006almost}
S.~Lloyd, \emph{Almost certain escape from black holes in final state
  projection models}, {\emph{Physical review letters} {\bfseries 96} (2006)
  061302}.

\bibitem{verlinde2013black}
E.~Verlinde and H.~Verlinde, \emph{Black hole entanglement and quantum error
  correction}, {\emph{Journal of High Energy Physics} {\bfseries 2013} (2013)
  107}.

\bibitem{gottesman2004comment}
D.~Gottesman and J.~Preskill, \emph{Comment on``the black hole final state''},
  {\emph{Journal of High Energy Physics} {\bfseries 2004} (2004) 026}.

\bibitem{bousso2014measurements}
R.~Bousso and D.~Stanford, \emph{Measurements without probabilities in the
  final state proposal}, {\emph{Physical Review D} {\bfseries 89} (2014)
  044038}.

\bibitem{gottesman1997stabilizer}
D.~Gottesman, \emph{Stabilizer codes and quantum error correction},
  {\emph{arXiv preprint quant-ph/9705052} (1997) }.

\bibitem{fattal2004entanglement}
D.~Fattal, T.~S. Cubitt, Y.~Yamamoto, S.~Bravyi and I.~L. Chuang,
  \emph{Entanglement in the stabilizer formalism}, {\emph{arXiv preprint
  quant-ph/0406168} (2004) }.

\bibitem{wolf2008area}
M.~M. Wolf, F.~Verstraete, M.~B. Hastings and J.~I. Cirac, \emph{Area laws in
  quantum systems: mutual information and correlations}, {\emph{Physical Review
  Letters} {\bfseries 100} (2008) 070502}.

\bibitem{van2010building}
M.~Van~Raamsdonk, \emph{Building up spacetime with quantum entanglement},
  {\emph{General Relativity and Gravitation} {\bfseries 42} (2010) 2323}.

\bibitem{qi2013exact}
X.-L. Qi, \emph{Exact holographic mapping and emergent space-time geometry},
  {\emph{arXiv preprint arXiv:1309.6282} (2013) }.

\bibitem{qi2017holographic}
X.-L. Qi, Z.~Yang and Y.-Z. You, \emph{Holographic coherent states from random
  tensor networks}, {\emph{Journal of High Energy Physics} {\bfseries 2017}
  (2017) 60}.

\end{thebibliography}\endgroup

%
%
%
%
%
%
%
%
%

\end{document}